\documentclass{aa}

\usepackage[utf8]{inputenc}  
\usepackage[dvipsnames]{xcolor}
\usepackage[T1]{fontenc}
\usepackage{lmodern}  
\usepackage{graphicx}
\usepackage{hyperref}
\usepackage{placeins}
\usepackage{float}

\usepackage{subfigure}
\usepackage{stfloats} 
\usepackage[normalem]{ulem}
\usepackage{amsmath}
\definecolor{elizabra1}{RGB}{10,200,200}

\definecolor{elizabra}{RGB}{120,100,180}
\hypersetup{
  colorlinks=true,    
  linkcolor=blue,
  citecolor=elizabra1,     
  urlcolor=Green       
}

\usepackage{txfonts}

\newcommand{\eliz}[1]{\textcolor{black}{#1}}

\numberwithin{subsection}{section}

\begin{document} 

  \titlerunning{Cosmic-ray ionization of low-excitation lines in AGN and starburst galaxies}
  \authorrunning{E. Koutsoumpou et al.}
 \title{Cosmic-ray ionization of low-excitation lines in active galactic nuclei and starburst galaxies}

\author{E. Koutsoumpou\inst{1}\fnmsep\thanks{\email{evkoutso@phys.uoa.gr}}, J. A. Fernández-Ontiveros\inst{2}, K. M. Dasyra\inst{1}, and L. Spinoglio\inst{3}
          }

   \institute{Section of Astrophysics, Astronomy \& Mechanics, Department of Physics, National and Kapodistrian University of Athens, University Campus Zografos, GR 15784, Athens, Greece
           \and
            Centro de Estudios de F\'isica del Cosmos de Arag\'on (CEFCA), Plaza San Juan 1, E--44001, Teruel, Spain
        \and
            Istituto di Astrofisica e Planetologia Spaziali (INAF--IAPS), Via Fosso del Cavaliere 100, I--00133 Roma, Italy
            }

\date{Received September 13, 2024; accepted November 25, 2024}



\abstract
{Cosmic rays (CRs) can significantly impact dense molecular clouds in galaxies, heating the interstellar medium (ISM) and altering its chemistry, ionization, and thermal properties. Their influence is particularly relevant in environments with high CR rates, such as starburst galaxies with supernova remnants or jets and outflows in active galactic nuclei (AGN). CRs also transfer substantial energy to the ionized phase of the ISM far from the ionization source, preventing gas cooling and driving large-scale winds. In this work, we use \textsc{Cloudy} photoionization models to investigate the effect of CRs on nebular gas which is an area of study that remains relatively under-explored, mainly focusing on cold molecular gas. Our models cover a broad range of density ($1$ to $10^4\,\rm{cm^{-3}}$), ionization parameter ($-3.5 \leq \log U \leq -1.5$), and CR ionization rate ($10^{-16}\, \rm{s^{-1}}$ to $10^{-12}\, \rm{s^{-1}}$). These are compared to VLT/MUSE observations of two prototypical AGN, Centaurus A (radio-loud) and NGC 1068 (radio-quiet), and the starburst NGC 253. We find that high CR rates ($\gtrsim 10^{-13}\, \rm{s^{-1}}$) typical of AGN and strong starburst galaxies can significantly alter the thermal structure of the ionized gas by forming a deep secondary low-ionization layer beyond the photoionization-dominated region. This enhances emission from low-ionization transitions, such as [\ion{N}{ii}]$\lambda$6584\AA, [\ion{S}{ii}]$\lambda \lambda$6716,6731\AA, and [\ion{O}{i}]$\lambda$6300\AA, affecting classical line-ratio diagnostics, metallicity, and ionization estimates. Unlike pure photoionization models, AGN simulations with high CR ionization rates reproduce the Seyfert loci in Baldwin, Phillips, and Terlevich (BPT) diagrams without requiring supersolar metallicities for the narrow-line region. Additionally, star-formation simulations with high CR ionization rates can explain line ratios in the LINER domain. We propose new maximum starburst boundaries for BPT diagrams in order to distinguish regions dominated by AGN photoionization from those that could be explained by star formation in conjunction with high CR ionization rates.}

 \keywords{Galaxies: jets -- Galaxies: active -- Galaxies: starburst -- ISM: cosmic rays --  ISM: clouds}
 
 \maketitle

\section{Introduction}\label{intro}

Several mechanisms can contribute to gas ionization in galaxies. First and foremost, there is photoionization caused by ultraviolet (UV) and X-ray emission from star-forming regions \citep[e.g.][]{Evans_1985, McKee_1989} and supermassive black holes in active galactic nuclei \citep[AGN; e.g.][]{2004aGroves,2004bGroves,Gallimore_2004,Wolfire_2022}. Additionally, AGN- and starburst-driven outflows produce shock waves through jets and winds that heat and compress the gas \citep{Bicknell_2000, Fragile_2004, Sutherland_2017}, resulting in excitation, ionization, and widening of the spectral line profiles. Diagnostic tools, such as the Baldwin, Phillips, and Terlevich diagrams (BPT; \citealt{1981BPT}), have been developed to analyze the gas excitation mechanisms in emission-line galaxies. The BPT diagram is built using the following optical line ratios: $[\ion{O}{iii}]\lambda5007\rm \mathring{A}$/H$\beta$, $[\ion{N}{ii}]\lambda6584\rm \mathring{A}$/H$\alpha$, $[\ion{S}{ii}]\lambda\lambda 6716,6731\rm \mathring{A}$/H$\alpha$, and $[\ion{O}{i}]\lambda6300\rm \mathring{A}$/H$\alpha$. These diagrams have since been complemented with the theoretical ``maximum starburst limit'' according to photoionization models \citep{2001Kewley} in order to identify AGN-dominated galaxies, an empirical separation between star-forming galaxies and Seyfert-\textsc{H\,ii} composite nuclei \citep{2003Kauf}, and an empirical division between Seyfert and low-ionization nuclear emission-line region (LINER) galaxies \citep{2006Kewley,Schawinski_2007}.

BPT diagrams are crucial in deciphering the complex physical processes of ionized nebulae and have been used to make detailed comparisons between observations and predictions from shock and photoionization models \citep{Evans_1985,Dopita_1986,Dopita_1995}.
\eliz{In spite of that, to reconcile the distribution of optical line ratios observed for AGN galaxies with photoionization models in BPT diagrams, supersolar metallicities ($2 - 4\, \rm{Z_\odot}$) have been usually invoked in the past \citep[e.g.][]{2004aGroves, 2004bGroves}. Other studies have shown that the observed line ratios could, at least partially, be explained by slightly supersolar metallicities ($1.6-2.6 \, \rm{Z_\odot}$), along with steeper UV slopes in the ionizing continuum \citep{Feltre_2016}, and a hotter accretion disk continuum, or a higher N/O abundance \citep{Zhu_2023}.
However, subsolar to solar abundances are obtained for the narrow-line region (NLR) in nearby AGN using photoionization models when additional optical line ratios, beyond those involved in the BPT, are considered \citep{Perez_2019,Perez_Diaz_2021}, or when infrared lines are included \citep{Perez_Diaz_2022}. These estimates also agree with semi-empirical calibrations of UV lines \citep{Dors_2019}.}

This discrepancy between models and observations can be caused by uncertainties in the ionizing continuum \citep{Zhu_2023} and/or additional sources of excitation, such as shocks \citep{Allen_2008}. In this regard, a more elusive source of gas excitation, whose influence on the BPT line ratios has not yet been thoroughly studied, are cosmic rays (CRs). These are highly energetic particles mostly produced by supernova remnants (SNRs) and accreting black holes \citep{Blasi_2013, Padovani_2017, Veilleux_2020, Wolfire_2022, Kantzas_2023} able to penetrate deeply into massive molecular clouds and deposit their energy at much greater depth than photons \citep{McKee_1989, Padovani_2018}, determining the chemistry and physics of the cold interstellar medium \citep[ISM;][]{Dalgarno_2006}. As they traverse through the ISM, the infiltration of low-energy CRs ($\lesssim 1\, \rm{GeV}$) is capable of producing excitation and ionization from secondary electrons due to collisions with atoms and molecules deep within the cloud \citep{Spitzer_1968, Gredel_1989, Gabici_2022}. Thus, CRs are expected to have a direct influence on nebular emission lines \citep{1984Ferland,Walker_2016}, especially in galaxies with strong star formation activity \citep{Phan_2024} or close to AGN jets where a dense flow of these particles is carried along \citep{Pacholczyk_1970, Guo_2011}. 

While other excitation mechanisms like UV and X-ray photoionization and shocks have attracted most of the attention in previous studies of ionized gas \citep[e.g.][]{Allen_2008,Ferland_2017,Chatzikos_2023}, the more difficult detection and characterization of CRs in galaxies has prevented a more detailed analysis of the impact that these particles have on the physical conditions and the stratification of the nebular gas. Nevertheless, understanding the effects that CRs have on the ISM of galaxies is essential in order to characterize their net contribution to feedback processes and therefore to galaxy evolution \citep{Ruszkowski_2017,Buck_2020,Peschken_2023,Lin_2023}.

In the present work, we use photoionization simulations to investigate the impact of CRs on the ionization of the nebular gas in different AGN and star-formation scenarios, covering a large parameter space in terms of CR ionization rate, gas density, and intensity of the ionizing radiation. For this purpose, three main prototypical galaxies were selected in order to compare the effects of CRs in different environments where they are expected to have a relevant contribution: an AGN with prominent radio jets, Centaurus A, an AGN with star-forming activity, NGC 1068, and the nearby starburst galaxy, NGC 253. Additionally, the photoionization-dominated NLR in NGC 1320 is included for comparison. The main properties and related data of the three selected galaxies are presented in Section \ref{GALAXIES}. We detail our models and the procedure for examining specific regions of interest on each galaxy in Section \ref{methods}. In Section \ref{results}, we present the main results of this study, and in Section \ref{discussion}, we discuss the implications and possible future directions. Finally, our conclusions are summarized in Section \ref{summary}.

\section{Observational data}\label{GALAXIES}

\subsection{Galaxy sample}
\begin{table}[!t]
\small
\caption{Observational parameters for the selected galaxies, including coordinates, redshift values, $z$, and redshift-independent distances, $D$, provided by the \textit{NASA/IPAC Extragalactic Database}\protect\footnotemark.}
\label{tab:sample}\centering          
\begin{tabular}{c c c c c} 
\hline\\[-0.2cm] 
\textbf{Galaxy} & \multicolumn{2}{c}{\textbf{Coordinates (J2000.0)}} &$\boldsymbol{z}$ &  $\boldsymbol{D}$ \\
                & \textbf{RA} & \textbf{Dec} & & \textbf{[Mpc]} \\
\hline               
Centaurus A & 13h 25m 27.6s & -43° 01' 09" & 0.0018 & 3.8 \\
NGC 1068    & 02h 42m 40.7s & -00° 00' 48" & 0.0038  & 10.6 \\
NGC 253     & 00h 47m 33.1s & -25° 17' 18" & 0.0008 & 3.2 \\
NGC 1320    & 03h 24m 48.7s & -03° 02' 32" & 0.0093 & 37.7 \\
\hline                  
\end{tabular}
\vspace{-0.3cm}
\end{table}

\footnotetext{The NASA/IPAC Extragalactic Database (NED)
is operated by the Jet Propulsion Laboratory at the California Institute of Technology under contract with NASA, \url{https://ned.ipac.caltech.edu}.}

The purpose of this study is to examine whether CRs are capable of producing the observed emission line ratios. For this analysis, we have carefully selected galaxies \eliz{with publicly available VLT/MUSE observations in the European Southern Observatory (ESO) Science Archive Facility (see Section \ref{muse_datacubes}),} with distinct characteristics related to CRs, originating either from strong radio-loud jets or SNRs associated with intense star formation. To effectively examine the influence of CRs, it is crucial to choose galaxies that are representative of these phenomena.

To strengthen our case, we examine Centaurus A, a well-studied jetted radio-loud AGN with a prominent jet \citep{Hardcastle_2003,Kraft_2008,Tingay_2009}. This is also the nearest jetted AGN, making it an ideal subject to study jet-induced CRs. We include in our study NGC 253, the nearest starburst galaxy which as of recently has direct CR rate measurements \citep{Behrens_2022, Holdship_2022, beck}, most likely associated with the CR component from SNRs. NGC 1068 has both an AGN and a starburst component whose composite photoionized spectrum from the UV to the far-IR has been successfully modeled \citep{spinoglio2005}. It serves as an intermediate example hosting both jets \citep{Bland_Hawthorn_1997, Mutie_2024} and star-forming regions \citep{Garcia_2016}, which makes it a particularly informative target for observing the effects of both jet and supernova related CRs. The effects of CRs in the far-IR molecular lines of $\rm H_{2}O^{+}$ and $\rm OH^{+}$ have been reported in the analysis of its submillimeter line spectrum \citep{spinoglio2012}. 

Furthermore, as a control source, NGC 1320, is used to counterbalance these cases. It stands out from the other galaxies, which are mostly influenced by CRs, because of its low CR-related behavior and characterized by a highly photoionized gas. For each of the aforementioned galaxies additional details are given in Table \ref{tab:sample} and additional relevant literature is provided in the Appendix \ref{appendix_gal}.

These galaxies offer an excellent basis to assess the effects of CRs, considering that the expected CR presence is significant in these systems. All of these sources, lying in close proximity (see Table \ref{tab:sample}), have been observed by the Very Large Telescope (VLT) Multi Unit Spectroscopic Explorer (MUSE) providing high spatial resolution optical line imaging, offering high-quality data necessary for this study (see Table \ref{datacubes} for observational details). All in all, the selected galaxy sample guarantees a thorough examination in a variety of galactic environments.

\subsection{MUSE data}\label{muse_datacubes}

Raw astronomical data have to be carefully processed to obtain verified and reliable products suitable for the final analysis.This is done via specific software, the pipeline, which involves several steps. VLT-MUSE data are processed by the ESO MUSE pipeline which performs a bias correction to remove camera electronic offsets, dark subtraction to compensate for the thermal noise of the detector, flat fielding to balance variability in detector sensitivity and illumination and background subtraction to remove ambient light and improve object visibility. 
The pipeline also includes flux calibration, which converts counts to standard flux units for comparability across different observations and wavelength calibration that ensures wavelength precision vital for spectral analysis. So as to get rid of artifacts that could impair image clarity, the removal of glitches due to the impact of CRs in the detector systems is also conducted. Finally, spectral extraction isolates the object's spectra from 2D images, allowing detailed scientific analysis. The MUSE data processing procedures in different operational stages are detailed in \cite{Weilbacher_2014, Weilbacher_2020}. 

In our analysis we make use of MUSE datacubes of Centaurus A, NGC 1068, NGC 253, and NGC 1320 which are publicly available in the ESO archive. For more specific information on the data reduction done on the galaxies of this work, see \cite{MING}, \cite{Venturi} and \cite{HUMIRE} for Centaurus A, NGC 1068 and NGC 253, respectively. An overview of the datasets utilized in this paper is provided in Table \ref{datacubes}.


\begin{table*}[!!!htb]
\caption{Description of MUSE datacubes used.}\label{datacubes}      
\centering          
\begin{tabular}{c c c c c c} 
\hline\\[-0.2cm] 
\textbf{Galaxy} & \textbf{Program Id} & \multicolumn{2}{c}{\textbf{FoV}} & \textbf{Angular Resolution} & \textbf{Observation Date} \\ 
                &                    & \textbf{["]}  & \textbf{[kpc]}  & \textbf{["]}            &  \\[0.05cm]
\hline\\[-0.3cm]                    
Centaurus A & 094.B-0321(A) & 63.4" x 63.2" & 2.47 x 2.47 & 1.4" & 2015-02-27 \\
NGC 1068    & 094.B-0321(A) & 64.6" x 63.8" & 5.29 x 5.16 & 0.9" & 2014-10-06 \\
NGC 253     & 0102.B-0078(A) & 87.4" x 87.2" & 1.51 x 1.51 & 0.9" & 2018-11-07 \\
NGC 1320    & 108.229J.001  & 63.6" x 63.8" & 12.51 x 12.55 & 0.7" & 2022-01-07 \\
\hline                  
\end{tabular}
\vspace{-0.3cm}
\end{table*}

\section{Methods}\label{methods}
\subsection{Spectrum extraction}\label{apertures}
\begin{figure*}[!t]
    \centering
    \subfigure[Centaurus A, H$\alpha$ line map.]{\includegraphics[width=0.33\textwidth]{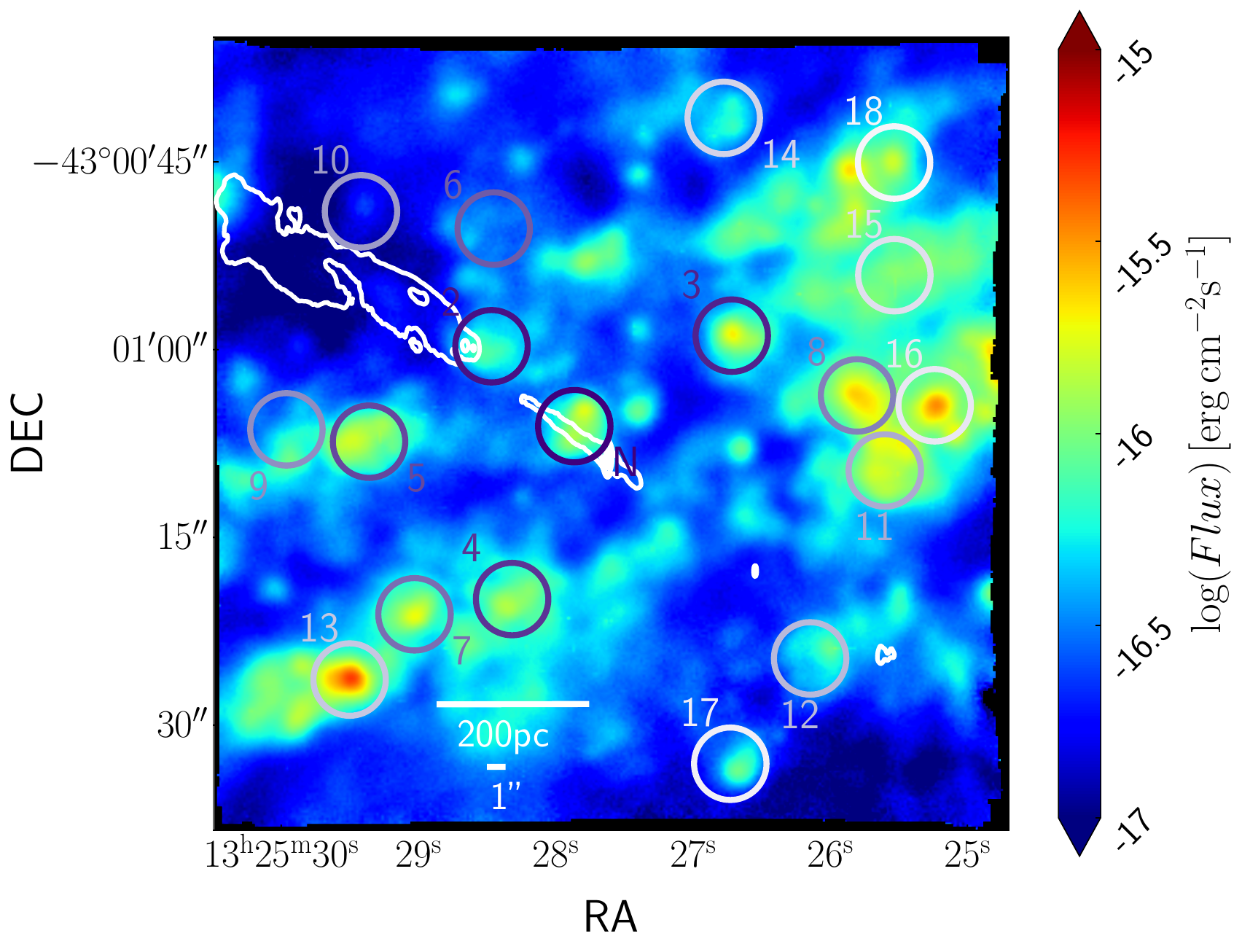}\label{subfig:ha_aper_cent}}~
    \subfigure[NGC 1068, H$\alpha$ line map.]{\includegraphics[width=0.33\textwidth]{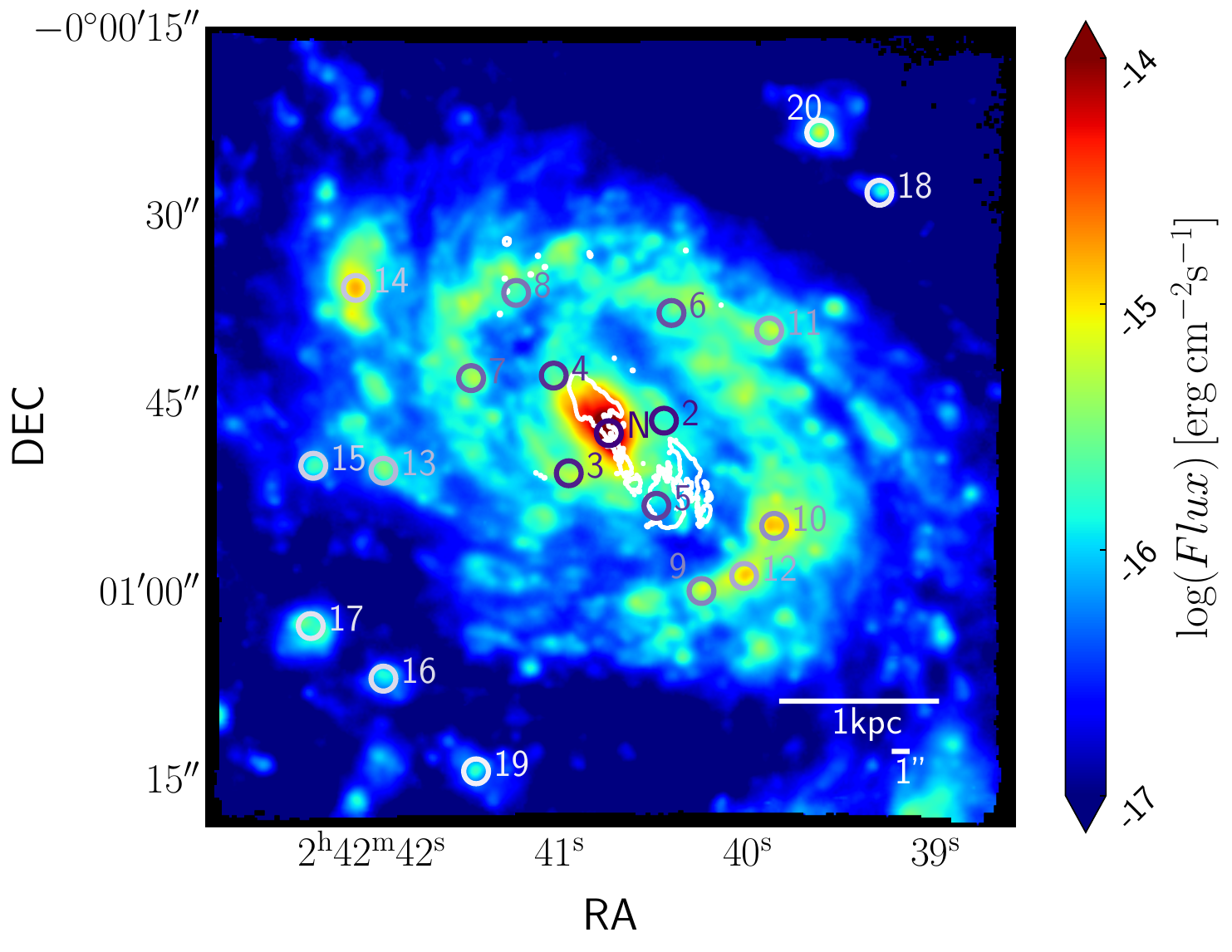}\label{subfig:ha_aper_68}}~
    \subfigure[NGC 253, H$\alpha$ line map.]{\includegraphics[width=0.33\textwidth]{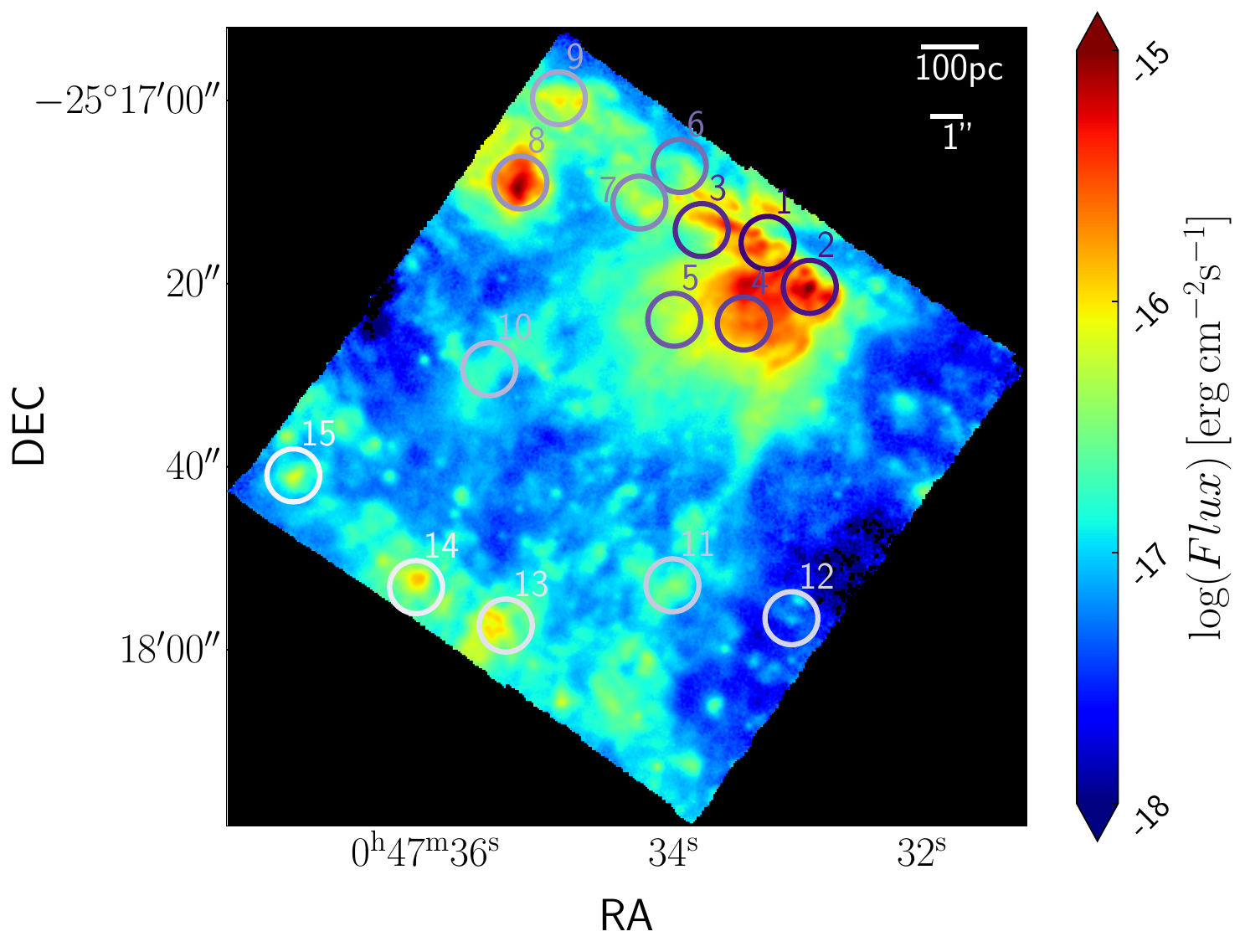}\label{subfig:ha_aper_253}}
    \subfigure[{Centaurus A, [\ion{O}{iii}]$\lambda$5007\AA \ line map.}]{\includegraphics[width=0.33\textwidth, keepaspectratio]{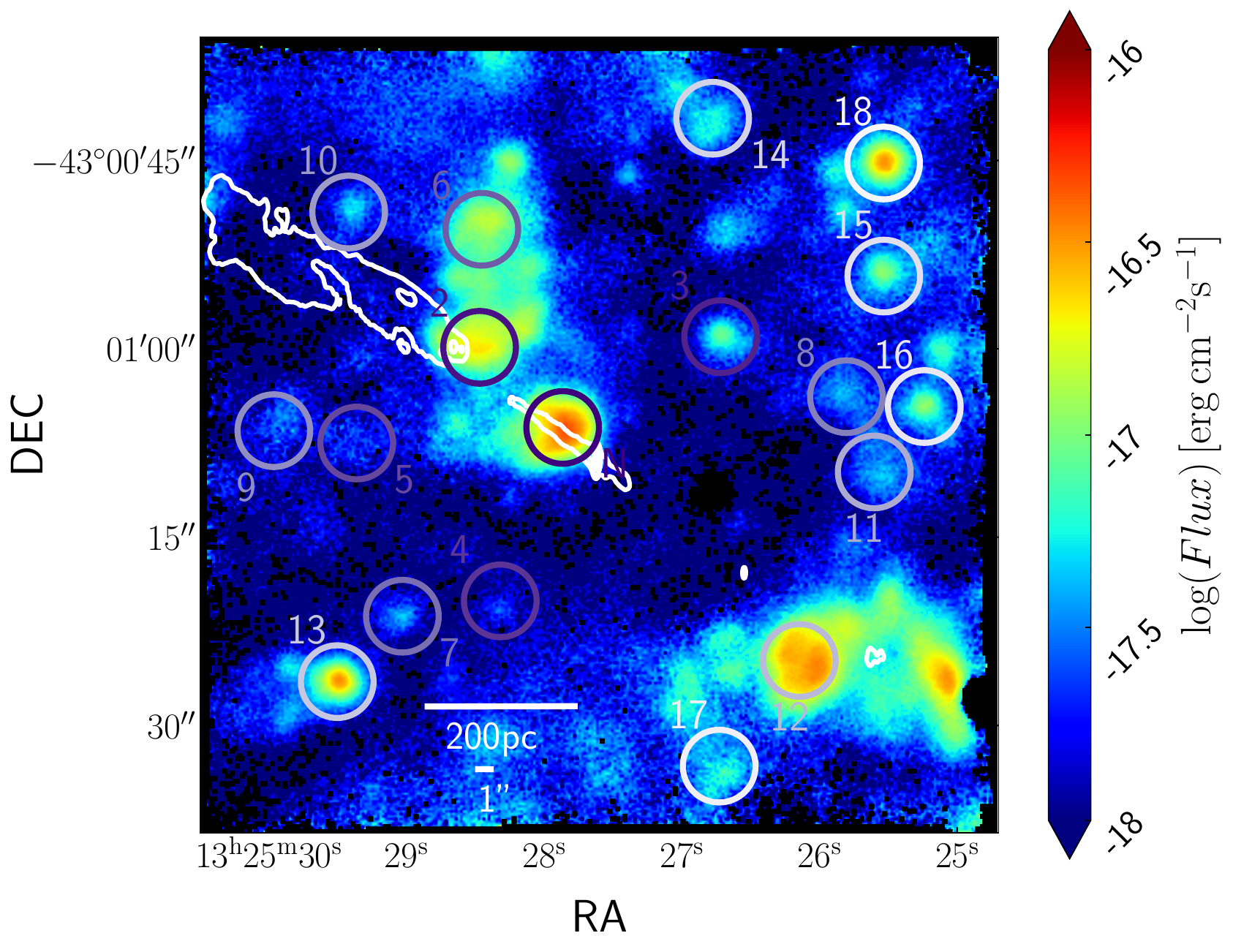}\label{subfig:o3_aper_cent}}~
    \subfigure[{NGC 1068, [\ion{O}{iii}]$\lambda$5007\AA \ line map.}]{\includegraphics[width=0.33\textwidth, keepaspectratio]{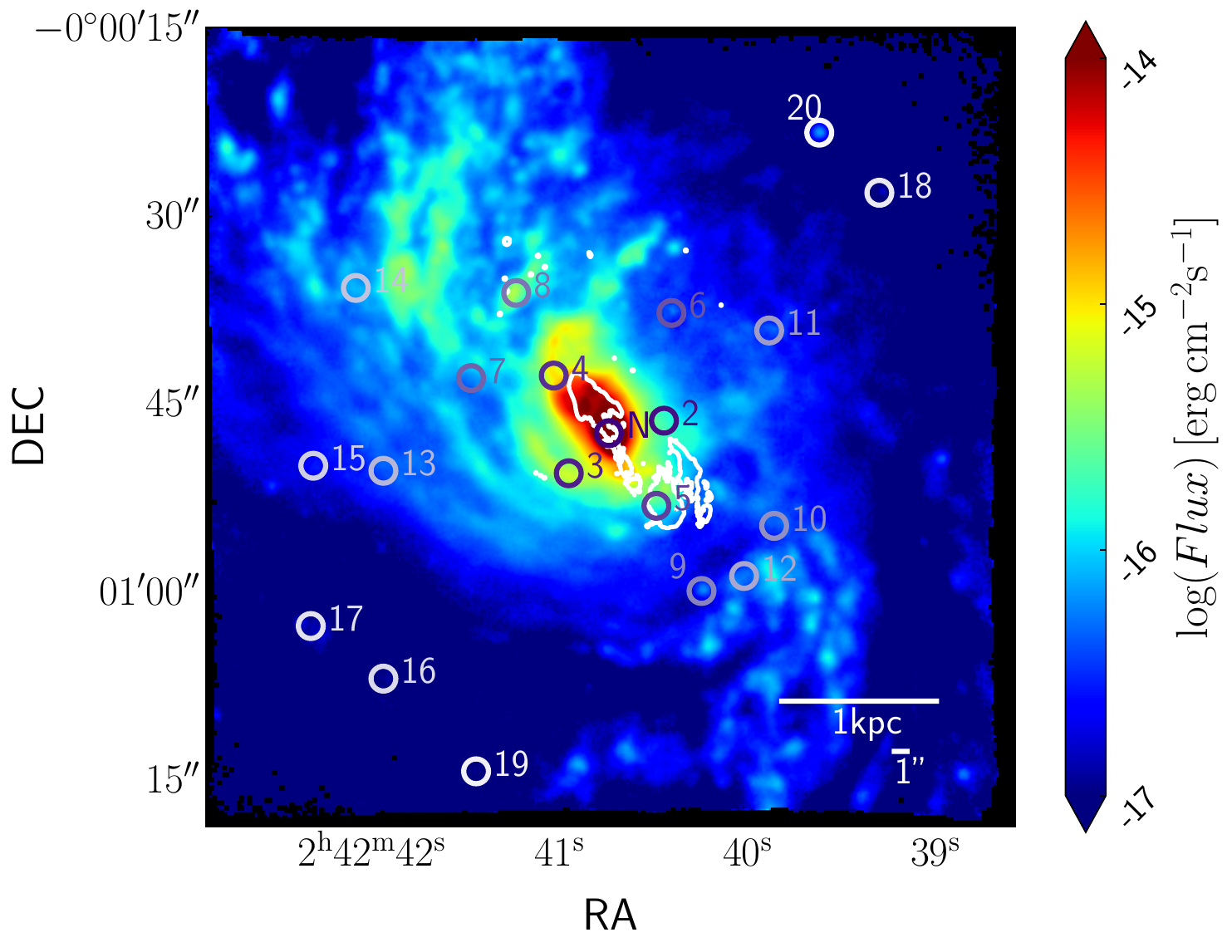}\label{subfig:o3_aper_68}}~
    \subfigure[{NGC 253, [\ion{O}{iii}]$\lambda$5007\AA \ line map.}]{\includegraphics[scale=0.25]{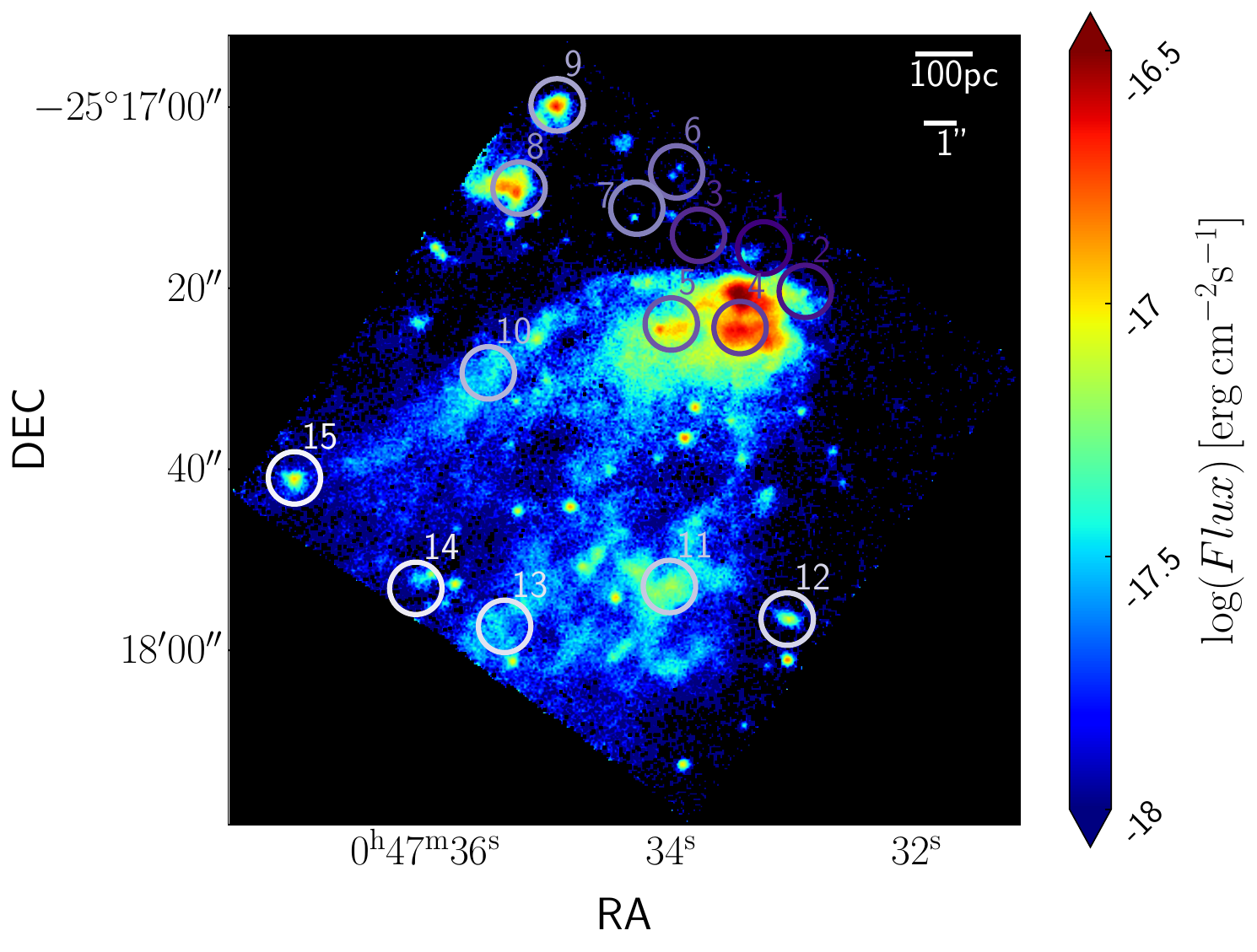}\label{subfig:o3_aper_253}}
    \caption{Apertures chosen to extract spectra with MPDAF. The different shades of purple going from deep purple to pale lilac represent the ascending distance, and are also noted with numbers. The apertures are drawn over H$\alpha$ and [\ion{O}{iii}]$\lambda$5007\AA \, emission with the stellar continuum subtracted. In Centaurus A, the jets are depicted by use of VLA data at 8.4 GHz \citep{Hardcastle_2003,Tingay_2009}, in white contours at levels  $7\times 10^{-4}\rm Jy/beam$, $ 10^{-2}\rm Jy/beam$, $0.1 \rm Jy/beam$, and $6 \,\rm Jy/beam$. Similarly, for NGC 1068, the jets are illustrated using combined data from e-MERLIN and the VLA at 5 GHz \citep{Gallimore_1996,Muxlow_1996,Mutie_2024}, in white contours at levels $10^{-4}\rm Jy/beam$, $5\times 10^{-2}\rm Jy/beam$, and $10^{-2}\rm Jy/beam$. All radio data were aligned with optical data based on astrometry.} \label{fig:chosen_apertures_o3}
    \vspace{-0.3cm}
\end{figure*}
The primary goal of this work is to explore the parameters that influence the excitation of gas across the various regions of galaxies, with a particular focus on the effects of CRs along with photoionization. Our method relies on selecting specific regions within the aforementioned galaxies and extracting the spectra of the MUSE datacubes from circular apertures. This technique allows an in depth examination of gas excitation within different galactic environments by use of the spatially resolved spectral information provided by the datacube of each galaxy. Each and every aperture is positioned strategically in order to understand the influence of CRs in star-forming regions and areas affected by the jets in the AGN cases.

\begin{figure*}[ht]
    \centering
    \subfigure[H$\alpha$.]{\includegraphics[width=0.31\textwidth]{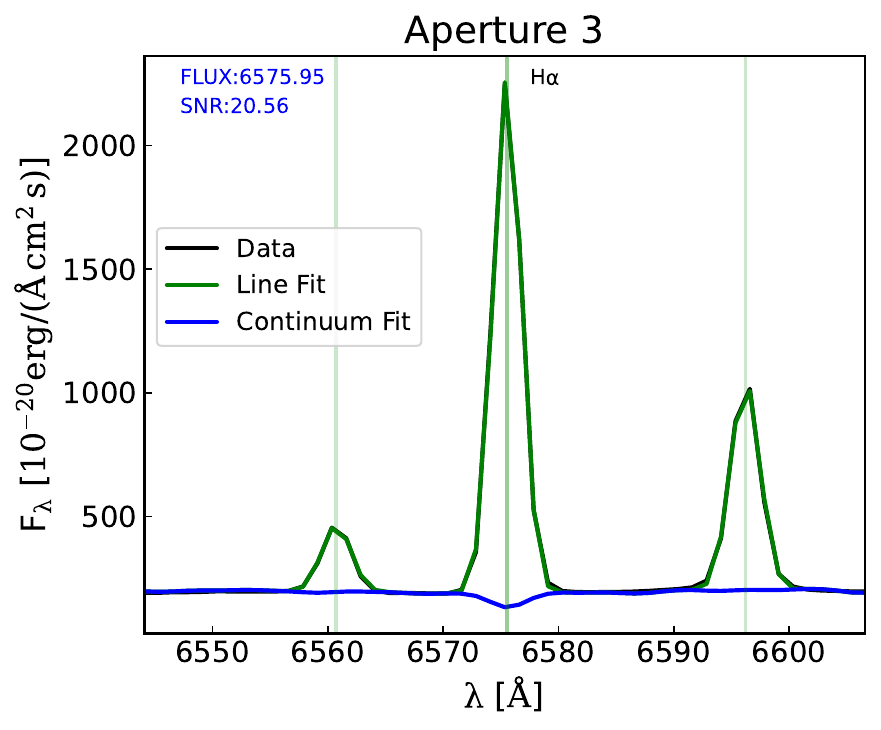}\label{subfig:ha_cent}}~
    \subfigure[{[\ion{N}{ii}]}$\lambda$6584\AA.]{\includegraphics[width=0.31\textwidth]{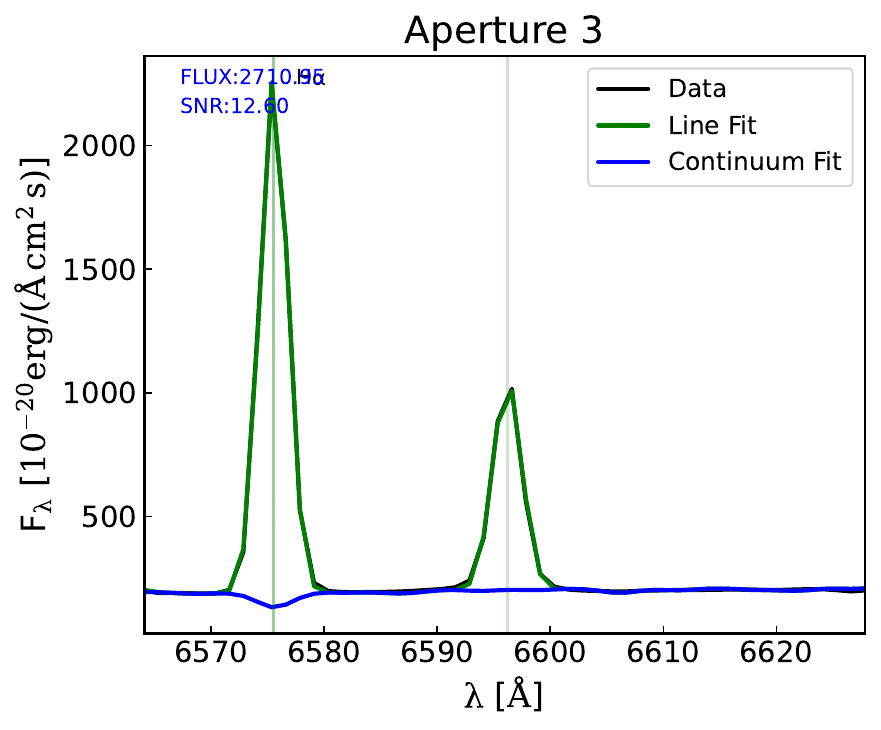}\label{subfig:n2_cent}}~
    \subfigure[{[\ion{S}{ii}]}$\lambda \lambda$6716,6731\AA.]{\includegraphics[width=0.31\textwidth]{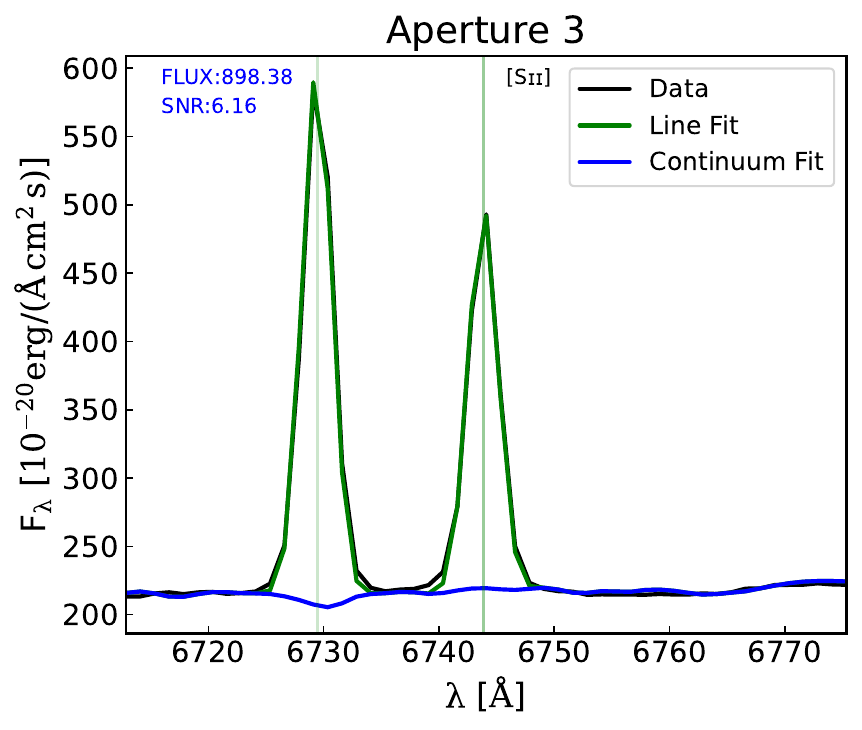}\label{subfig:s2_cent}}
    \subfigure[{[\ion{O}{i}]}$\lambda$6300\AA.]{\includegraphics[width=0.31\textwidth]{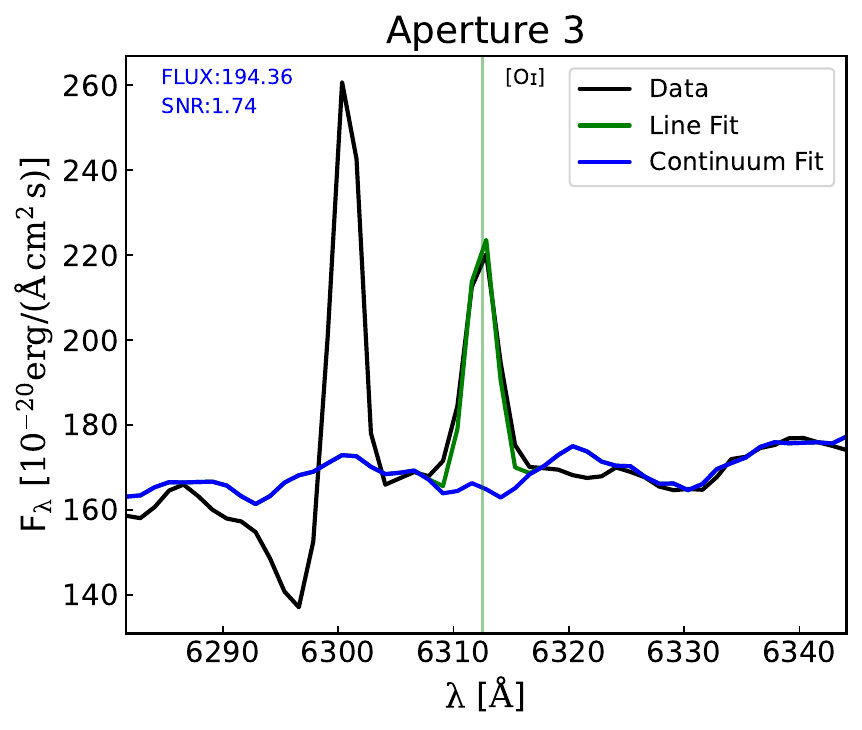}\label{subfig:o1_cent}}~
    \subfigure[{[\ion{O}{iii}]}$\lambda$5007\AA.]{\includegraphics[width=0.31\textwidth]{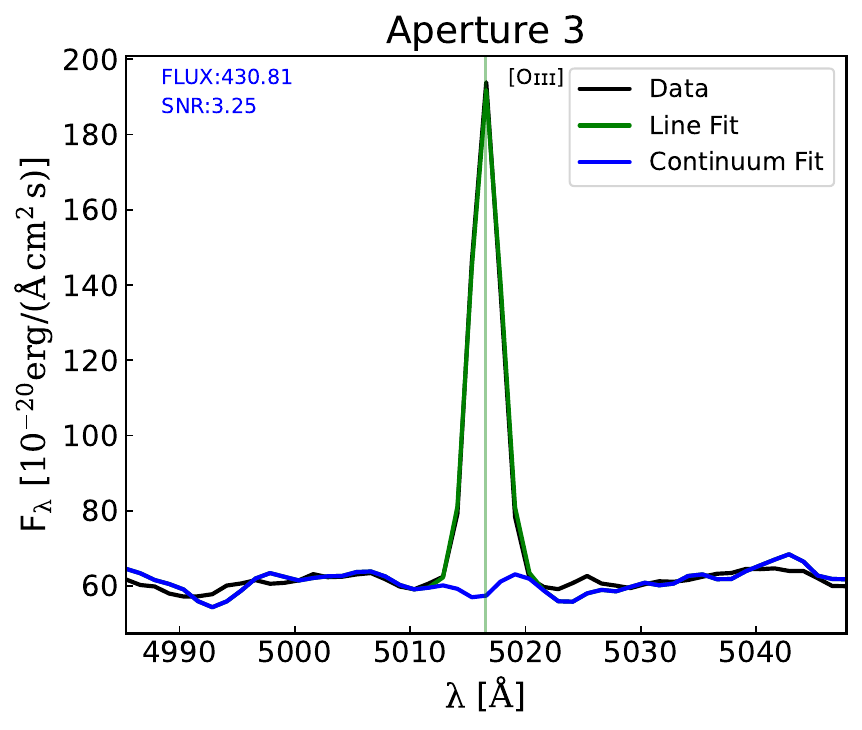}\label{subfig:o3_cent}}
    \caption{BPT emission-line fits, using Pyplatefit \citep{pypla}, in the rest frame of Centaurus A.}\label{fig:cent_bpt_lines_apertures}

    \centering
    \subfigure[Centaurus A.]{\includegraphics[width=0.31\textwidth]{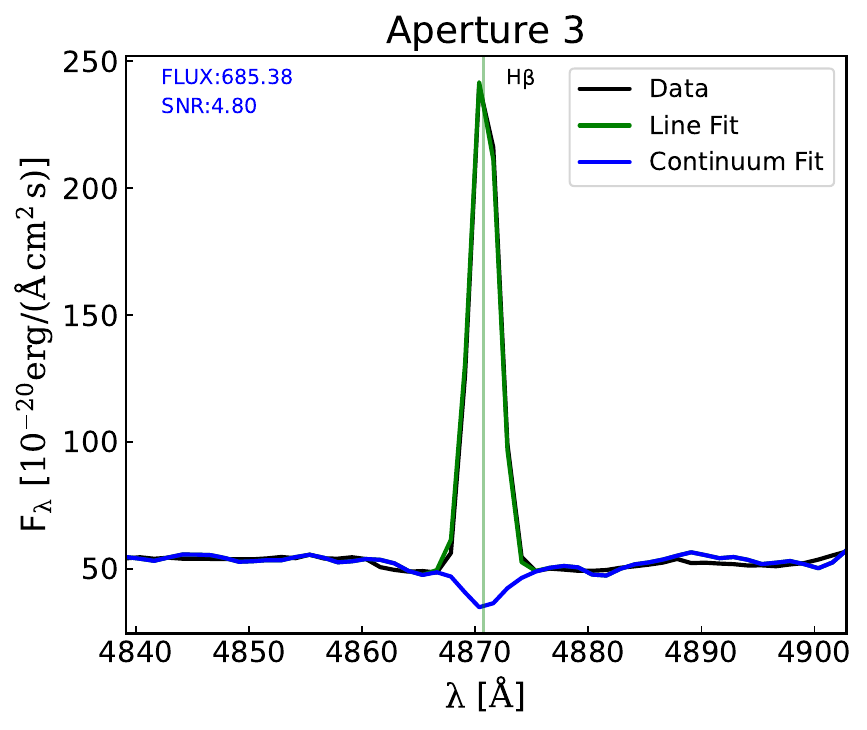}\label{subfig:hb_aper_cent}}~
    \subfigure[NGC 1068.]{\includegraphics[width=0.31\textwidth]{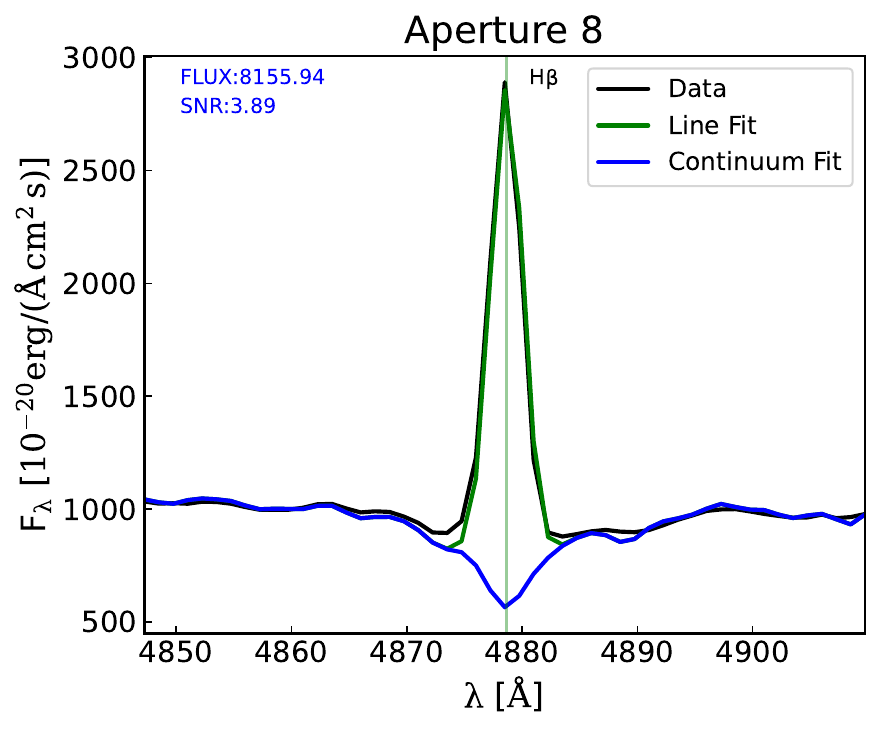}\label{subfig:hb_aper_1068}}~
    \subfigure[NGC 253.]{\includegraphics[width=0.31\textwidth]{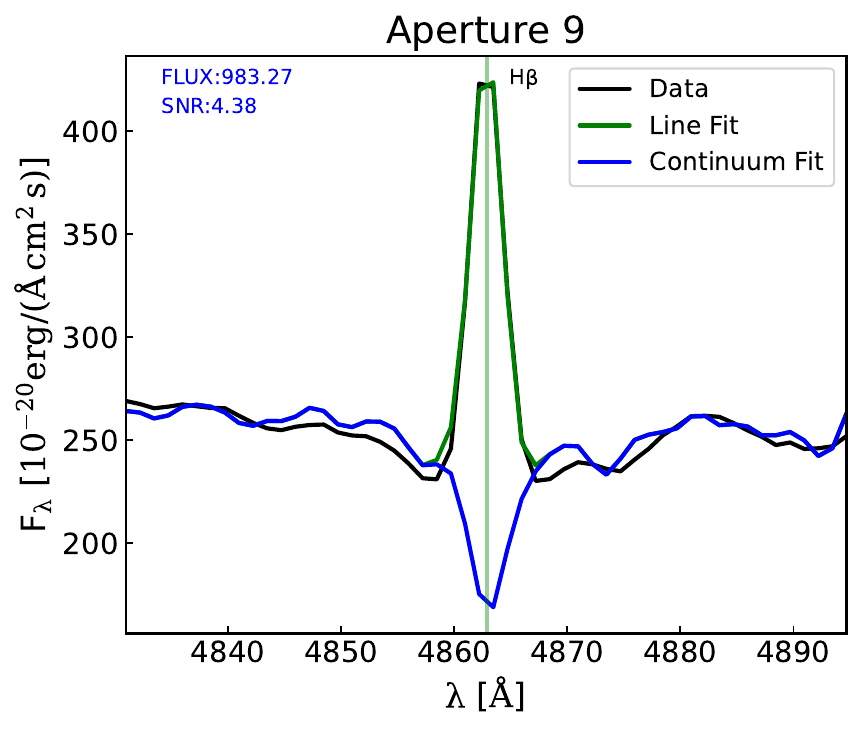}\label{subfig:hb_aper_253}}
    \caption{H$\beta$ emission line in each galaxy's rest frame, fitted with Pyplatefit \citep{pypla}.}\label{fig:hb_apertures}
    \vspace{-0.3cm}
\end{figure*}

We predict that areas in proximity of the AGN jets are mainly impacted by the CRs related to these phenomena. As opposite, regions situated near star-forming regions and farther from the galactic center are presumed to be dominated by photoionization from UV radiation produced by hot young stars. We intend to distinguish the corresponding contributions of photoionization and CRs in various regions of each galaxy by collecting and analyzing the spectra associated within the different apertures selected. 

In the cases of Centaurus A and NGC 1068, we select apertures containing gas that interacts with the jets/outflows, as well as potential sites for star formation. In NGC 253 we inspect regions close to the central area of the galaxy, some of which are affected by strong stellar winds and others are potential sites of star formation. To achieve this, we identify regions that exhibit strong emission in both [\ion{O}{iii}]{$\lambda$5007\AA} and H$\alpha$ along with the other emission lines required to create the BPT diagrams, namely [\ion{N}{ii}]$\lambda6584$\AA, [\ion{S}{ii}]$\lambda\lambda6716,6731$\AA~ and [\ion{O}{i}]$\lambda6300$\AA.~ The BPT diagrams and their applications are introduced in Section \ref{intro} and fully detailed in Section \ref{results}.

In particular, for our analysis we implemented 18 apertures of a $\sim 50$\,pc radius in the vicinity of Centaurus A, 15 of the same radius in the vicinity of NGC 253, and 20 apertures of a $\sim 80$pc radius in the vicinity of NGC 1068, as the MUSE datacube field of view of NGC 1068 in kpc is the largest among the three galaxies considered (see Table \ref{datacubes}). From all the apertures of each galaxy we extracted the convolved spectrum using MPDAF \citep{MPDAF}. We also chose one aperture of $\sim 200$\,pc radius from the central area of NGC 1320, to establish the position of highly photoionized gas on the BPT diagrams to compare with the rest of our models and observations. 

We present the selected apertures overplotted on the H$\alpha$ and [\ion{O}{iii}]$\lambda$5007\AA \ emission line maps in Fig.~\ref{fig:chosen_apertures_o3}. Figs.~\ref{subfig:ha_aper_68} and \ref{subfig:o3_aper_68} show the apertures in NGC 1068, Figs.~\ref{subfig:ha_aper_cent} and \ref{subfig:o3_aper_cent} correspond to Centaurus A, while Figs.~\ref{subfig:ha_aper_253} and \ref{subfig:o3_aper_253} correspond to NGC 253.
We note that NGC 253 lacks of the ``N'' aperture, since this is not an AGN. In this case, the apertures are numbered from 1 to 15, with 1 being the closest to the central region aperture, as shown in Figs.~\ref{subfig:ha_aper_253} and \ref{subfig:o3_aper_253}. In Centaurus A and NGC 1068, the jet direction is very well defined by radio continuum maps \citep{Dufour_1978,McKinley_2018,Joseph_2022,Gallimore_1996, Gallimore_2004}. In Figs.~\ref{subfig:ha_aper_cent} and \ref{subfig:o3_aper_cent} we show the VLA map at $8.4\, \rm{GHz}$ provided by \citet{Tingay_2009}. In the case of NGC 1068, Figs.~\ref{subfig:ha_aper_68} and \ref{subfig:o3_aper_68} show the combined $5\, \rm{GHz}$ continuum map from VLA and e-Merlin, provided by \citet{Mutie_2024}.

The emission lines detected in the convolved spectra are fitted with \textsc{Pyplatefit} \citep{pypla}, a Python package that fits the emission and absorption lines of MUSE spectra, inspired by the IDL routine \textsc{Platefit} \citep{Tremonti_2004, Brinchmann_2004}. \eliz{In this package, continuum subtraction is performed by fitting and subtracting simple stellar population models \citep{Bruzual_2003,Brinchmann_2013}.} Subsequently, the key emission lines utilized in the BPT diagrams are fitted in each galaxy's rest frame wavelengths. We present the respective fit of the BPT lines coming from Aperture 2 in Centaurus A as shown in Fig. \ref{fig:cent_bpt_lines_apertures}. The fit of H$\beta$ for the 3 galaxies is also presented in Fig. \ref{fig:hb_apertures}. As H$\beta$ is a highly absorbed line and from the fits presented, it is clear that there is attenuation that needs to be addressed (see the discussion in Section  \ref{Cloudy_models}).

In particular, when examining the central region of NGC\,1068, we found broad-line components associated with $\sim100$--$200\, \rm{km\,s^{-1}}$ outflows known in the vicinity of the nucleus \citep{Garcia_2016}. In order to address this issue in the nuclear aperture N, we applied a multicomponent fitting using the \textsc{lmfit} package \citep{newville_2015_11813}. \eliz{In this case, the continuum was subtracted after fitting a Chebyshev polynomial with \textsc{Specutils} \citep{earl2024astropyspecutils}, which additionally smooths the spectrum using a median filter prior to the fit, to mitigate the effects of noise and remove significant stellar and AGN contributions.} We fitted both narrow and wide emission line components with Gaussian functions. This procedure enabled us to acquire reliable measurements of the flux for every emission-line component, and specifically of the emission-line fluxes used in BPT diagrams that are necessary for the study of the central region in NGC 1068.

\begin{table}[!t] 
\caption{Range of \textsc{HII-CHI-Mistry} estimated values for the ionization parameter, the oxygen abundance, and the N/O relative abundance of the extracted regions, including the physical size of the aperture.}\label{tab:parameters}   
\centering
\setlength{\tabcolsep}{5.pt}
\begin{tabular}{c c c c c} 
\hline\\[-0.2cm] 
\textbf{Galaxy} & $\mathbf{\log  U}$ & $\mathbf{Z/Z_{\odot}}$ & $\mathbf{\log (N/O)}$ & \textbf{Radius} \\	
& &  &  & \textbf{[pc]} \\	
\hline                    
Centaurus A   & [-3.5, -3.2] & [0.4, 1.2] & [-1.3, -0.8] & 50 \\ 
NGC 1068      &  [-3.4,-2.8]  & [0.5, 1.1] & [-1.2, -0.7] & 80 \\ 
NGC 253       & [-3.1,-2.6]  & [0.4, 0.8] & [-0.9, -0.7] & 50 \\ 
NGC 1320      & [-3.4, -2.9]   & [0.6, 1.1]   & [-1.2, -0.8]   & 200 \\ 
\hline                  
\end{tabular}
\vspace{-0.3cm}
\end{table}

\subsection{Constraining the parameter space}\label{subsec:param_space}

Given that the parameter space is fairly large, we have to restrict the main parameters before simulating the cloud physical conditions. One of the main parameters under study in the present work is the ionization parameter, $U$ defined as the ratio of ionizing photons to the gas density and expressed as \citep{AGN3}:
\begin{equation}
    U = \frac{Q}{4 \pi r^2 n_{\rm H} c}, 
\end{equation}
where \( Q \) is the photon emission rate, \( r \) is the distance to the source, \( n_{\rm H} \) is the gas density, and \( c \) is the speed of light. The other main parameters relevant to this study are the initial hydrogen density $n_{\rm H}$, the metallicity $Z$, and the CR ionization rate $\zeta_\mathrm{CR}$.  

As mentioned in Section~\ref{intro}, studying the excitation mechanisms is a strenuous task especially when having in mind all these factors that can affect the emission lines.
Constraining the ionization parameter is necessary for grasping the roles that photoionization and its competitive excitation gas process, that of CRs, play in our models.
With respect to $n_{H}$, our models cover a wide range of values in order to cover a variety of regions with different densities. 
It is, moreover, important to constrain the metallicity and have realistic values of it, as it has been a usual practice to use unrealistic supersolar metallicities \eliz{to effectively model observations of Seyfert galaxies or NLR \citep{2004bGroves}.}

Metallicity is an essential parameter in understanding galaxy evolution through the enrichment of the ISM. Metals are created in the cores of stars, transported by convection, and dispersed by outflows and supernovae (SNe). The abundance ratios of oxygen or nitrogen relative to hydrogen are considered indicators of metallicity. Since oxygen is abundant, as can be detected by prominent optical emission lines, the most popular choice to measure metallicity is using: $12\, + \,\log  \rm (O/H)$ \citep{Asplund, Maiolino_2019}. The N/O ratio also serves to differentiate between metals coming from helium burning in stellar cores and already existent in the ISM components \citep{Van_Zee_1998}. 

We are able to constrain the ionization parameter, O/H, and N/O ratios for each of the regions measured in the sample of galaxies, by employing \eliz{\textsc{HII-CHI-Mistry},\footnote{\textsc{HII-CHI-Mistry} is publicly available at:\\ \url{https://home.iaa.csic.es/~epm/HII-CHI-mistry.html}} developed by \cite{Perez_2014}, uses Bayesian-like statistics to estimate these parameters by comparing emission-line ratios with a large grid of photoionization models. The code has been adapted to use metallicity tracers in different wavelength ranges; in particular, we employed the optical versions for star-forming galaxies \citep{Perez_2014} and Seyfert nuclei \citep{Perez_2019}.}

In the optical wavelengths, \textsc{HII-CHI-Mistry} employs the following emission lines, [\ion{O}{ii}]$\lambda3727$\AA, \, [\ion{Ne}{iii}]$\lambda3868$\AA, \, [\ion{O}{iii}]$\lambda4363$\AA, \, [\ion{O}{iii}]$\lambda4959$\AA, \, [\ion{O}{iii}]$\lambda5007$\AA, \, [\ion{N}{ii}]$\lambda6584$\AA, \, and [\ion{S}{ii}]$\lambda\lambda6716,6731\,$\AA \, to determine the chemical composition and physical conditions of ionized gas. However, [\ion{O}{ii}]$\lambda3727$\AA \ and [\ion{Ne}{iii}]$\lambda3868$\AA \ are not within the MUSE spectral range, which covers 4800 to 9300\AA. Therefore, the MUSE datacubes used (Section \ref{datacubes}), limit the lines we use in the \textsc{HII-CHI-Mistry} routines to [\ion{O}{iii}]$\lambda4959$\AA, [\ion{O}{iii}]$\lambda5007$\AA, [\ion{N}{ii}]$\lambda6584$\AA, and [\ion{S}{ii}]$\lambda\lambda6716,6731$\AA. These estimates result in metallicities from slightly subsolar to $\sim 1.2 \, Z_{\odot}$ in accordance to metallicities found in already published literature \citep{Perez_Diaz_2021} making the hypothesis of $1\, \rm{Z_\odot}$ realistic. The N/O abundance ratio is also a valuable diagnostic tool to analyze galactic processes. \eliz{In our dataset, we derive relative abundances in the range of $-1.5 < \log(\text{N/O}) < -0.8$ (Table~\ref{tab:parameters}), which are very close to the solar value of $\log \text{(N/O)}_\odot \sim -0.86\, \rm{dex}$ \citep{Asplund}.}

\subsection{{\sc Cloudy} models}

\label{Cloudy_models}
 
We model the line fluxes with the photoionization and radiative transfer code \textsc{Cloudy} \citep{Ferland_2013, Ferland_2017} and make also use of the py\textsc{Cloudy} \citep{Morisset_2013}. We examine as possible excitation mechanisms 
photons, in a wide spectral range, depending on the ionizing spectrum, and CRs. With \textsc{Cloudy} we examine CRs in the sense of the CR ionization rate $\zeta_\mathrm{CR}$ integrating an electron CR component in the energy range $ 5\, \rm MeV - 10\rm\, GeV$. \textsc{Cloudy} does not include extinction and to compare the results of our simulations with observational data we transform the \textsc{Cloudy} modeled fluxes to attenuated fluxes using Calzetti's law \citep{calzetti}. 

Initially, we develop a grid of \textsc{Cloudy} models for five discrete CR ionization rates $10^{-16}\, \rm{s^{-1}}$, $10^{-15}\, \rm{s^{-1}}$, $10^{-14}\, \rm{s^{-1}}$, $10^{-13}\, \rm{s^{-1}}$, and $\,10^{-12}\, \rm{s^{-1}}$ for hydrogen densities in the $1 \leq n_{\rm H} \leq 10^4 \, \rm{cm^{-3}}$ range, with a step of $10^{0.1}\, \rm cm^{-3}$ and an ionization parameter $-3.5 \leq \log  U \leq -1.5$ with a step of 0.1 dex. \eliz{Regarding the ionization parameter $U$ values, we chose this range motivated by the estimations derived with \textsc{HII-CHI-Mistry} (see Table \ref{tab:parameters}), in agreement with values previously reported in the literature for AGN and star-forming regions \citep{Perez_2009,Perez_2019,Carvalho_2020}.} The model grids have a solar chemical composition \citep{Asplund}. However, the models with CR ionization rates $10^{-15}\, \rm{s^{-1}}$ and $ 10^{-14}\, \rm{s^{-1}}$ do not show substantial differences; therefore, we present only the $10^{-14}\, \rm{s^{-1}}$ case. Moreover, the effect of CRs in the case of $10^{-16}\, \rm{s^{-1}}$ is small and will be presented for comparison in Sec. \ref{obs_metal}. \eliz{The CR ionization rates values used in our models is supported by indirect measurements for both AGN and star-forming galaxies reported in the literature \citep{Gonz_2013,Gonz_Alf_2018,Holdship_2022,Behrens_2022}. A further justification of these CR values can be found in Section \ref{subsec:high_CRs}.}



\begin{figure}[ht!]
    \centering
    \includegraphics[width=0.9\linewidth]{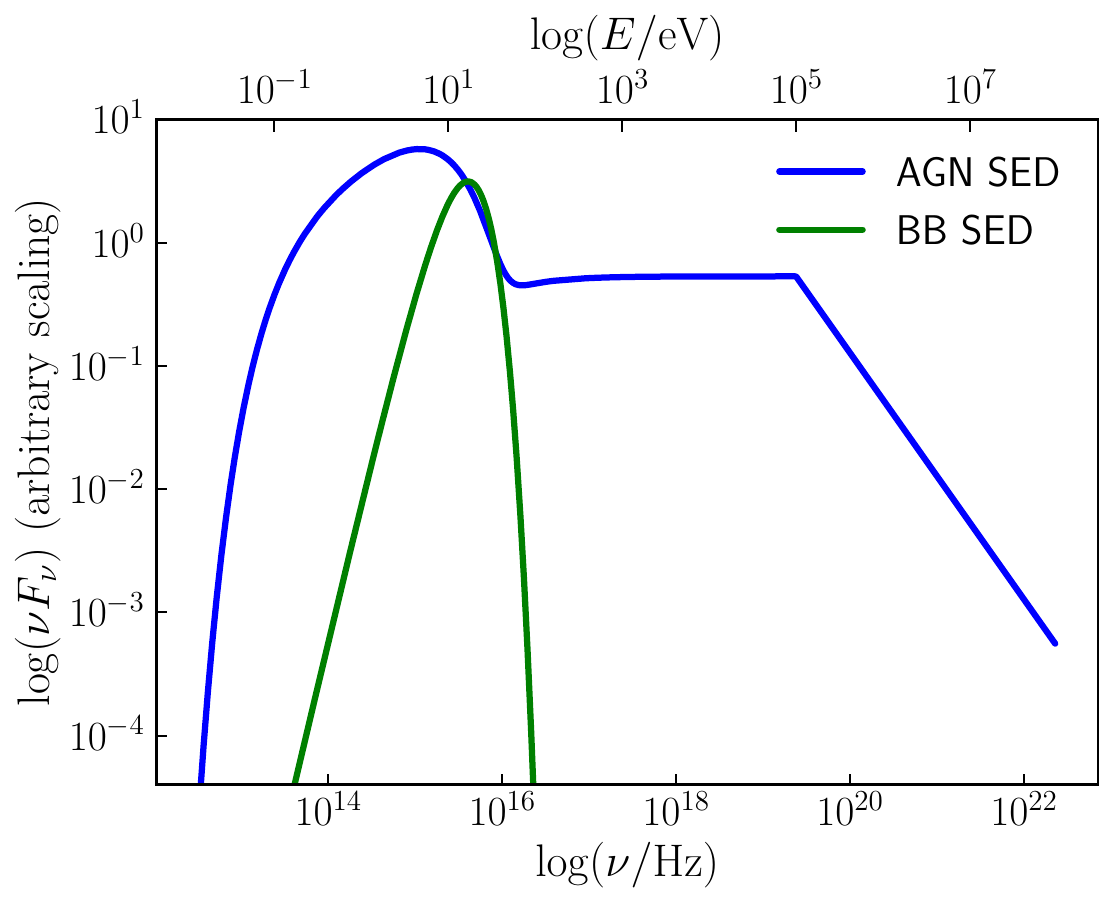}
  	\caption{AGN and BB SEDs simulated with \textsc{Cloudy}. The blue solid line represents the AGN SED used in the AGN models, while the green solid line represents the BB SED used in the SF models. The y-axis is $\nu F_\nu$ in arbitrary scaling.}
    \label{fig:SEDs}
    \vspace{-0.4cm}
\end{figure}

We employ two different models as input for the \textsc{Cloudy} simulations. AGN models are used for the cases of Centaurus A and NGC 1068, while star-forming models are used for NGC 253. The star-forming models, generated with the \textsc{blackbody} command, provide less energetic photons, as expected for a thermal continuum from stellar populations \citep{Ferland_2013}. \eliz{For simplicity, and to have better control over the shape of the incident continuum through the temperature, we adopted a black body \citep[e.g.][]{Perez_Fern_2024} instead of more sophisticated stellar population models. Nevertheless, we tested our predictions using Starburst99 models \citep{leitherer99}, and found that the differences in the BPT line ratios have a median relative difference of $\sim 7\%$, indicating a marginal deviation.} 
The AGN models, on the other hand, are simulated using \textsc{Cloudy}'s AGN command and assume a ionizing continuum typical of AGN \citep{Ferland_2013}. AGN models can effectively reproduce the intense radiation and ionizing processes in the two jetted galaxies. The shapes of both the AGN and SF ionizing continua are shown in Fig.~\ref{fig:SEDs}, with arbitrary scaling for $\nu F_\nu$.

More precisely, in the grid of models used for Centaurus A and NGC 1068, characterized as AGN models for the rest of the paper, the spectral energy distribution (SED) of the central photoionizing continuum is assumed to be consisting of a large blue bump from the accretion disc, radio emission from synchrotron radiation in the jet, infrared emission from dust, a soft X-ray excess, and a power-law component at hard X-rays. This may be expressed mathematically as the following expression: 
\begin{equation}
 F_\nu = \nu^{\alpha_{\text{uv}}} \exp\left(-\frac{h\nu}{kT_{\text{bb}}}\right)\exp\left(-\frac{kT_{\text{IR}}}{h\nu}\right) + \alpha_{\text{ox}}\nu^{\alpha_{\text{x}}},   
\end{equation}
which is typical for AGN. In this formula \(F_\nu\) is the flux density as a function of frequency \(\nu\), \(\alpha_{\text{uv}}\) and \(\alpha_{\text{x}}\) are the UV and X-ray spectral indexes respectively, \(T_{\text{bb}}\) is the representative temperature of the big blue bump, \(T_{\text{IR}}\) the temperature related to the infrared cutoff for the big blue bump, and \(\alpha_{\text{ox}}\) describes the optical to X-ray spectral index.  As a SED for our AGN models, we employed the following values, as they are both typical for AGN and are in accordance with the observational data of the two cases we study. We used \(\alpha_{\text{uv}} = -0.5\), \(\alpha_{\text{x}} = -1.0\), \(\alpha_{\text{ox}} = -1.4\), \(T_{\text{BB}} = 10^5\) K, and \(T_{\text{IR}} = 1.6 \times 10^3\) K. 

In the case of NGC 253, we simulate star-forming (hereafter SF) models using a black body (BB) with a characteristic temperature of $T_{\text{BB}} = 5 \times 10^4\, \rm K$ as ionizing continuum. Such a high temperature is representative of the hot, massive O-type stars that dominate the emission in massive young star clusters ($\sim 10^5\, \rm{M_\odot}$, $\lesssim 6\, \rm{Myr}$) as those found in the nuclear starburst of this galaxy \citep{Watson_1996,Fernandez2009,Mills_2021}. 

\eliz{In all cases, we adopted solar elemental abundances, including $12\, + \,\log \rm (O/H)_\odot = 8.69$, $\log \rm (N/O)_\odot = -0.86$, $\log \rm (C/O)_\odot = -0.3$, and $\log \rm (He/H)_\odot = 0.085$ \citep{Asplund}. These values are consistent with \textsc{HII-CHI-Mistry} abundances obtained for the different regions in our sample (Table~\ref{tab:parameters}) and with estimates obtained for the NLR of AGN in the local Universe from previous studies \citep[e.g.][]{Perez_Diaz_2021,Dors_alone_2021,Dors_2022, Perez_Diaz_2022,Perez_Diaz_2024}. On the other hand, the presence of dust in NLR clouds has been a topic of debate, with studies both supporting \citep[e.g.][]{2004aGroves,2004bGroves,Zhu_2023} and opposing it \citep[e.g.][]{Ferguson_1997,Nagao_2006,Richardson_2014}. Nevertheless, \citet{Feltre_2016} show that dust depletion in models with $1.5\, \rm{Z_\odot}$ has a small impact on BPT line ratios, while differences between dusty and dust-free models in \citet{Zhu_2023} are significant primarily at supersolar metallicities. We opted for dust-free models in our simulations, which are suggested to be more appropriate for radio galaxies \citep{Matsuoka_2009}, where the effects of CR ionisation are expected to be more important. The presence of dust in NLR clouds is expected to increase the electron temperature through photoelectric heating and a lower cooling efficiency caused by the depletion of metals from the gas phase. Predictions for dusty models will be explored in a future study discussing the effect of CRs on mid-infrared transitions.}


\section{Results}\label{results}
\subsection{BPT diagrams}\label{subsec:bpts}



Our study intends to bridge the knowledge gap and offer an alternative method to model the observational data shown in the upper AGN area on the BPT diagrams. Our primary objective is to better understand the physical mechanisms that underlie ionized nebulae and their implications on BPT plots. We compare the location of the extracted spectra for the different regions on the BPT diagrams with the predictions obtained from the photoionization along with CRs models. We present the BPT diagrams for all three galaxies in the local universe that are examined in this study: Centaurus A, NGC 1068, and NGC 253. 

\begin{figure*}[!ht]
    \centering
    \subfigure[$\zeta_\mathrm{CR}=10^{-14}\,\rm s^{-1}$.]{\includegraphics[width=\textwidth]{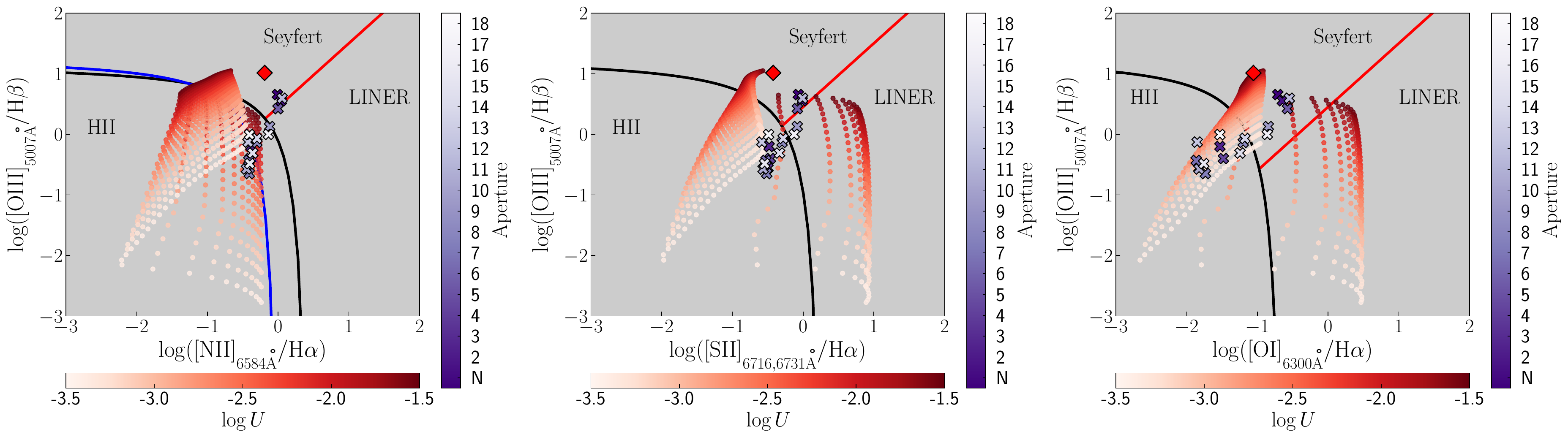}\label{subfig:U_cent_14}}
    \subfigure[$\zeta_\mathrm{CR}=10^{-13}\,\rm s^{-1}$.]{\includegraphics[width=\textwidth]{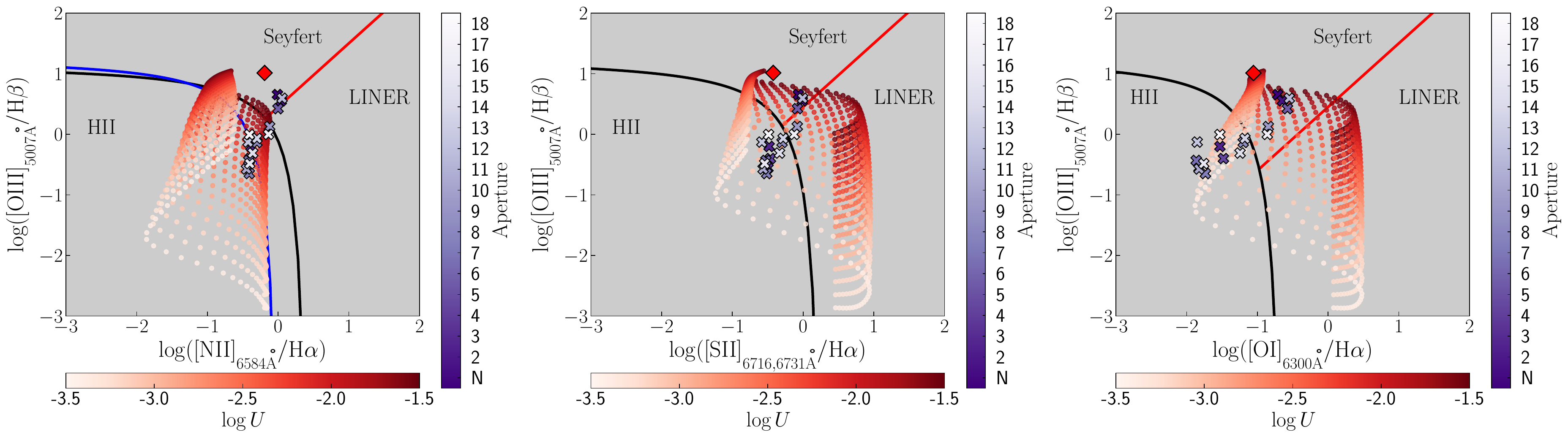}\label{subfig:U_cent_13}}
    \subfigure[$\zeta_\mathrm{CR}=10^{-12}\,\rm s^{-1}$.]{\includegraphics[width=\textwidth]{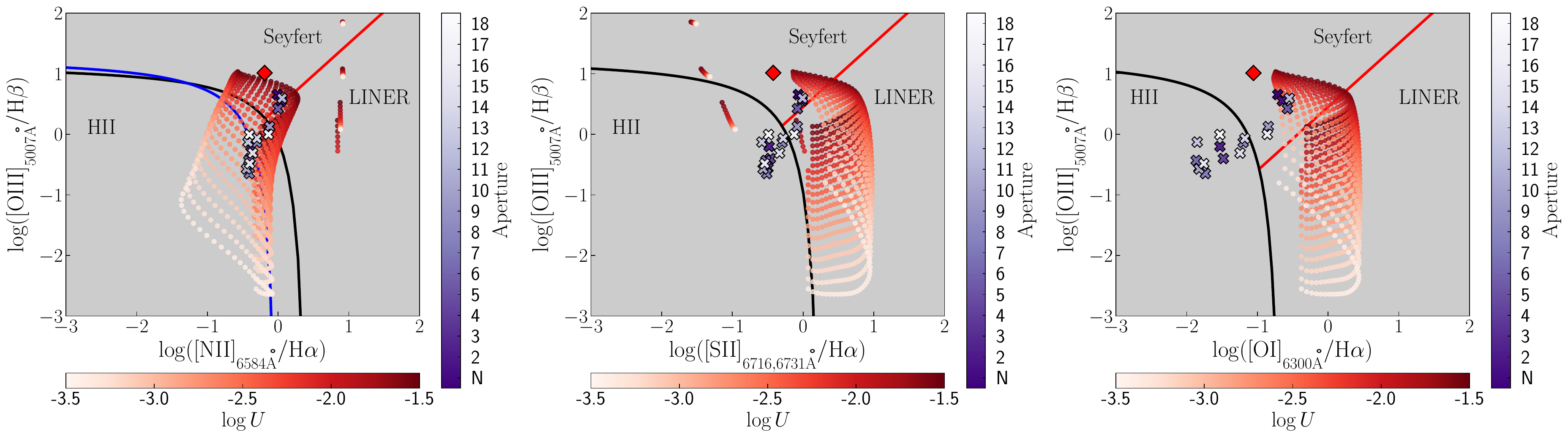}\label{subfig:U_cent_12}}
    \caption{BPT diagrams with the AGN photoionization models compared with the observations from the selected apertures in Centaurus A (Fig.~\ref{subfig:ha_aper_cent}). The BPT diagrams for [\ion{N}{ii}], [\ion{S}{ii}], and [\ion{O}{i}] are shown on the left, middle, and right, respectively. The different shades of purple going from deep purple to pale lilac/white represent the ascending distance from the nucleus, as also noted with numbers, with "N" being the closest aperture. Also, from white to deep red, the different shades of red represent the range of ionization parameter values, $-3.5\leq \log U\leq -1.5$. All the models shown have solar abundances. The red diamonds represent the measured line ratios for the photoionization-dominated Seyfert 2 nucleus in NGC 1320. The Kewley, Kauffmann, and Schawinski lines correspond to the black, blue, and red solid lines, respectively.}\label{fig:cent_BPTS_U}
    \vspace{-0.2cm}
\end{figure*}

\begin{figure*}[!ht]
    \centering
    \subfigure[$\zeta_\mathrm{CR}=10^{-14}\,\rm s^{-1}$.]{\includegraphics[width=\textwidth]{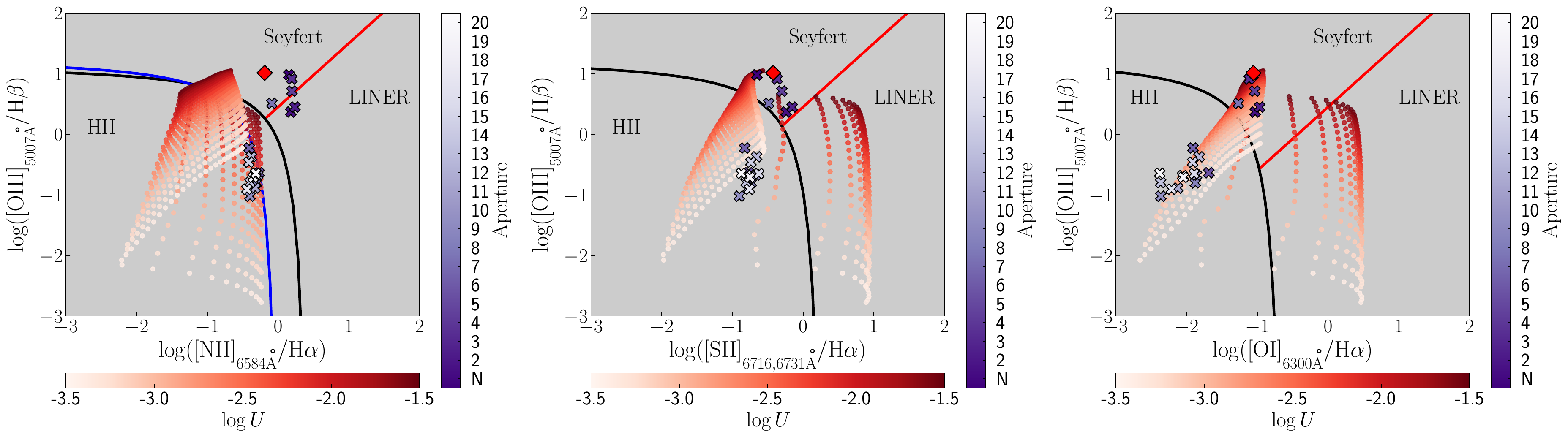}\label{subfig:U_1068_14}}
    \subfigure[$\zeta_\mathrm{CR}=10^{-13}\,\rm s^{-1}$.]{\includegraphics[width=\textwidth]{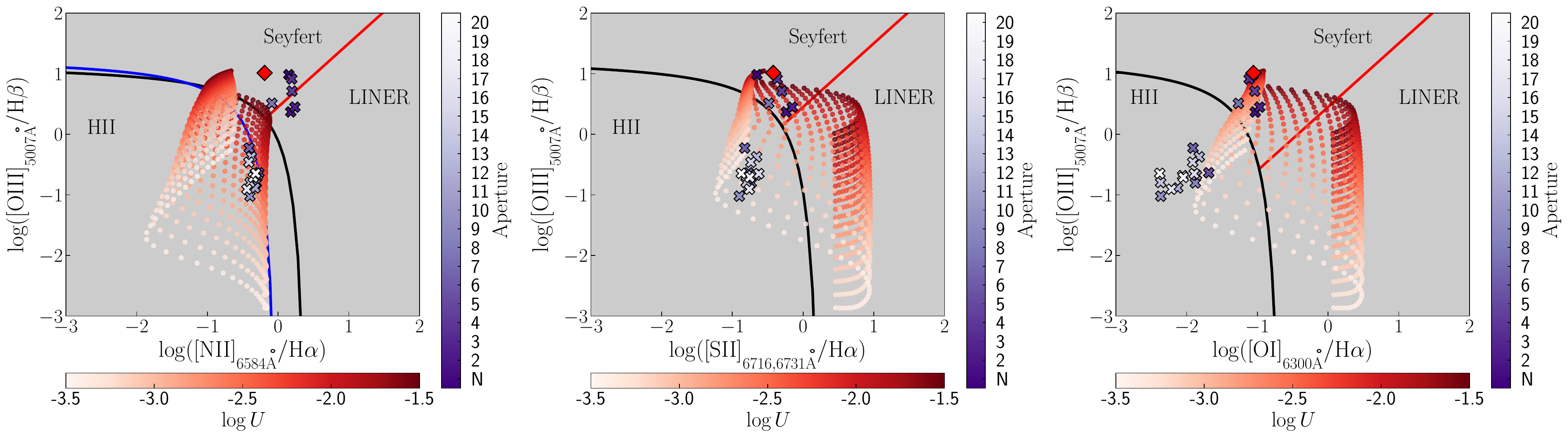}\label{subfig:U_1068_13}}
    \subfigure[$\zeta_\mathrm{CR}=10^{-12}\,\rm s^{-1}$.]{\includegraphics[width=\textwidth]{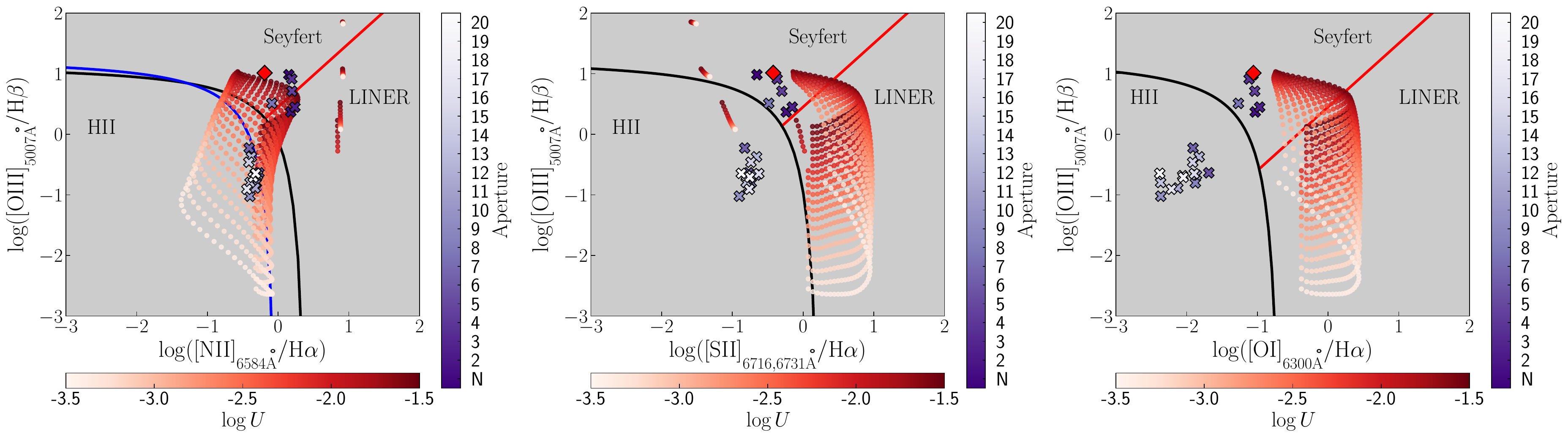}\label{subfig:U_1068_12}}
    \caption{BPT diagrams with the AGN photoionization models compared with the observations from the selected apertures in NGC 1068 (Fig.~\ref{subfig:ha_aper_68}). 
    The BPT diagrams for [\ion{N}{ii}], [\ion{S}{ii}], and [\ion{O}{i}] are shown on the left, middle, and right, respectively. The different shades of purple going from deep purple to pale lilac/white represent the ascending distance from the nucleus, as also noted with numbers, with "N" being the closest aperture. Also, from white to deep red, the different shades of red represent the range of ionization parameter values, $-3.5\leq \log U\leq -1.5$. All the models shown have solar abundances. The red diamonds represent the measured line ratios for the photoionization-dominated Seyfert 2 nucleus in NGC 1320. The Kewley, Kauffmann, and Schawinski lines correspond to the black, blue, and red solid lines, respectively.}\label{fig:1068_BPTS_U}
        \vspace{-0.2cm}
\end{figure*}

\begin{figure*}[!ht]
    \centering
    \subfigure[$\zeta_\mathrm{CR}=10^{-14}\,\rm s^{-1}$.]{\includegraphics[width=\textwidth]{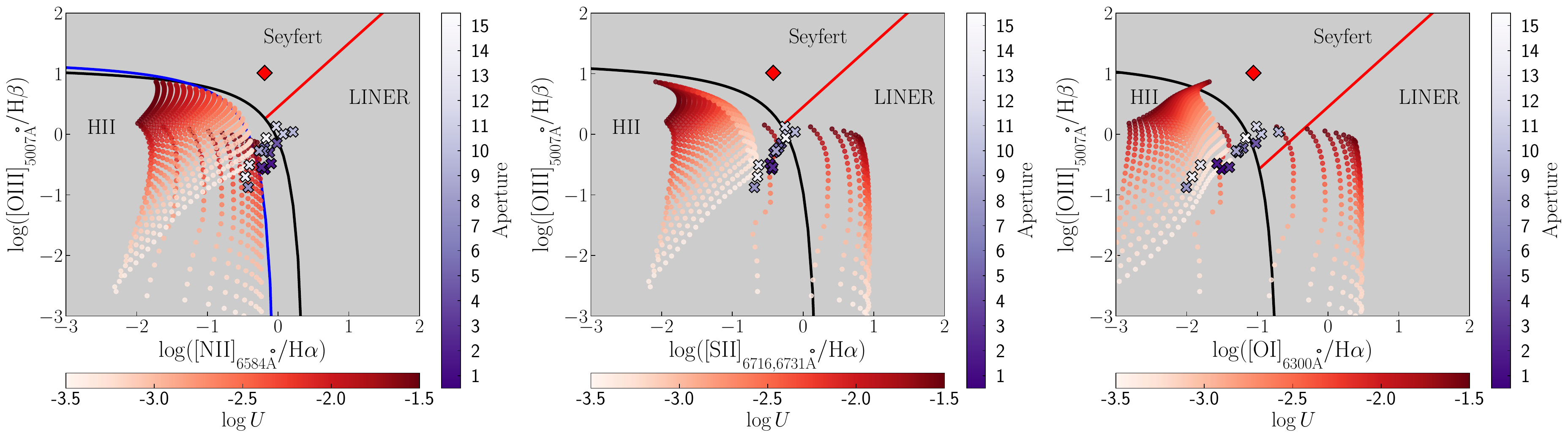}\label{subfig:U_253_14}}
    \subfigure[$\zeta_\mathrm{CR}=10^{-13}\,\rm s^{-1}$.]{\includegraphics[width=\textwidth]{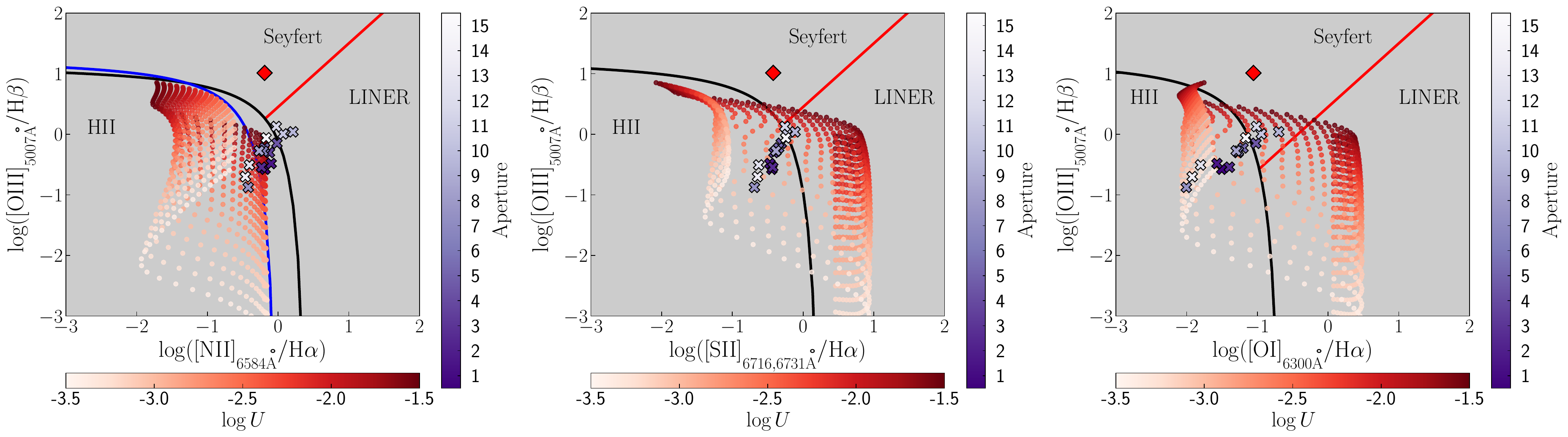}\label{subfig:U_253_13}}
    \subfigure[$\zeta_\mathrm{CR}=10^{-12}\,\rm s^{-1}$.]{\includegraphics[width=\textwidth]{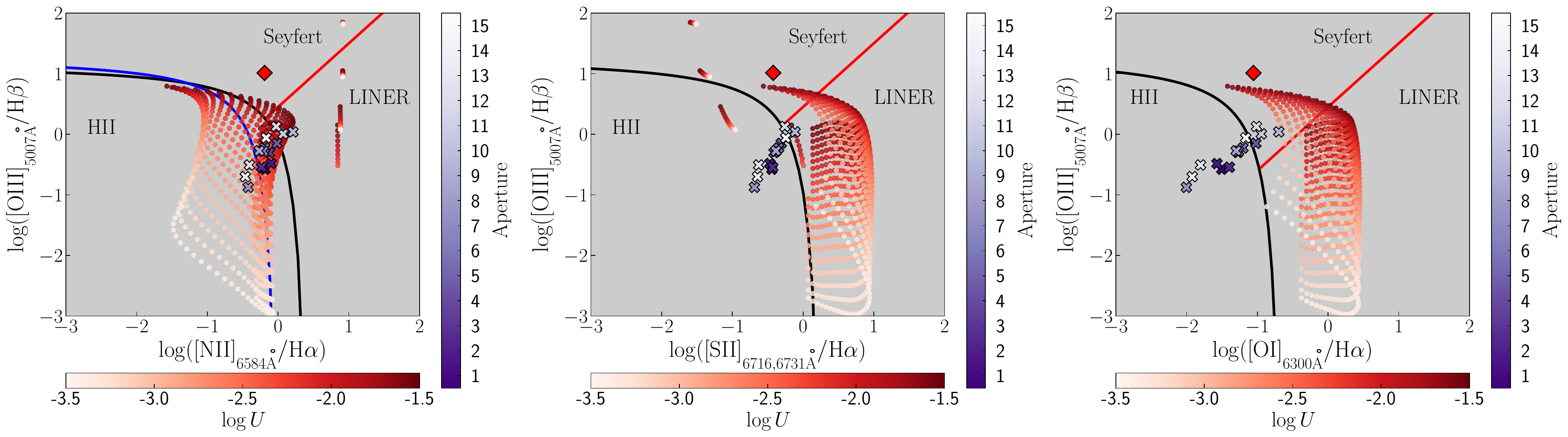}\label{subfig:U_253_12}}
    \caption{BPT diagrams with the SF photoionization models compared with the observed line ratios from the selected apertures in NGC 253 (Fig.~\ref{subfig:ha_aper_253}). The BPT diagrams for [\ion{N}{ii}], [\ion{S}{ii}], and [\ion{O}{i}] are shown on the left, middle, and right, respectively. The different shades of purple going from deep purple to pale lilac/white represent the ascending distance, as also noted with numbers, with "1" being the most central aperture. Also, from white to deep red, the different shades of red represent the range of ionization parameter values, $-3.5\leq \log U\leq -1.5$. All the models shown have solar abundances. The red diamonds represent the measured line ratios for the photoionization-dominated Seyfert 2 nucleus in NGC 1320. The Kewley, Kauffmann, and Schawinski lines correspond to the black, blue, and red solid lines, respectively.}\label{fig:253_BPTS_U}
        \vspace{-0.2cm}
\end{figure*}

The observational data of Centaurus A and NGC 1068 are compared to the AGN models, while those of NGC 253 are compared to the SF models; all models are presented in Section \ref{Cloudy_models}. The emission line ratios of all galaxies create data points on the BPT diagrams depicted as deep purple to white crosses that are located both in star-forming and composite regions, under the Kewley division line \citep{2006Kewley}, as well as above it, in the Seyfert/LINER area of the diagrams. The color of the crosses indicates their physical distance from either the nucleus of their host AGN or from the most central area in the starburst case, going from deep purple being the closest one to white being the most distant one and accompanied as well by the ascending aperture number. The apertures closest to the nucleus, depicted in darker shades of purple,  tend to be in the Seyfert/LINER area of the BPT above the Kewley line, whereas the areas further away from the nuclei of both AGN, have emission line ratios that locate crosses in the star-forming area of the BPT diagrams (Figs. \ref{fig:cent_BPTS_U}, \ref{fig:1068_BPTS_U}). In NGC 253, we find that even though this is not an AGN, there are apertures with line ratios similar to those of AGN (Fig. \ref{fig:253_BPTS_U}), which means that the gas in the vicinity of NGC 253 has similar properties to gas in AGN. Additionally, there are several areas that exhibit star-forming characteristics in the two jetted AGN, Centaurus A and NGC 1068 (Figs. \ref{fig:cent_BPTS_U}, \ref{fig:1068_BPTS_U}).

In Fig. \ref{fig:cent_BPTS_U} as well as in Fig. \ref{fig:1068_BPTS_U}, we see the parameter space covered by our models. These models cover the ionization parameter range $-3.5 \leq \log  U \leq -1.5$ from white, being the lower ionization parameter, to dark red, being the highest (Figs. \ref{fig:cent_BPTS_U}, \ref{fig:1068_BPTS_U}, and \ref{fig:253_BPTS_U}). These models also span between $0 \leq \log  n_{\rm H}\leq 4$ from white,  being the lower initial hydrogen density, to forest green, being the highest, as shown in Appendix \ref{appendix_den}.
These particular diagrams are produced for different CR ionization rates, from the top to the bottom in the figures: $-14 \leq {\rm log~\zeta_\mathrm{CR}\leq -12}$ for all galaxies; the instance of Centaurus A can be found in Figs. \ref{subfig:U_cent_14}, \ref{subfig:U_cent_13}, and \ref{subfig:U_cent_12}, respectively. 

Not only the AGN models shown in Figs. \ref{fig:cent_BPTS_U}, and \ref{fig:1068_BPTS_U} but the SF models as well, shown in Figs. \ref{fig:253_BPTS_U}, are relocated in the right area, the AGN area, of the BPT diagrams as the CR rate grows higher. These models mainly fall on the LINER part of the AGN region. As a general result (Figs. \ref{fig:cent_BPTS_U}, \ref{fig:1068_BPTS_U}, and \ref{fig:253_BPTS_U}), analyzing in detail the distribution of the models, it is clear that as the CR rate grows, there is more emission on [\ion{N}{ii}]$\lambda$6584\AA, [\ion{S}{ii}]$\lambda \lambda$6716,6731\AA, and  [\ion{O}{i}]$\lambda $6300\AA. This extra emission is behind the relocation of models in the AGN locus of the BPT diagrams. 

It is also evident that for each of the three aforementioned emission lines there is a density over which the emission of the line falls and then begins to increase again. This can be seen in \ref{fig:cent_BPTs_nh}, \ref{fig:1068_BPTs_nh}, and  \ref{fig:253_BPTs_nh} where the models seem to fold and change direction over a certain density. This is due to the effect of the CRs that can penetrate the clouds deep enough and ionize the gas in these regions but over a certain column density where the gas is also denser and more shielded they gradually lose their ability to excite.


The impact of CRs on ionized gas seems to be a plausible mechanism behind the emission line ratios in the AGN locus of the BPT diagrams in the absence of a strong radiation field. In the case of starburst nuclei high enough CR rates are able to model AGN-like emission in the vicinity of its host without assuming an AGN continuum. For CR ionization rate $\sim 10^{-13}\, \rm s^{-1}$ we acquire AGN and SF models capable of reproducing quite well the observations in the BPT with [\ion{S}{ii}]$\lambda \lambda$6716,6731\AA, for all the cases studied (Figs. \ref{subfig:U_cent_13}, \ref{subfig:U_1068_13}, and \ref{subfig:U_253_13}). Then, for $10^{-12}\rm s^{-1}$ there are models in the far right area of the [\ion{N}{ii}]$\lambda$6584\AA \ BPT diagrams for both AGN and SF models (Figs.~\ref{subfig:U_cent_12}, \ref{subfig:U_1068_12}, and \ref{subfig:U_253_12}). Moreover, we find that the highest CR rates $10^{-12}\, \rm s^{-1}$, fit the observed emission in the [\ion{N}{ii}]$\lambda$6584\AA \ BPT for both AGN and SF models in all galaxies (Figs.~\ref{subfig:U_cent_12}, \ref{subfig:U_1068_12}, and \ref{subfig:U_253_12}).

Finally, in the [\ion{O}{i}]$\lambda $6300\AA \ BPT diagram, AGN models match the observations for $10^{-13}\, \rm s^{-1}$ in the case of Centaurus A (Fig.~\ref{subfig:U_cent_13}), while the observed line ratios in NGC 1068 could be explained with $10^{-14}\, \rm s^{-1}$ (Fig.~\ref{subfig:U_1068_13}). SF models also require $10^{-13}\, \rm s^{-1}$ to extend the simulated [\ion{O}{i}]/H$\alpha$ ratios beyond the star-forming limit and reproduce the observed ratios in NGC 253 (Fig.~\ref{subfig:U_253_13}). Most of the SF models with $10^{-14}\, \rm s^{-1}$ do not reach Seyfert/LINER loci in the BPT, which are occupied by some of the regions in NGC 253. At $10^{-12}\, \rm s^{-1}$, both AGN and SF models overestimate the observed [\ion{O}{i}]/H$\alpha$ ratios. 
The ionization of low-excitation lines by CRs will be further discussed in Section~\ref{subsec:NOS_lines}.

It is also noteworthy that models with the highest CRs ionization rates, together with $\log  U \gtrsim -2.0$ and $ \log  n_{\rm H} \gtrsim 1 $, drive both the AGN and SF models in the Seyfert/LINER area of the [\ion{N}{ii}]$\lambda$6584\AA \ BPT diagrams (Figs. \ref{subfig:U_cent_12}, \ref{subfig:U_1068_12}, \ref{subfig:U_253_12}, \ref{subfig:nh_cent_12}, \ref{subfig:nh_1068_12}, and \ref{subfig:nh_253_12}). Overall, our photoionization models show that even the highest ionization parameter values, depicted with dark red circles, are not sufficient to reach the Seyfert/LINER domain of the BPT without the presence of a high enough CR ionization rate. This behavior is closely related to the sensitivity of low-ionization emission lines, such as [\ion{S}{ii}]$\lambda \lambda$6716,6731\AA, [\ion{N}{ii}]$\lambda$6584\AA, and [\ion{O}{i}]$\lambda $6300\AA \ to the low-energy CRs. The effects of CRs on these emission lines will be discussed in Section~\ref{subsec:structure_plots}, \eliz{while a comparison of these results with other studies using photoionization models is presented in Section~\ref{obs_metal}.}

Finally, in Centaurus A and NGC 1068 BPT diagrams (see Figs. \ref{fig:cent_BPTS_U}, and \ref{fig:1068_BPTS_U}), apertures located closer to the nucleus and the jets, in darker shades of purple, are mainly positioned in the LINER/Seyfert area and better reproduced by models of the higher CR ionization rates $10^{-13}\rm s^{-1}$--$10^{-12}\rm s^{-1}$. This seems logical considering that due to their position, they are probably affected by the energetic particles originating from the jets, something that it is to be further discussed in Sec. \ref{subsec:high_CRs}.

\subsection{Gas stratification diagrams} \label{subsec:structure_plots}

\begin{figure*}[!htp]
    \centering
    \subfigure[Temperature, AGN, $n_{\rm H}=100\,\rm{cm^{-3}}$.]{\includegraphics[width=0.33\textwidth] {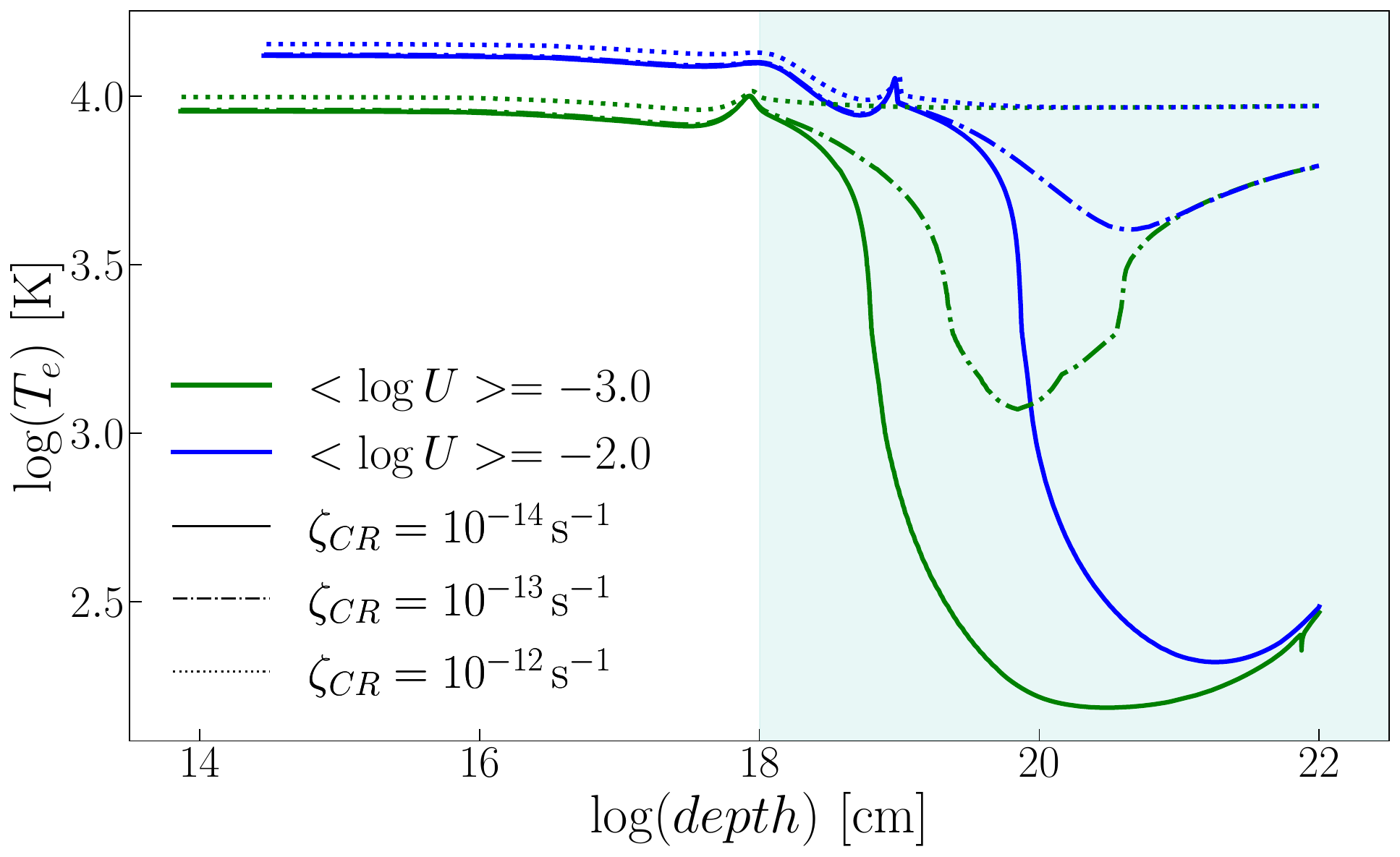}\label{subfig:temp_nh2_agn}}~
    \subfigure[Temperature, AGN, $n_{\rm H}=10^3\,\rm{cm^{-3}}$.]{\includegraphics[width=0.33\textwidth]{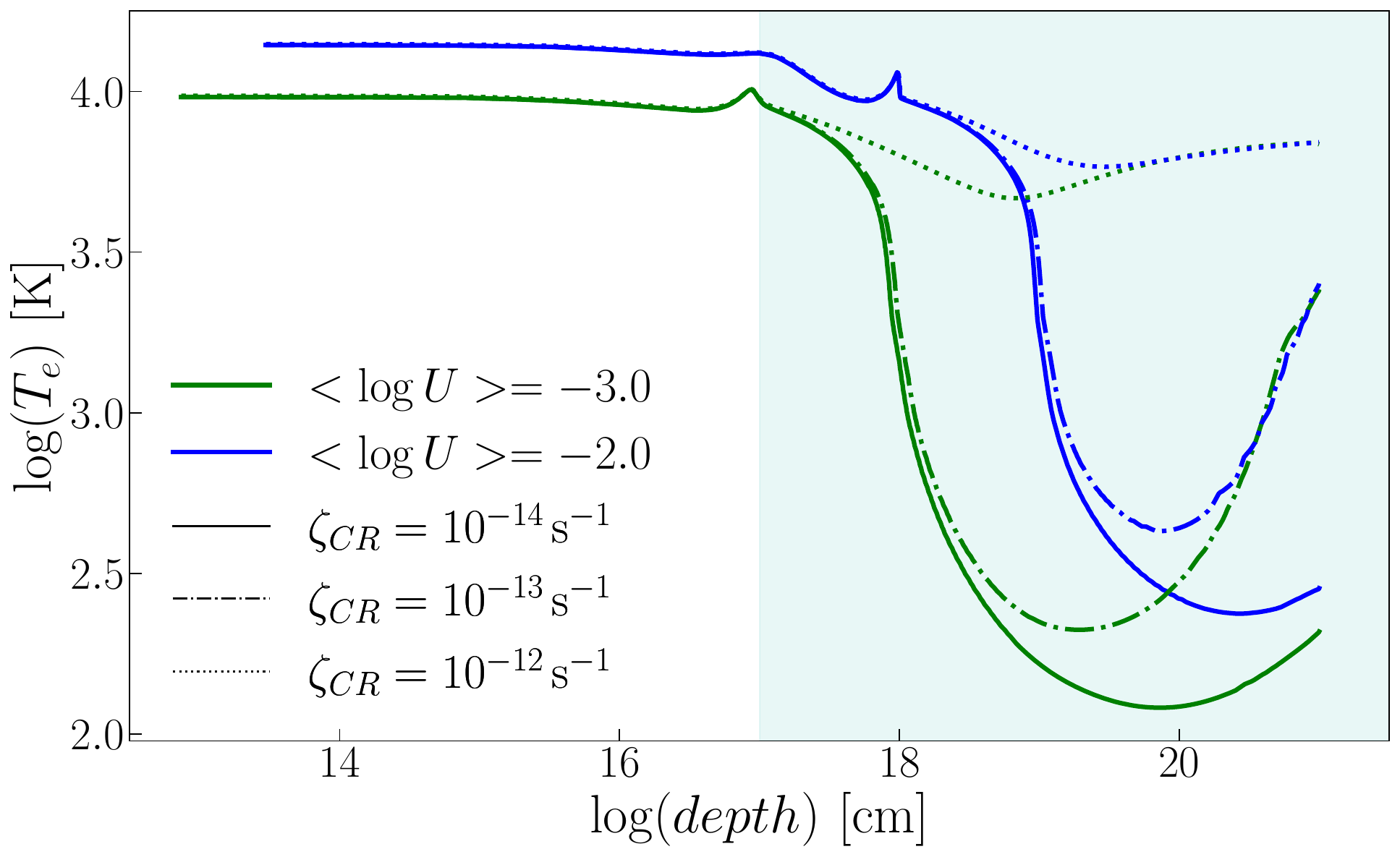}\label{subfig:temp_nh3_agn}}~
    \subfigure[Temperature, SF, $n_{\rm H}=100\,\rm{cm^{-3}}$.]{\includegraphics[width=0.33\textwidth]{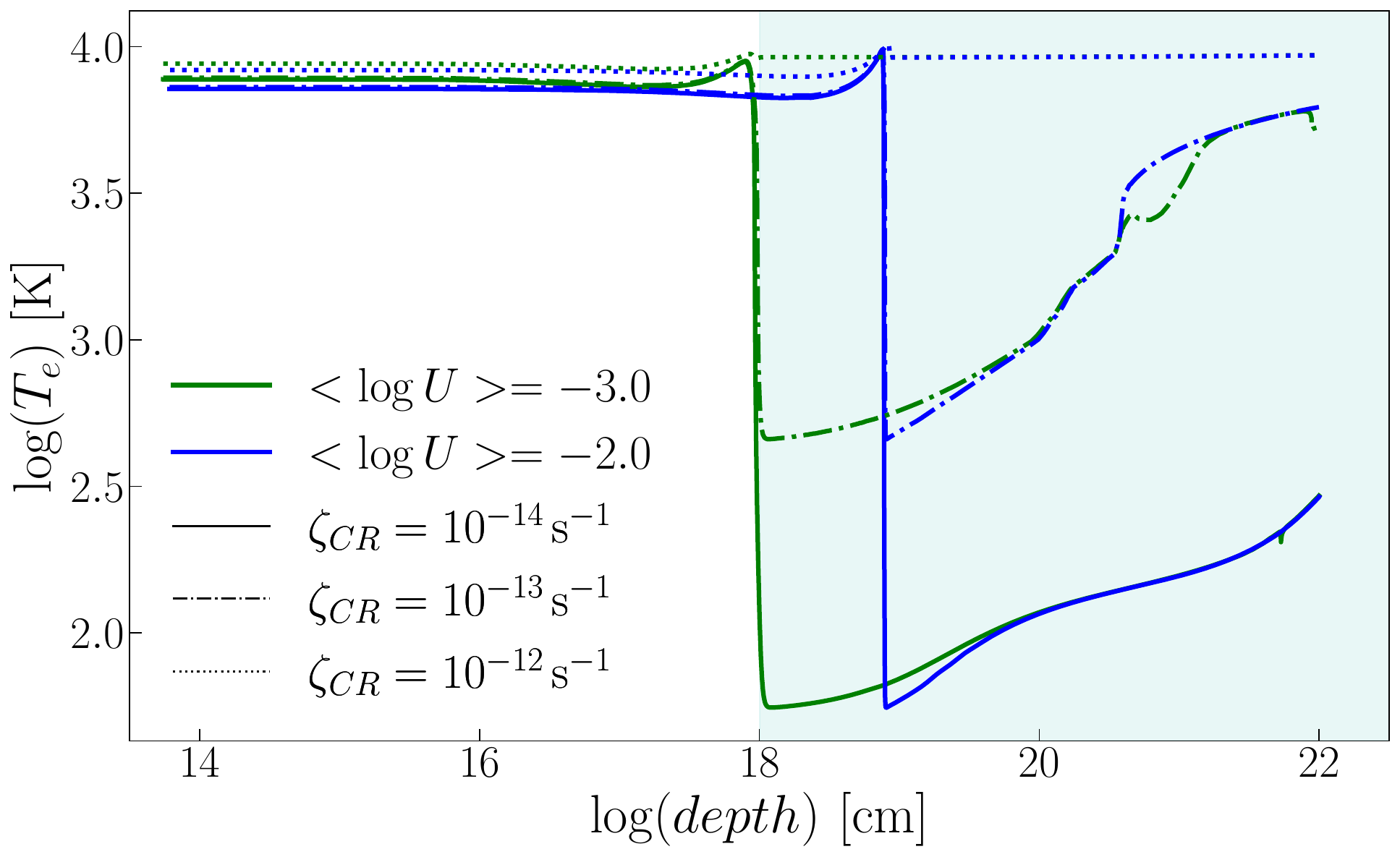}\label{subfig:temp_nh2_sf}}
    \caption{Temperature vs. depth in the simulated cloud for AGN and SF models, for a given initial density (see subcaption), three $\zeta_\mathrm{CR}$ values, $10^{-14}\, \rm  s^{-1}$ (solid line), $10^{-13}\, \rm  s^{-1}$ (dash-dotted), and $10^{-12}\, \rm  s^{-1}$ (dotted), and two $\log U$ values of $-3.0$ (green) and $-2.0$ (blue). The teal-shaded area 
    indicates the approximate region where CR heating becomes dominant.}\label{fig:temp}
    \centering
    \subfigure[{[\ion{N}{ii}]}$\lambda$6584\AA, AGN, $n_{\rm H}=100\,\rm{cm^{-3}}$.]{\includegraphics[width=0.33\textwidth]{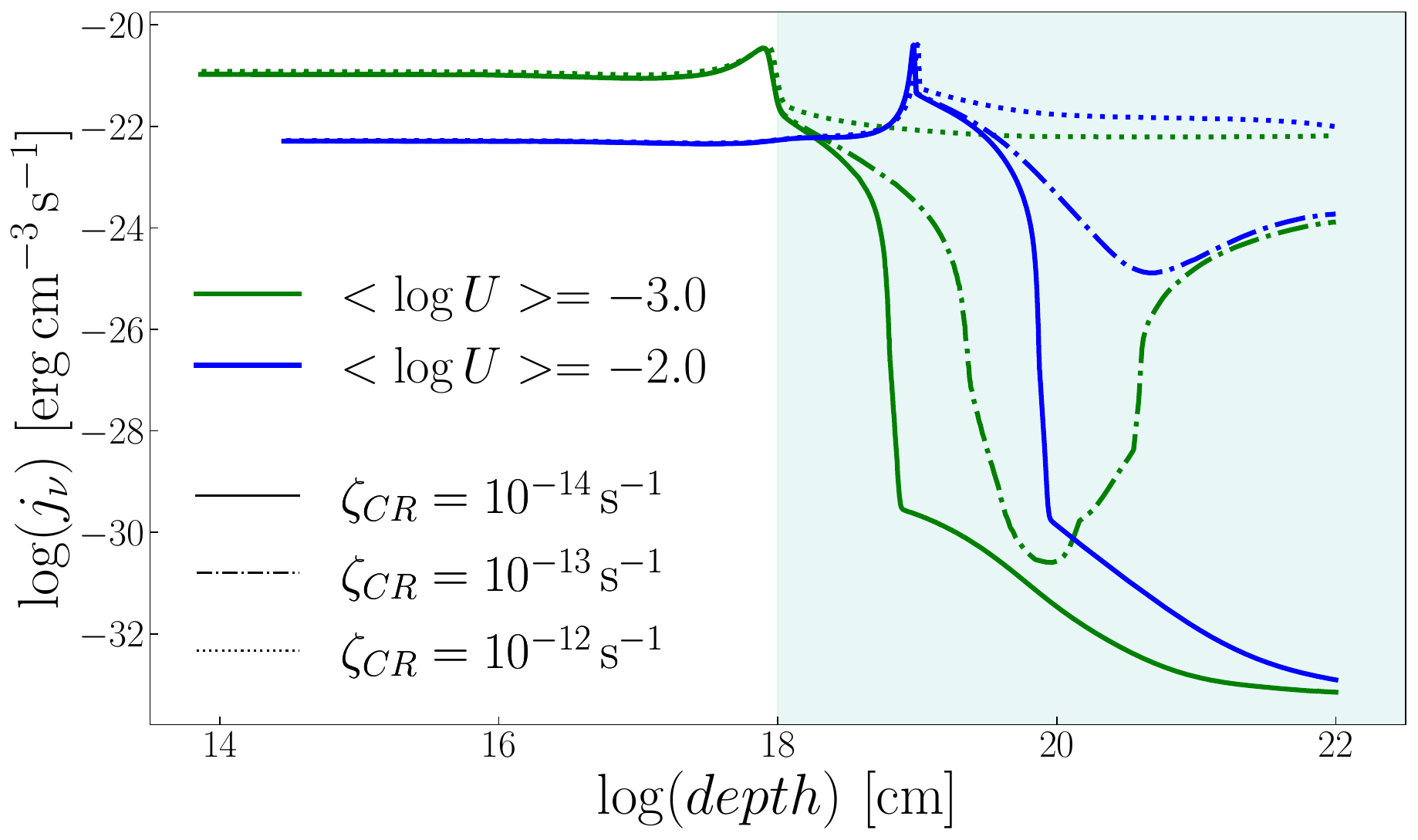}\label{subfig:n28_nh2_agn}}~
    \subfigure[{[\ion{N}{ii}]}$\lambda$6584\AA, AGN, $n_{\rm H}=10^3\,\rm{cm^{-3}}$.]{\includegraphics[width=0.33\textwidth]{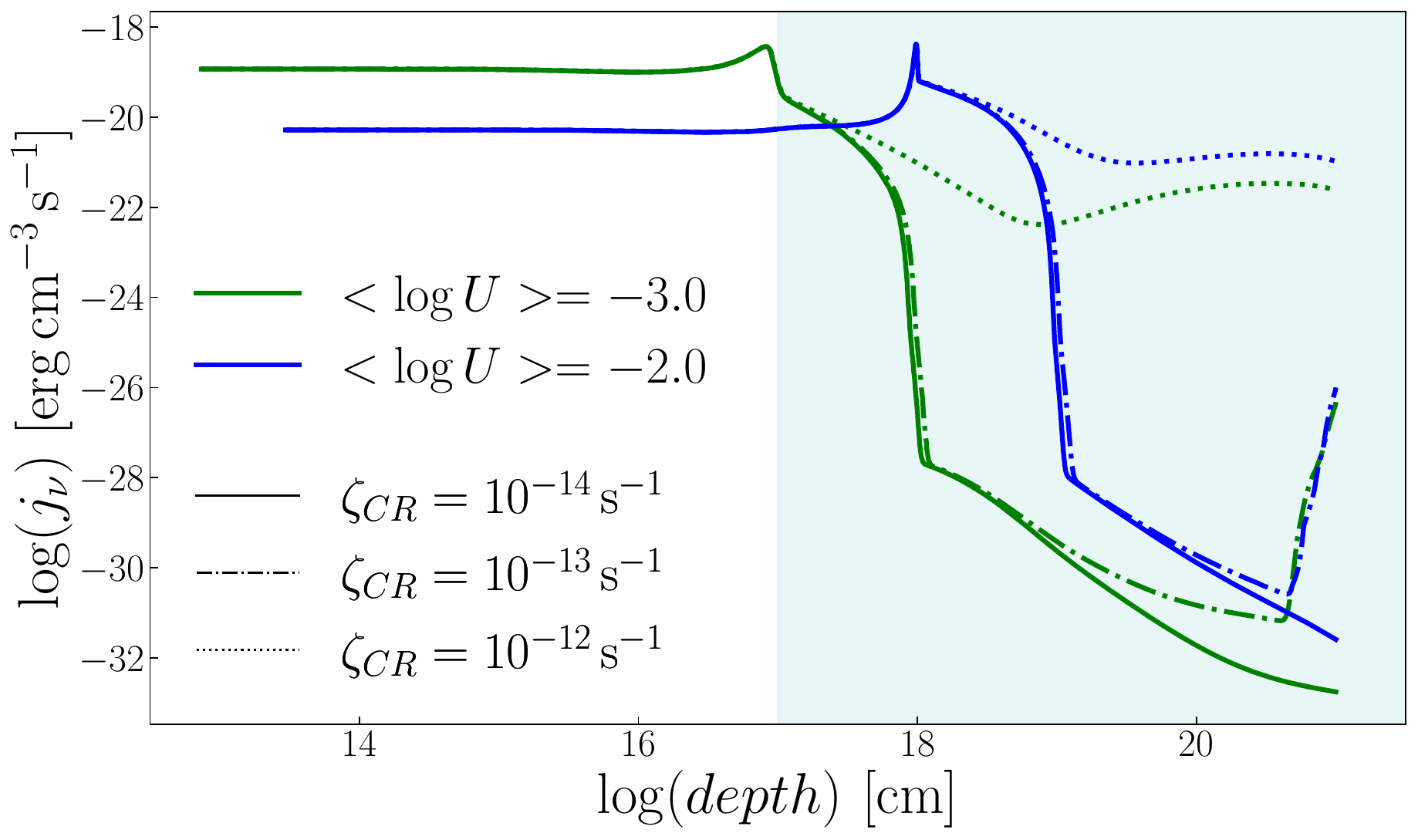}\label{subfig:n28_nh3_agn}}~
    \subfigure[{[\ion{N}{ii}]}$\lambda$6584\AA, SF, $n_{\rm H}=100\, \rm{cm^{-3}}$.]{\includegraphics[width=0.33\textwidth]{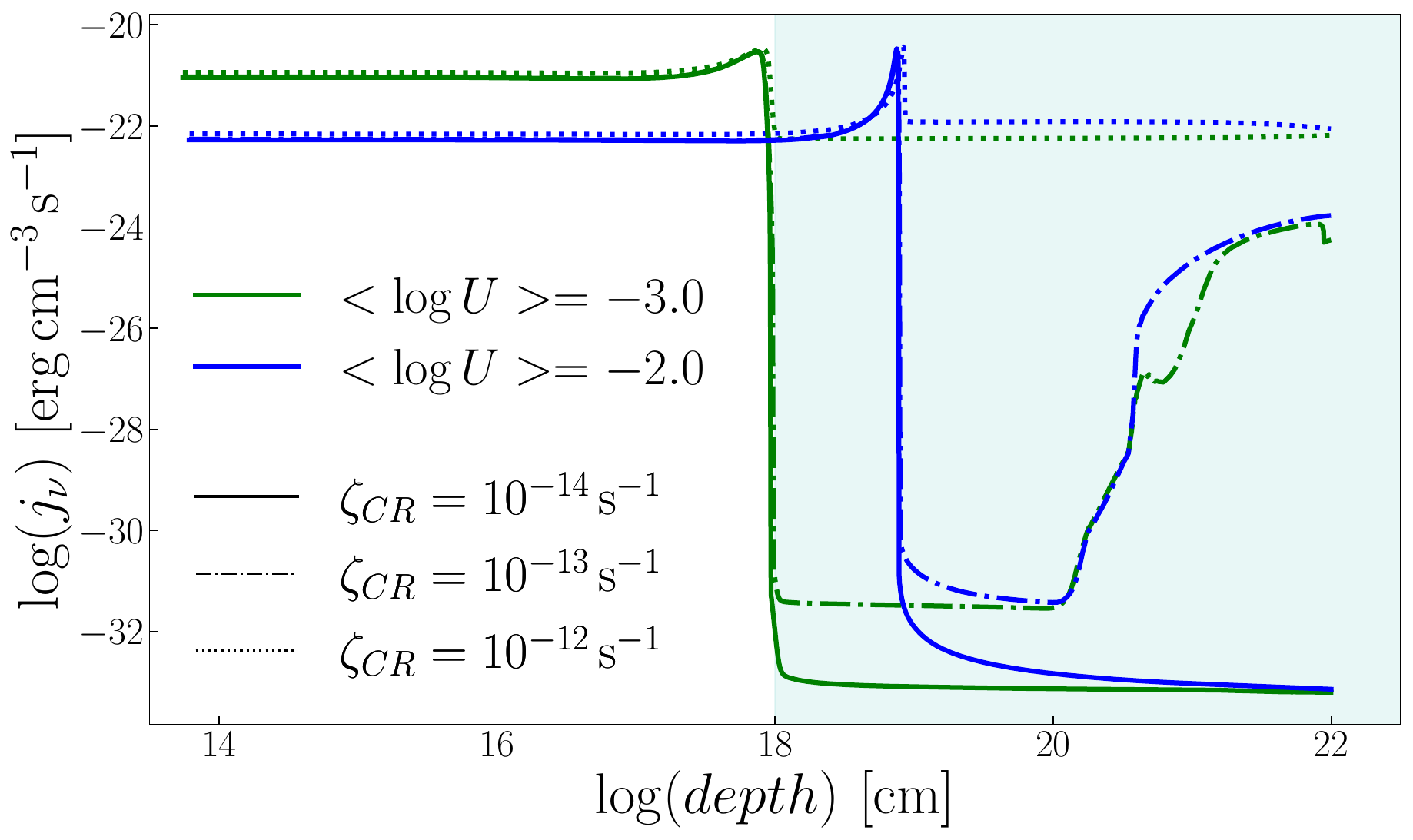}\label{subfig:n28_nh2_sf}}
    \subfigure[{[\ion{S}{ii}]}$\lambda\lambda$6716,6731\AA, AGN, $n_{\rm H}=100\,\rm{cm^{-3}}$.]{\includegraphics[width=0.33\textwidth]{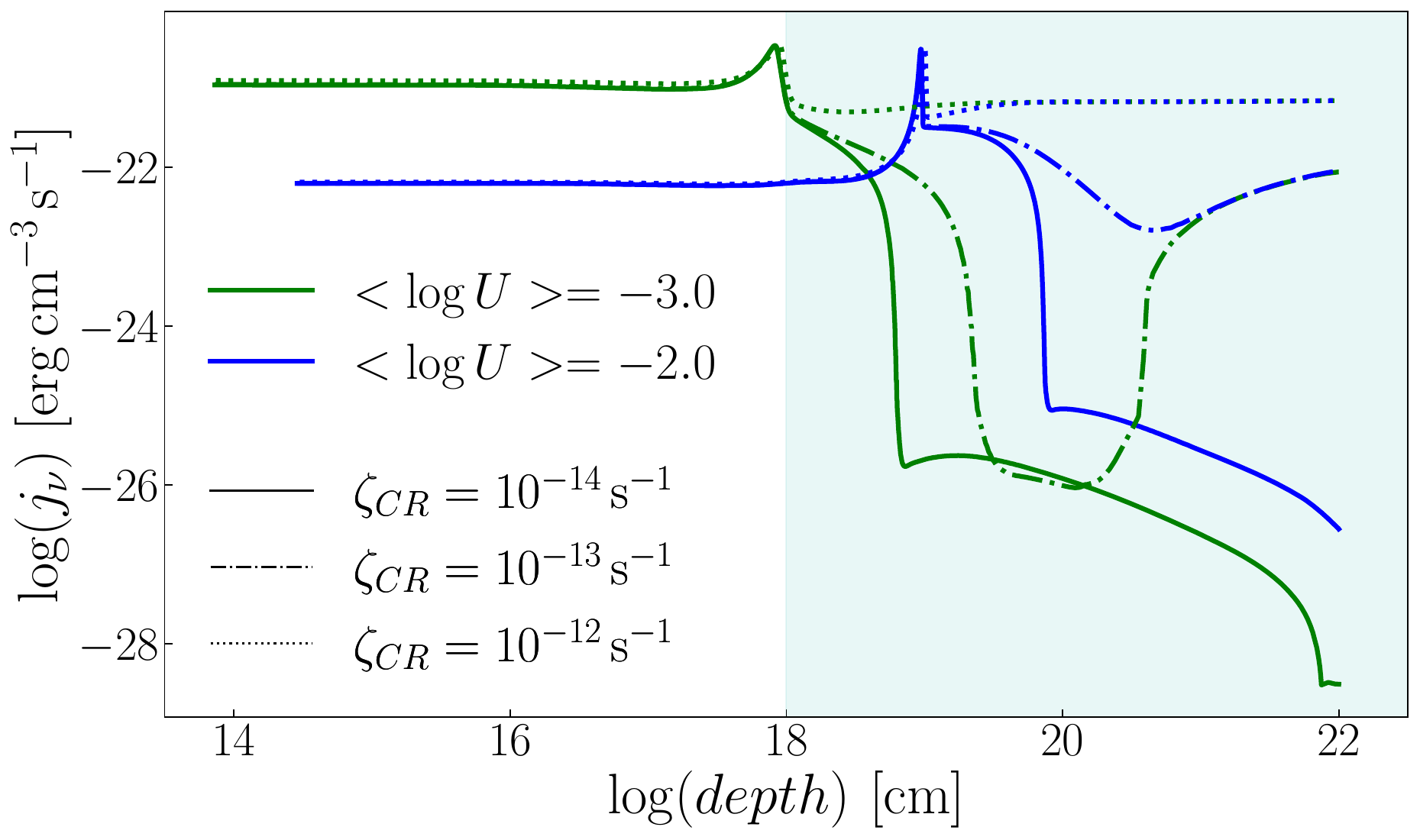}\label{subfig:s2_nh2_agn}}~
    \subfigure[{[\ion{S}{ii}]}$\lambda\lambda$6716,6731\AA, AGN, $n_{\rm H}=10^3\,\rm{cm^{-3}}$.]{\includegraphics[width=0.33\textwidth]{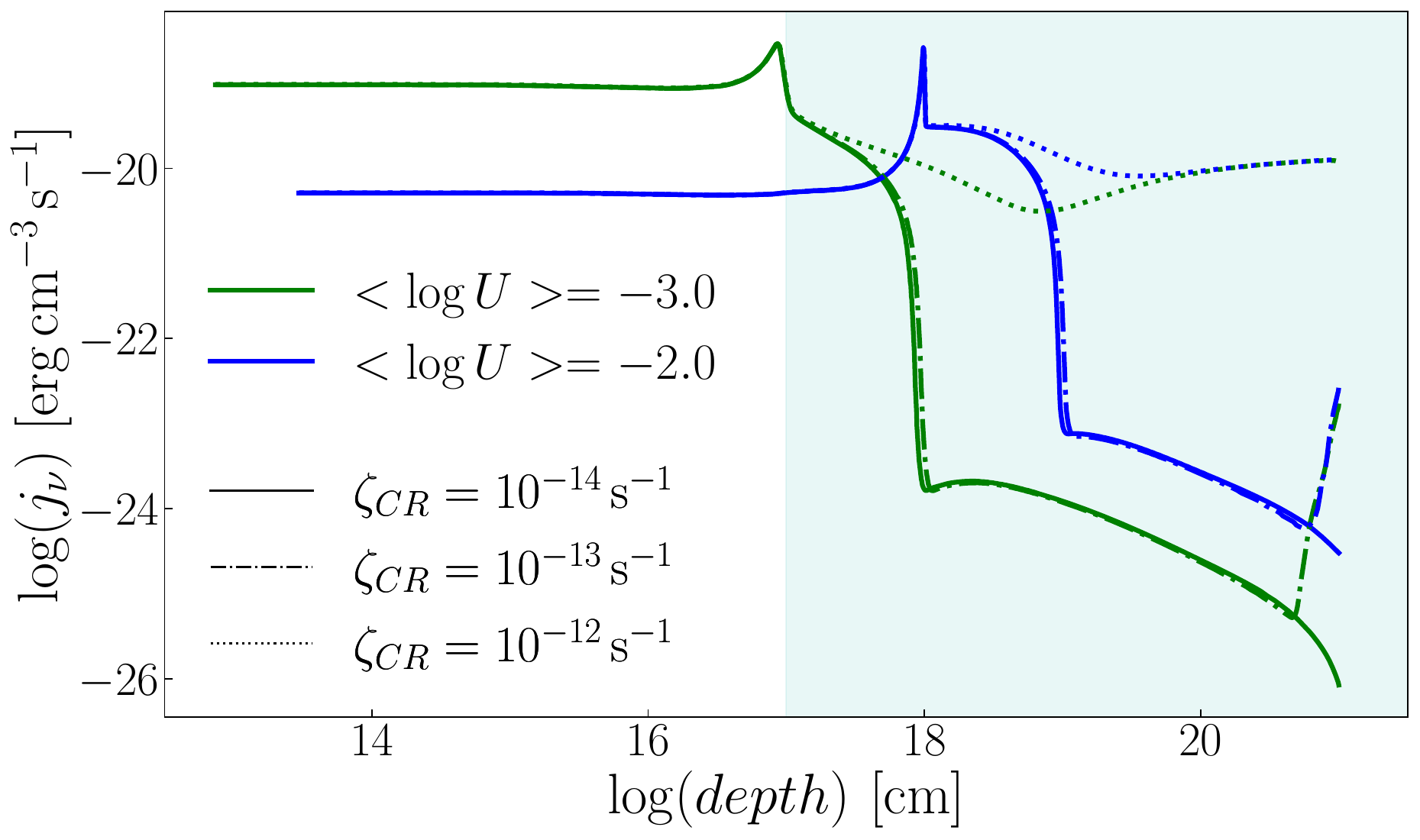}\label{subfig:s2_nh3_agn}}~
    \subfigure[{[\ion{S}{ii}]}$\lambda \lambda$6716,6731\AA, SF, $n_{\rm H}=100\,\rm{cm^{-3}}$.]{\includegraphics[width=0.33\textwidth]{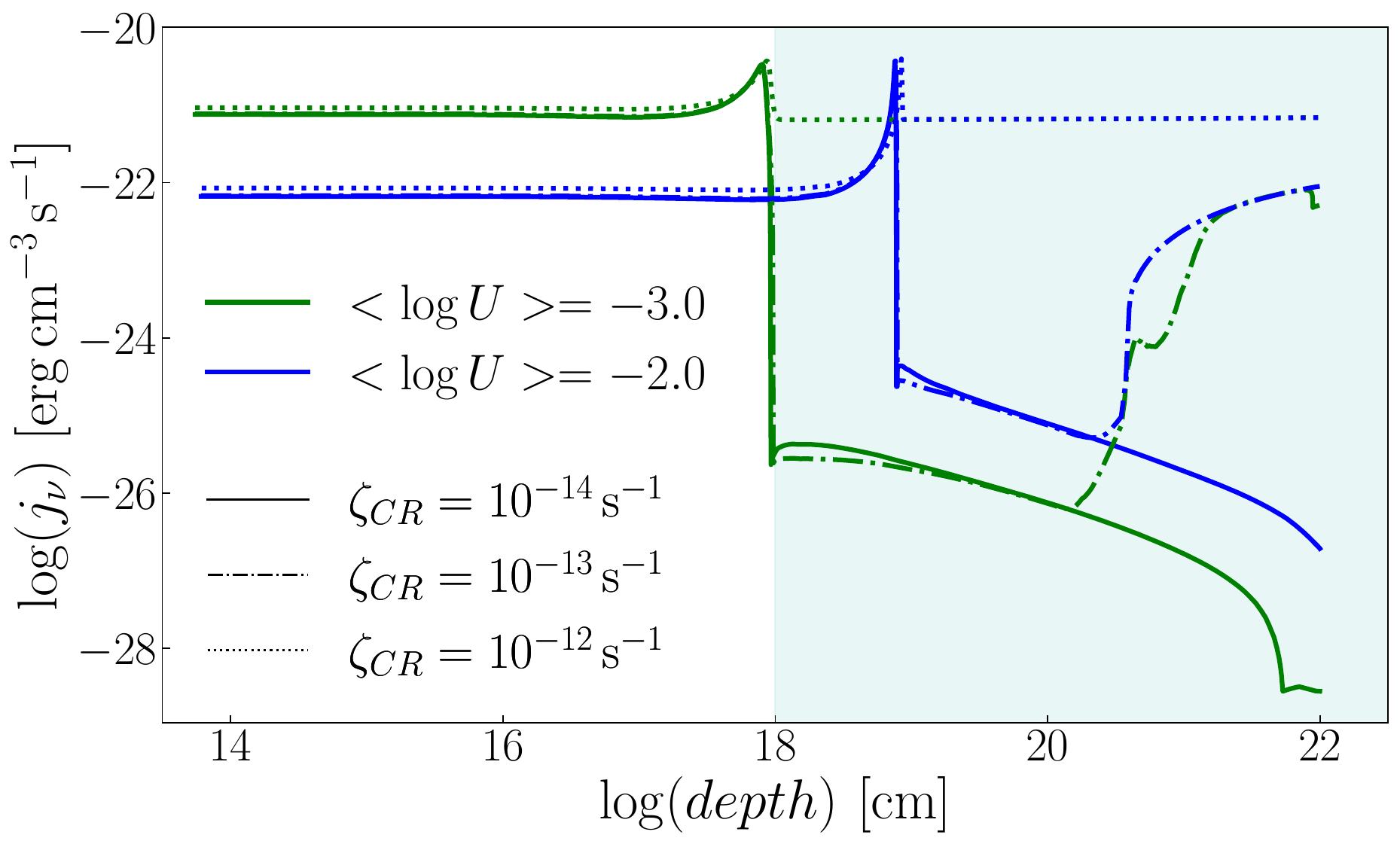}\label{subfig:s22_nh2_sf}}
    \subfigure[{[\ion{O}{i}]}$\lambda$6300\AA, AGN, $n_{\rm H}=100\,\rm{cm^{-3}}$.]{\includegraphics[width=0.33\textwidth]{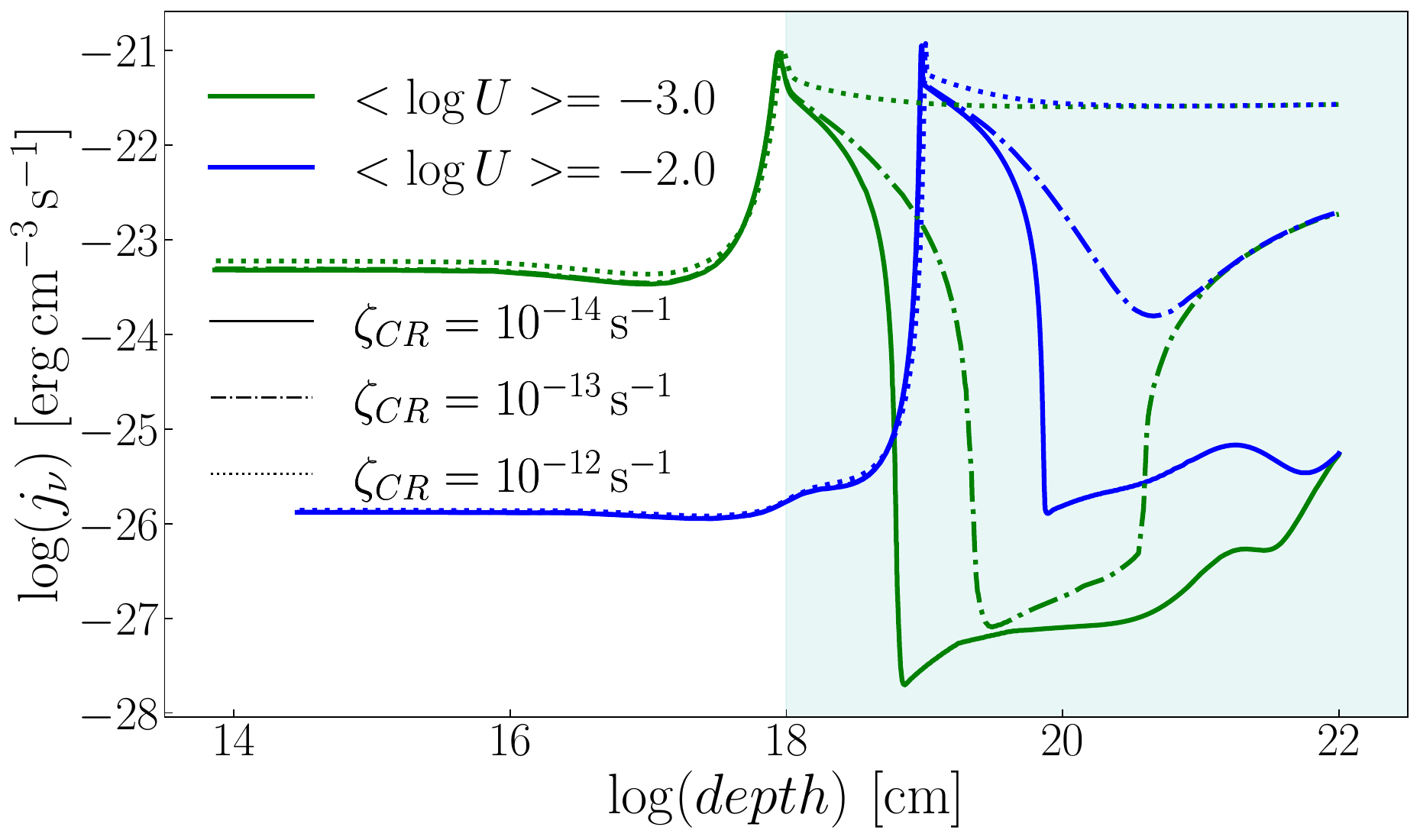}\label{subfig:o1_nh2}}~
    \subfigure[{[\ion{O}{i}]}$\lambda$6300\AA, AGN, $n_{\rm H}=10^3\,\rm{cm^{-3}}$.]{\includegraphics[width=0.33\textwidth]{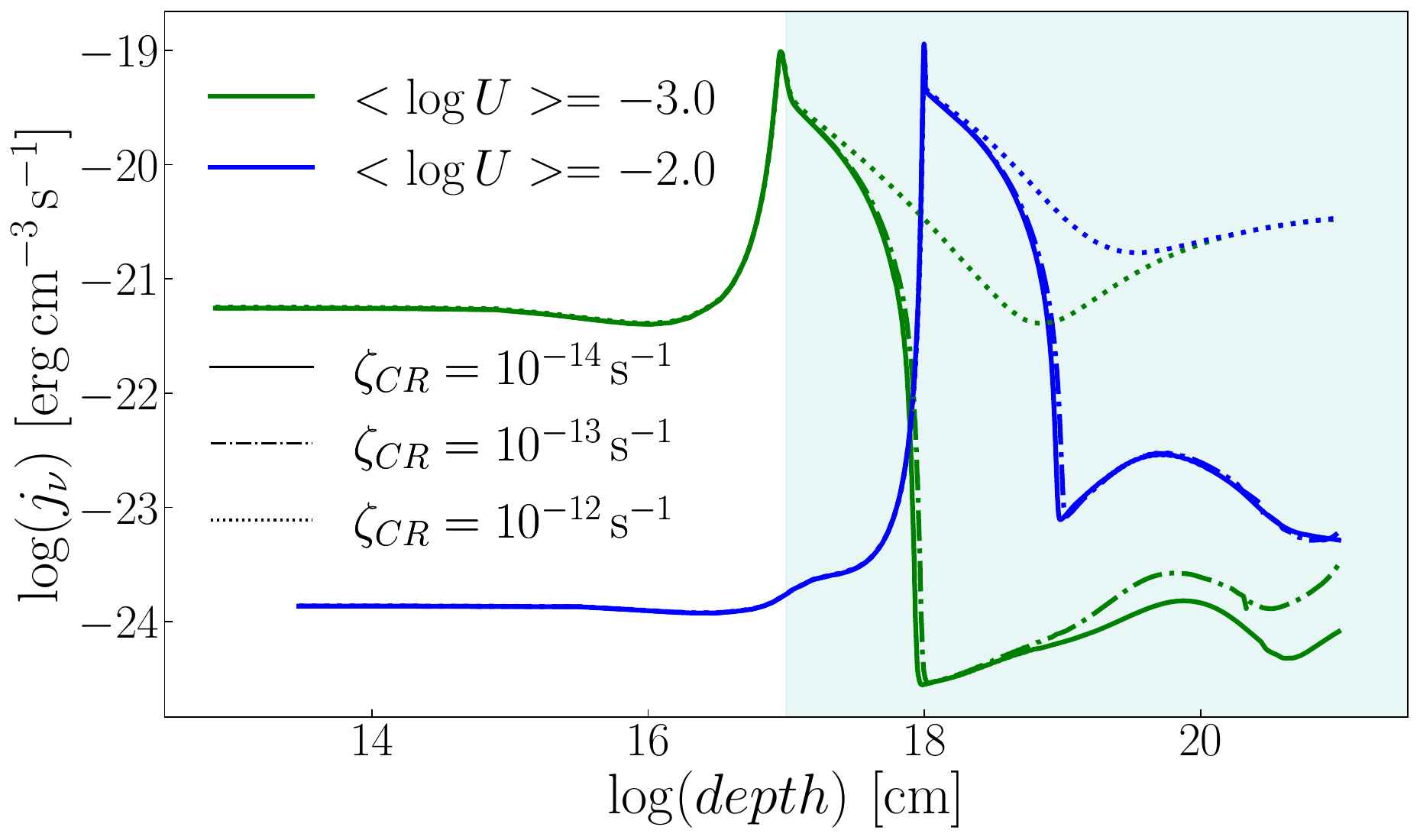}\label{subfig:o1_nh3_agn}}~
    \subfigure[{[\ion{O}{i}]}$\lambda$6300\AA, SF, $n_{\rm H}=100\,\rm{cm^{-3}}$.]{\includegraphics[width=0.33\textwidth]{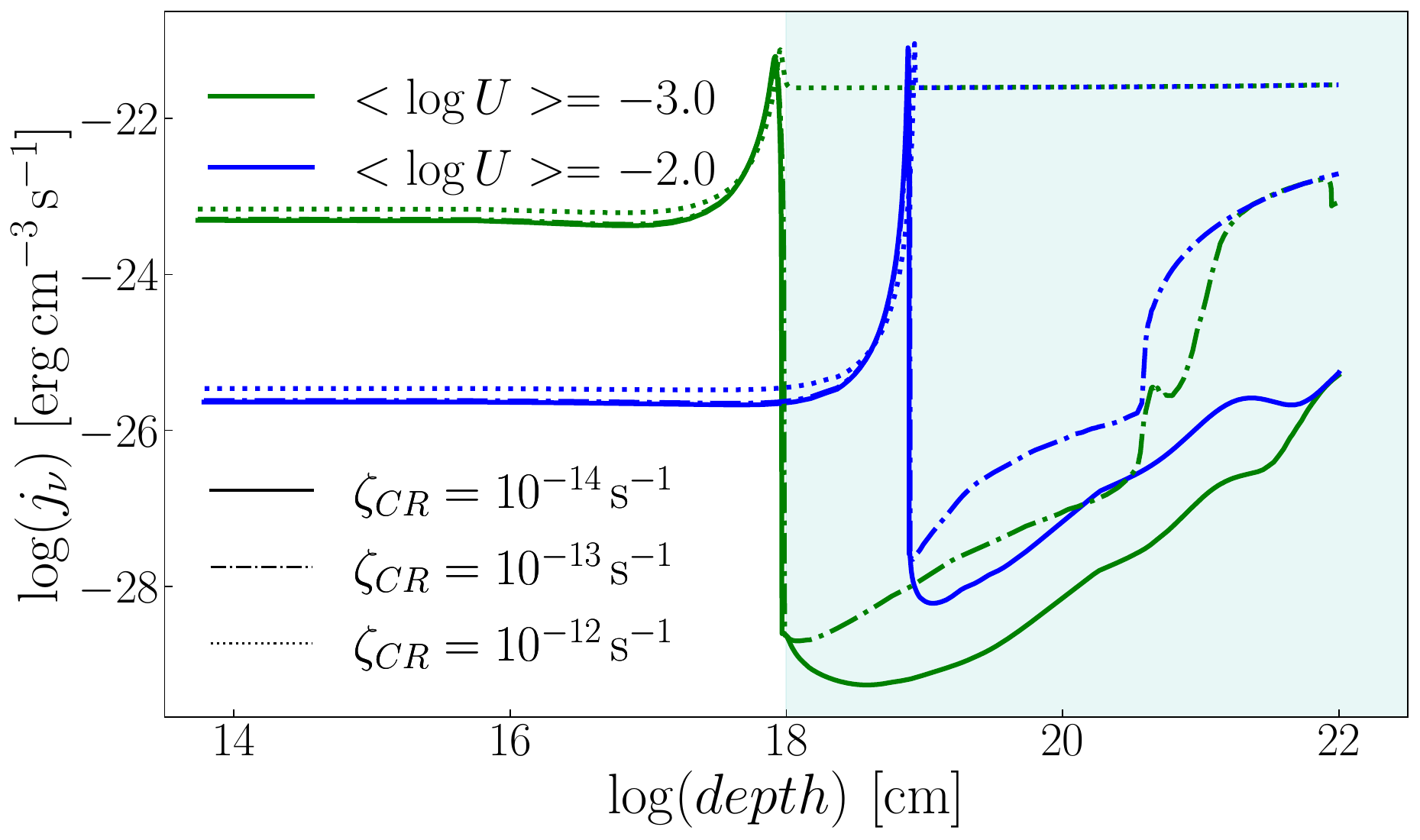}\label{subfig:o1_nh2_sf}}
    \subfigure[H$\alpha$, AGN, $n_{\rm H}\, =\, 100 \, \rm{cm^{-3}}$.]{\includegraphics[width=0.33\textwidth]{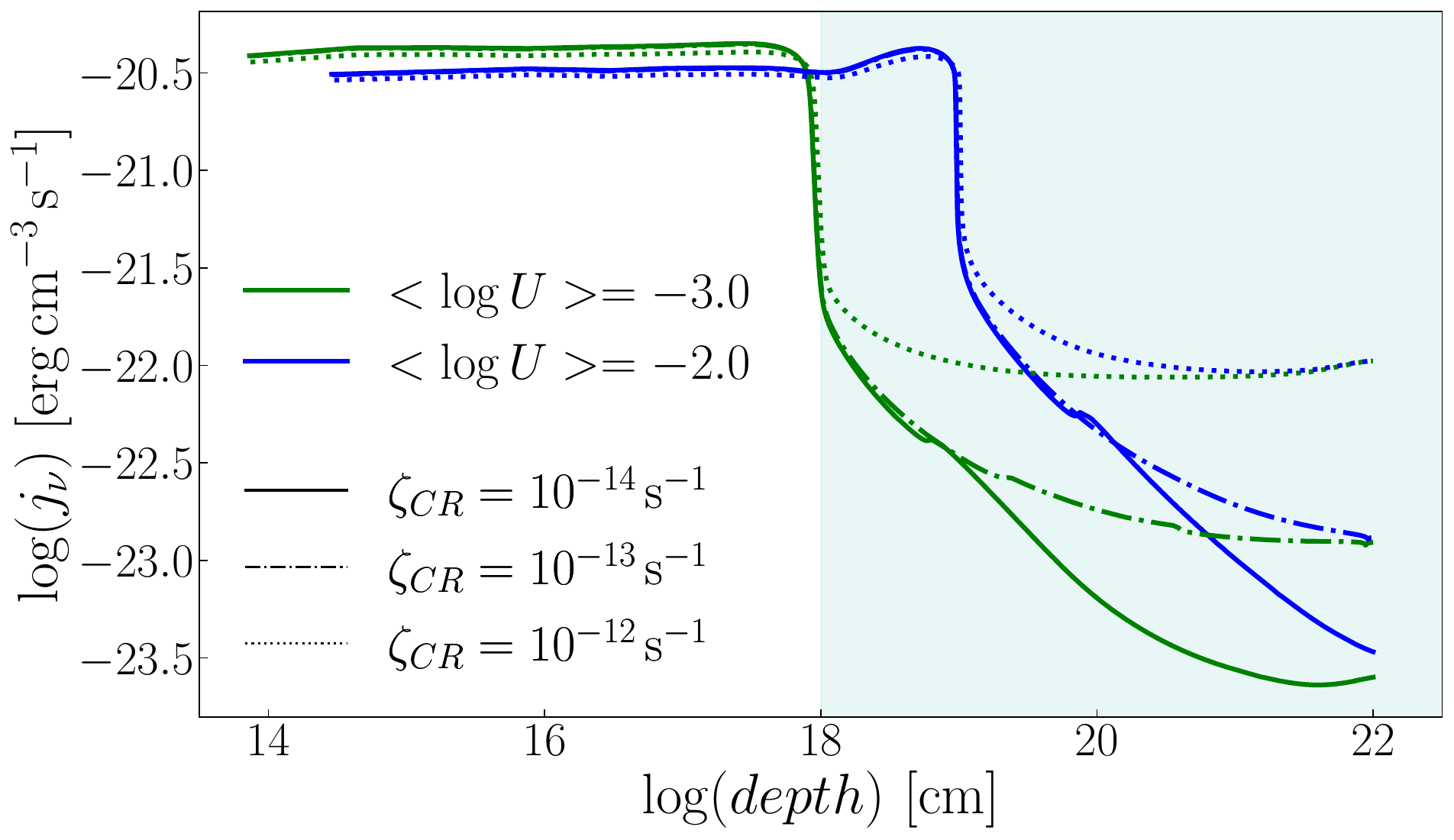}\label{subfig:ha_nh2_agn}}~
    \subfigure[H$\alpha$, AGN, $n_{\rm H}\, =\, 10^3 \, \rm{cm^{-3}}$.]{\includegraphics[width=0.33\textwidth]{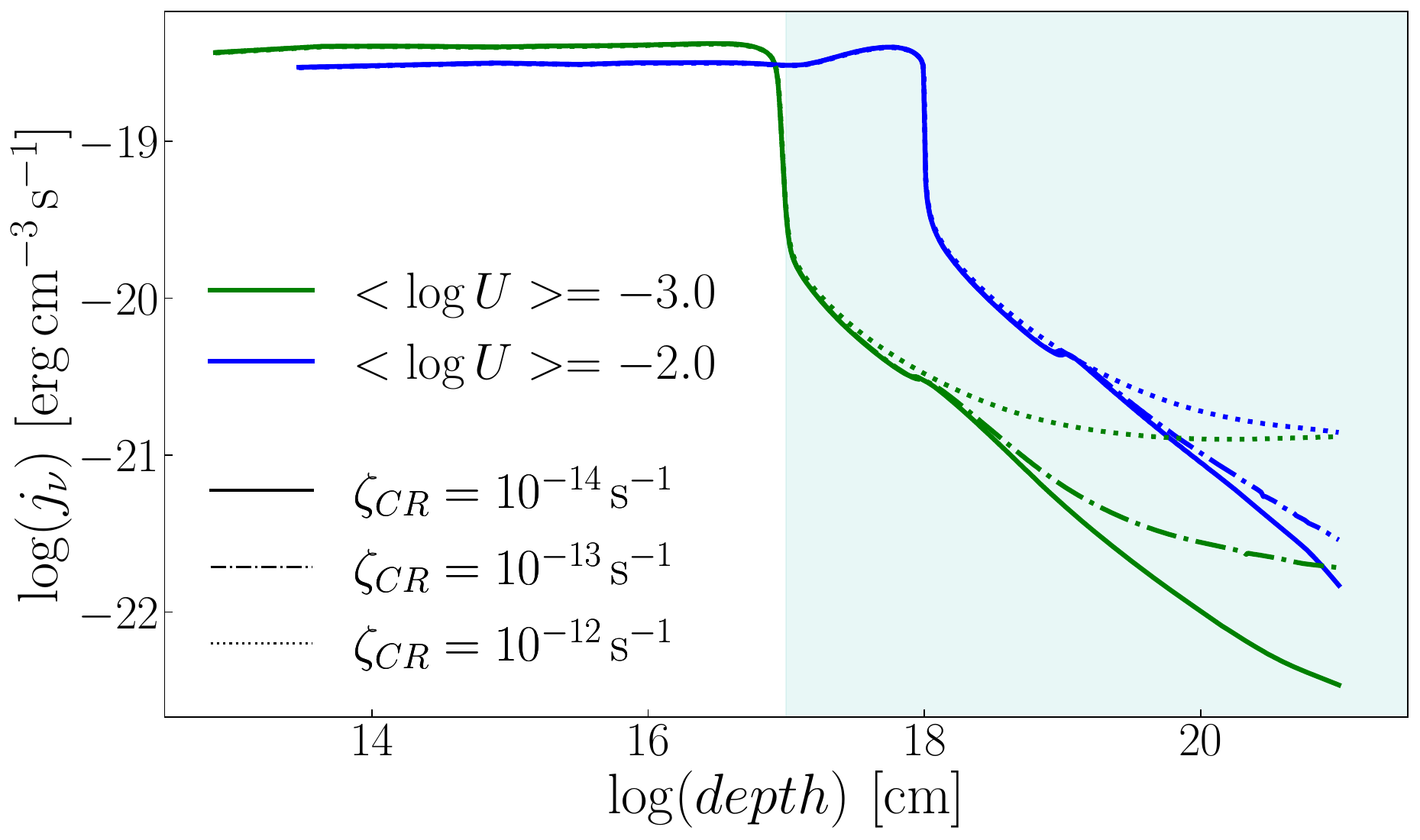}\label{subfig:ha_nh3_agn}}~
    \subfigure[H$\alpha$, SF, $n_{\rm H}\, =\, 100 \, \rm{cm^{-3}}$.]{\includegraphics[width=0.33\textwidth]{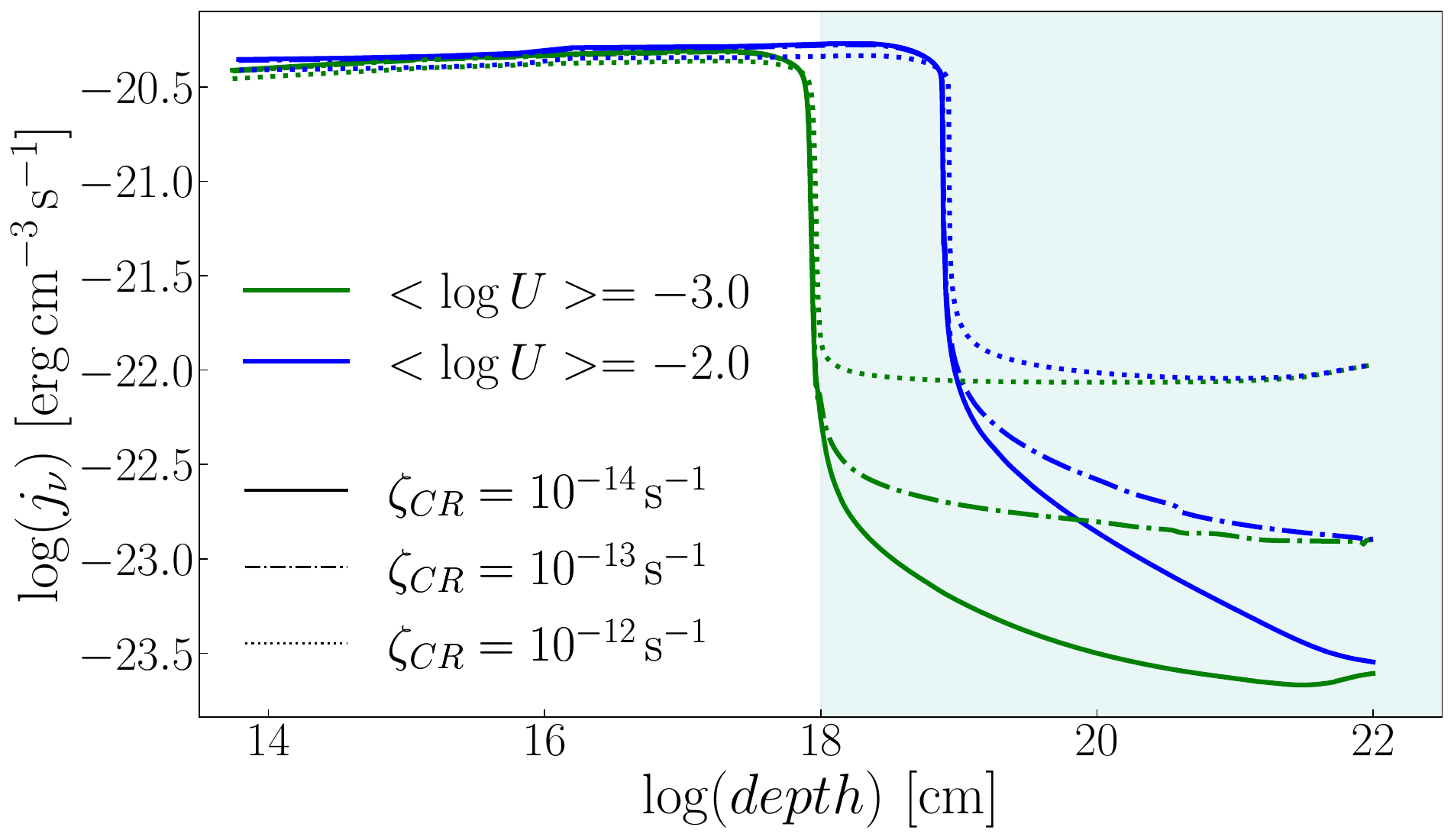}\label{subfig:ha_nh2_sf}}
    \caption{Line emissivities vs. depth for AGN and SF models, for three $\zeta_\mathrm{CR}$ values of $10^{-14}\,\rm  s^{-1}$ (solid line), $10^{-13}\, \rm  s^{-1}$ (dash-dotted), and $10^{-12}\, \rm  s^{-1}$ (dotted), and two $\log U$ values of $-3.0$ (green) and $-2.0$ (blue). The teal-shaded area 
    indicates the region where CRs heating becomes dominant.}\label{fig:struc_bpt1}
\end{figure*}

\begin{figure*}[!ht]
    \centering
    \subfigure[{[\ion{O}{iii}]}$\lambda$5007\AA, AGN, $n_{\rm H}=100\,\rm{cm^{-3}}$.]{\includegraphics[width=0.33\textwidth]{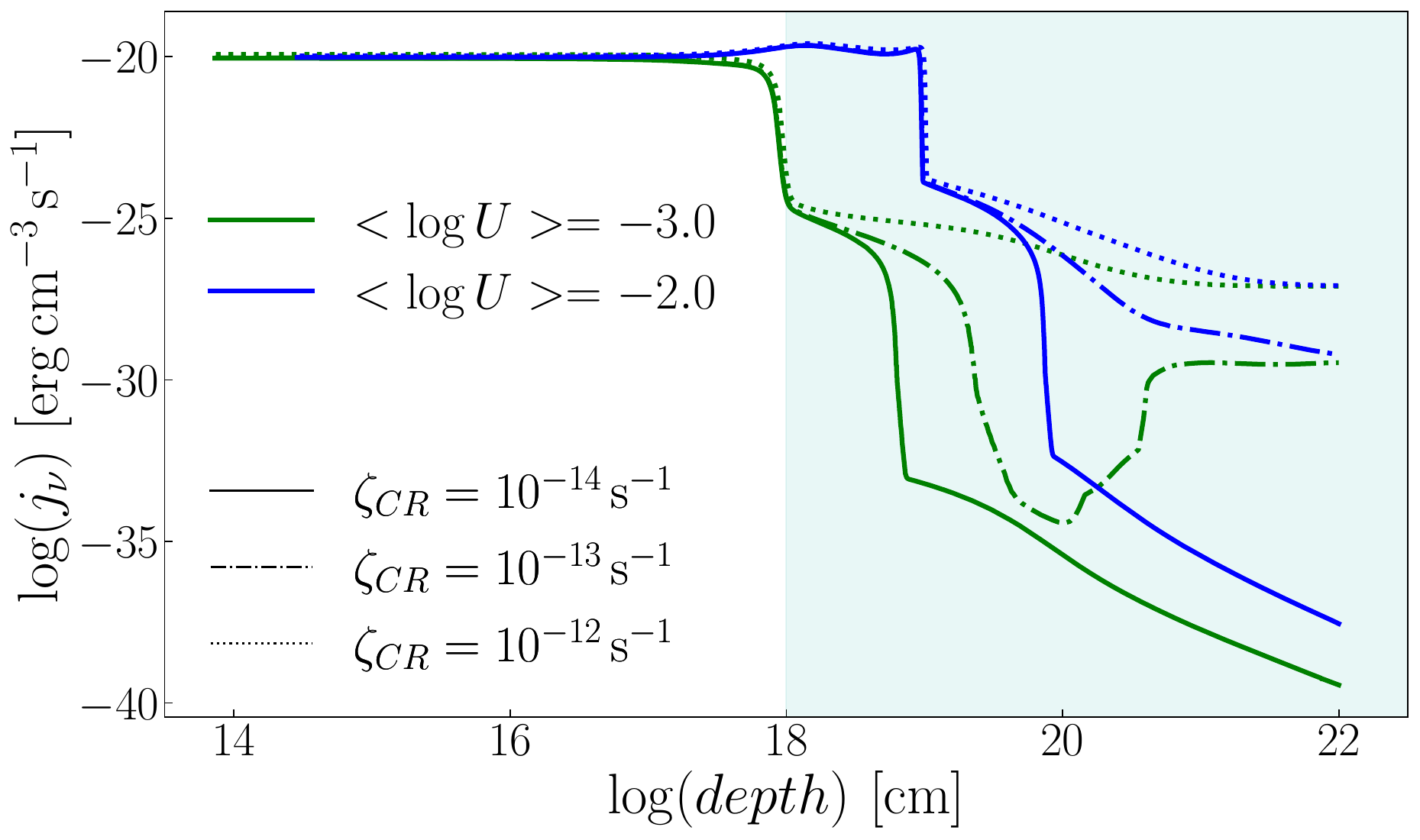}\label{subfig:o3_nh2_agn}}~
    \subfigure[{[\ion{O}{iii}]}$\lambda$5007\AA, AGN, $n_{\rm H}=10^3\,\rm{cm^{-3}}$.]{\includegraphics[width=0.33\textwidth]{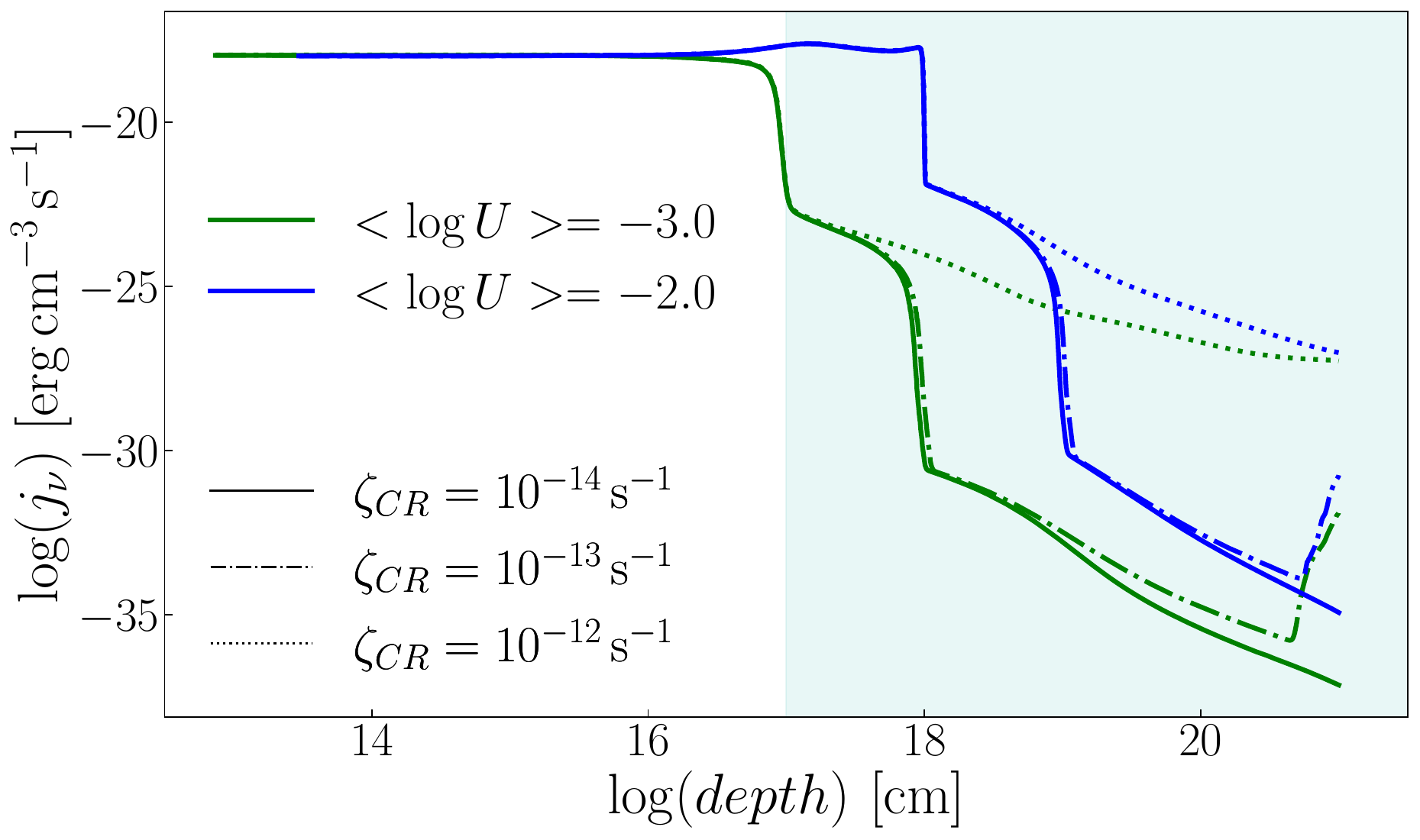}\label{subfig:o3_nh3_agn}}~
    \subfigure[{[\ion{O}{iii}]}$\lambda$5007\AA, SF, $n_{\rm H}=100\,\rm{cm^{-3}}$.]{\includegraphics[width=0.33\textwidth]{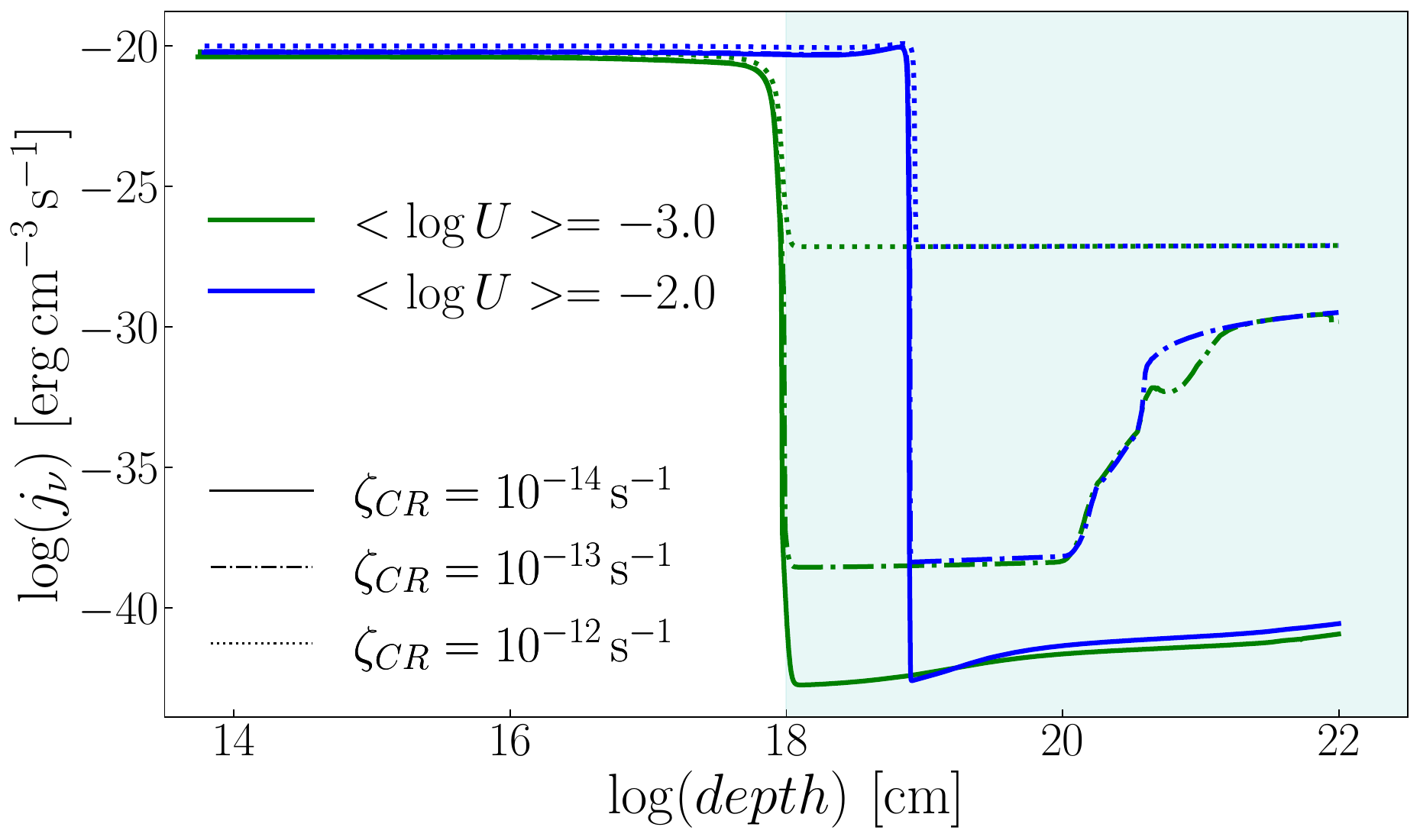}\label{subfig:o3_nh2_sf}}
    \subfigure[H$\beta$, AGN, $n_{\rm H}\, =\, 100 \, \rm{cm^{-3}}$.]{\includegraphics[width=0.33\textwidth]{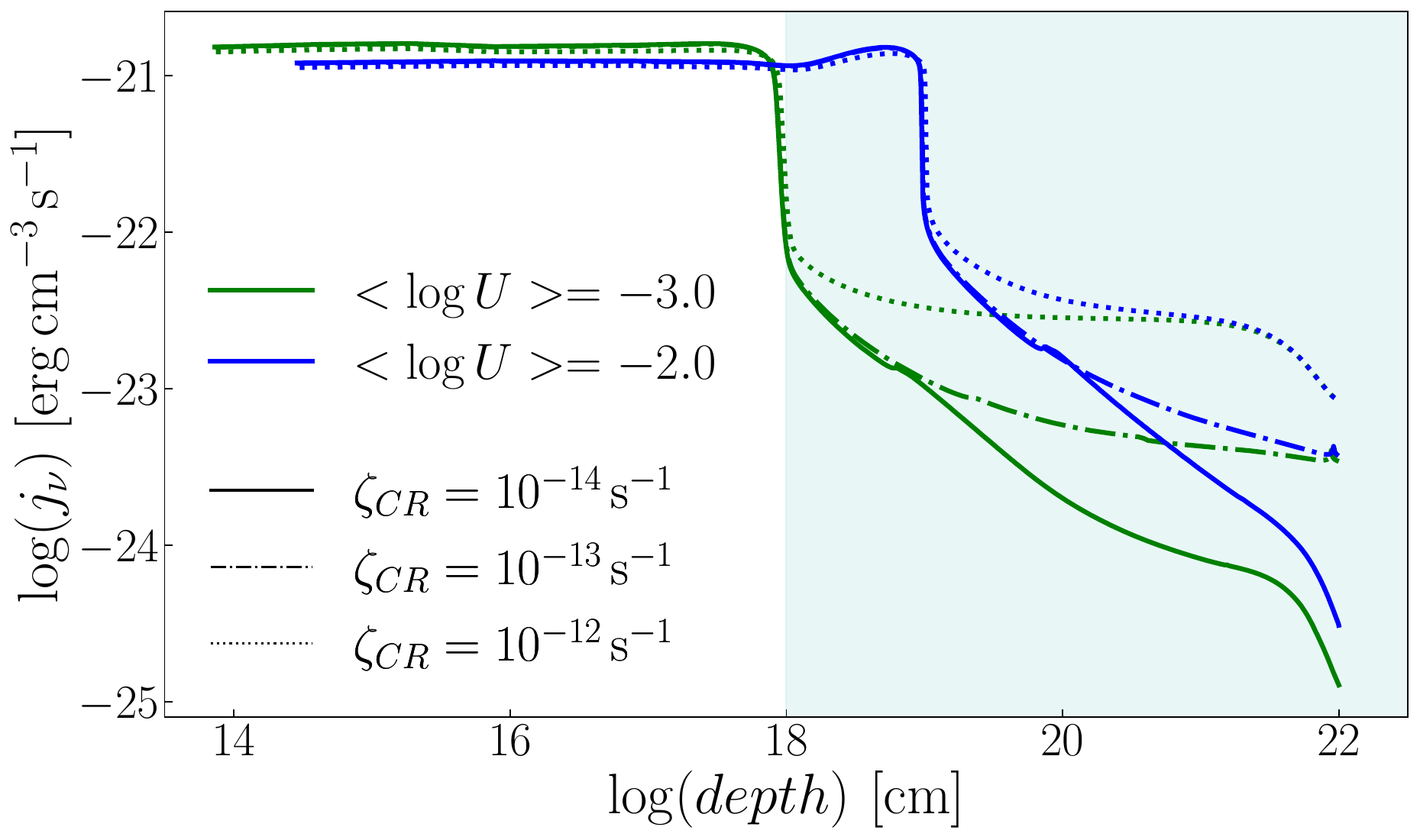}\label{subfig:hb_nh2_agn}}~
    \subfigure[H$\beta$, AGN, $n_{\rm H}\, =\, 10^3 \, \rm{cm^{-3}}$.]{\includegraphics[width=0.33\textwidth]{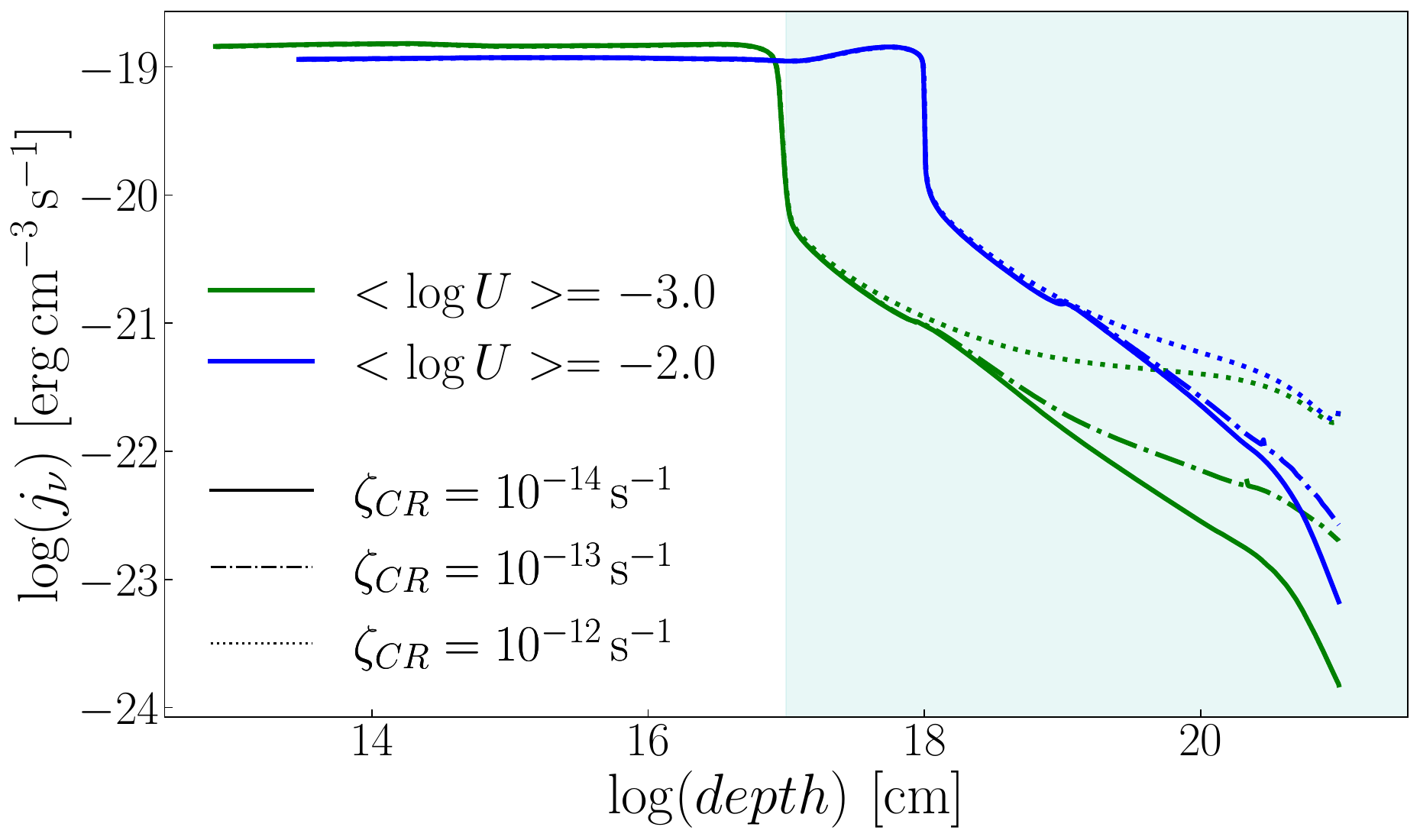}\label{subfig:hb_nh3_agn}}~
    \subfigure[H$\beta$, SF, $n_{\rm H}\, =\, 100 \, \rm{cm^{-3}}$.]{\includegraphics[width=0.33\textwidth]{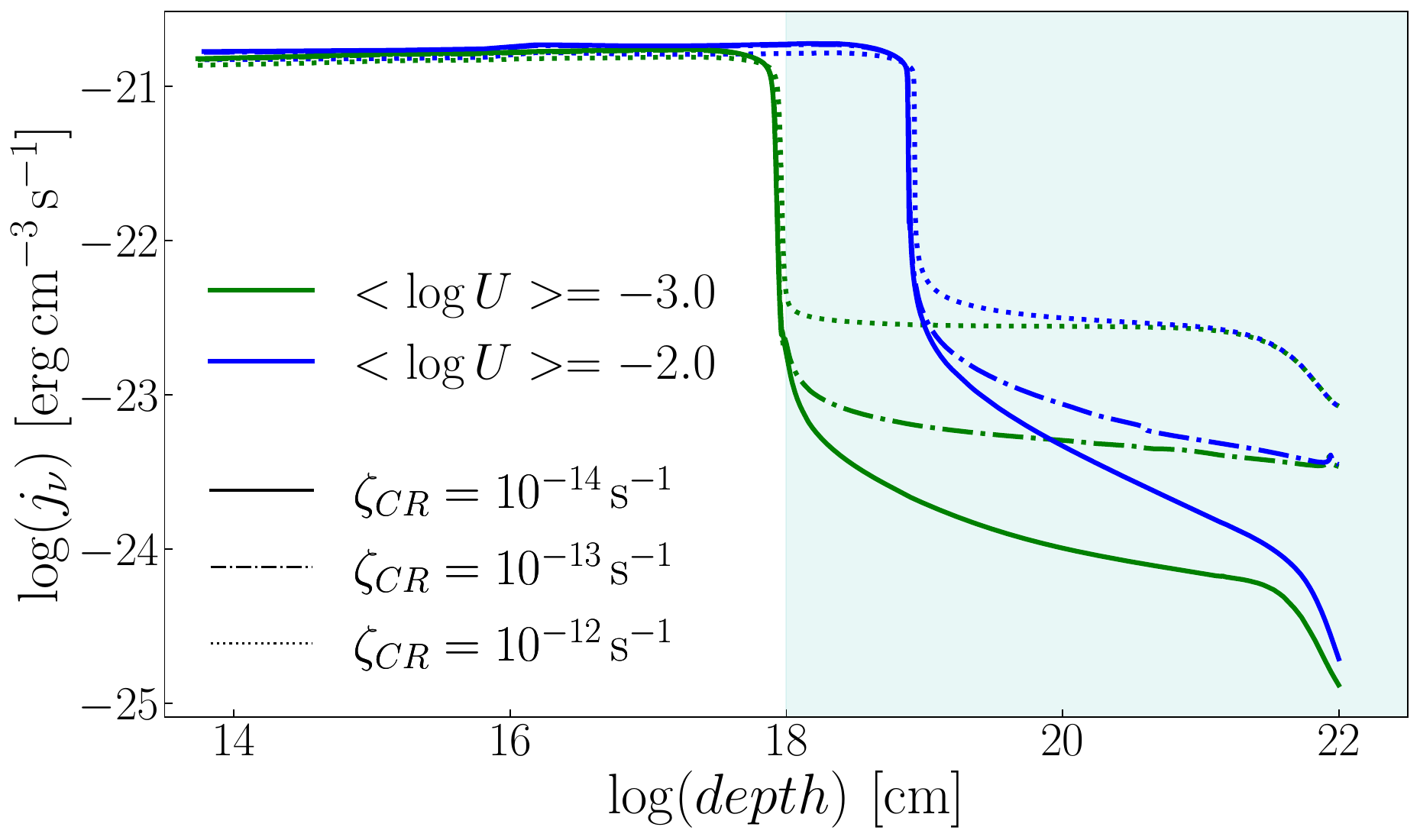}\label{subfig:hb_nh2_sf}}
    \caption{Line emissivities vs. depth for AGN and SF models, for three $\zeta_\mathrm{CR}$ values of $10^{-14}\,\rm  s^{-1}$ (solid line), $10^{-13}\, \rm  s^{-1}$ (dash-dotted), and $10^{-12}\, \rm  s^{-1}$ (dotted), and two $\log U$ values of $-3.0$ (green) and $-2.0$ (blue). The teal-shaded area 
    indicates the region where CRs heating becomes dominant.}\label{fig:struc_bpt2}
\end{figure*}


The ionization and excitation of the gas is most likely achieved with UV radiation coming from young hot stars \citep{Jeong_2018}. Nevertheless, it is interesting to note that UV photons of energies more than 13.6 eV, which is the ionization potential of hydrogen, are quickly absorbed and cannot pass through all the layers of the clouds \citep{Gabici_2022}. The UV radiation is significantly attenuated for column densities $N_{\rm H} \geq 8 \times 10^{21}\,\rm{cm^{-2}}$ and CRs become the prevailing ionizing mechanism \citep{McKee_1989}. 

The effects of CRs are deeper in the clouds where examining the emissivity of emission lines explains why our models move to the right area on the BPT diagrams. In order to analyze this, we produce the following gas stratification diagrams (hereafter referred to as structure plots) for two ionization parameter values, that is, $\log U= \{-3.0, \, -2.0\}$ and for densities $n_{\rm H} = \{100, \, 10^3 \} \, \rm{cm^{-3}}$ in the case of AGN models and the same for  $n_{\rm H} = {100} \, \rm{cm^{-3}}$ in the case of SF models. Fig. \ref{fig:temp} makes quite evident the fact that the electron temperature at the illuminated face of the cloud is mainly affected by photoionization and the deeper layers show increase in temperature in the cases of the higher CR ionization rates $10^{-13}\,\rm{s^{-1}}$ and $10^{-12}\,\rm{s^{-1}}$.  The higher the ionization parameter is, the bigger is the depth up to which photons dominate. However, after a certain layer CRs are the main excitation mechanism. 

Similarly when investigating simulations with higher initial hydrogen density $n_{\rm H}$ in the AGN models in Figs. \ref{subfig:temp_nh2_agn}, and \ref{subfig:temp_nh3_agn} the photoionization stops in the shallow parts of the cloud and the shielded layers, due to the higher densities, are again reached and excited only by high enough CR rates $ \sim 10^{-13}-10^{-12}\,\rm{s^{-1}}$. This warm secondary ionized layer reaches electron temperature of $\rm{T_{e}} \sim 8000\,\rm{K}$, and enhances the emissivity of the low ionization lines. Similar work by \cite{beck} on ionized, atomic and molecular emission lines is consistent with this result. In particular, in the structure plots (i.e., Figs. \ref{subfig:temp_nh2_agn}, and \ref{subfig:temp_nh3_agn},  \ref{subfig:n28_nh2_agn}, and \ref{subfig:n28_nh3_agn}, \ref{subfig:s2_nh2_agn}, and \ref{subfig:s2_nh3_agn}) it is also highlighted that when the gas density increases, CRs release their energy via collisions and thus heat the gas and boost emission lines in a less deep layer. For \eliz{the lower ionization parameter, $\log \rm{U} =-3.0$, photoionization} dominates the heating of the gas up to $\sim 10^{17}\, \rm{cm}$ for $n_\text{H} = 10^3\, \rm{cm^{-3}}$ and then the CRs govern also shallower layers as shown in the teal shadowed area (Fig. \ref{subfig:temp_nh3_agn}). On the other hand, the same ionization parameter value combined with CRs, for less dense gas of $n_\text{H} = 100\, \rm{cm^{-3}}$ penetrates and excites the gas up to $\sim 10^{18}\, \rm{cm}$ (Fig. \ref{subfig:temp_nh2_agn}) and then the CRs dictate the ionization in deeper cloud areas. This behavior is consistent with the fact that the mean free path of a CR is inversely proportional to the number density of the target particles in the gas clouds.


The behavior of the electron temperature is mirrored in lower ionization lines such as [\ion{N}{ii}]$\lambda$6584\AA, [\ion{S}{ii}]$\lambda \lambda$6716,6731\AA, and [\ion{O}{i}]$\lambda$6300\AA, with ionization potentials of 14.53 eV, 10.36 eV, and 13.62 eV, respectively. The sensitivity of the aforementioned emission lines is also reflected on the behavior of \textsc{Cloudy} models in the BPT diagrams, due to the increase of the CR ionization rate, as presented in Section \ref{subsec:bpts}, and it is evident that the lines with the lower ionization potentials are immediately affected by the increase in the CR ionization rate (Figs.~\ref{subfig:n28_nh2_agn}--\ref{subfig:o1_nh2_sf}).


Furthermore, we notice that photoionization prevails in the illuminated layers, while CRs become more significant in the deeper layers of the cloud, where the photoionization does not reach. All of the aforementioned emission lines' structure diagrams, (Figs. \ref{fig:struc_bpt1}, \ref{fig:struc_bpt2}) show this deeper additional emission originating from inside of the cloud. The results of the presence of low-energy CRs are not the same on all emission lines. [\ion{O}{iii}]$\lambda$5007\AA, H$\alpha$, and H$\beta$ are mildly affected by the higher CR rates (Figs. \ref{subfig:ha_nh3_agn}--\ref{subfig:hb_nh2_sf}), confirming the relocation of the models towards the upper right area of the BPT diagrams and not in other direction.

When comparing the effect of different ionizing sources on the emission lines, including high or low CR ionization rates, we find that AGN and SF models exhibit both distinct differences and similarities. The ionizing radiation in AGN originates in the accretion disk around supermassive black holes, emitting a broad spectrum from X-rays to UV light, often including a significant nonthermal contribution at high energies. \eliz{In contrast, the UV radiation in star-forming galaxies comes from hot and young massive stars, 
characterized by a thermal continuum spectrum that drops sharply above the helium double ionization edge at $54\, \rm{eV}$ \citep[e.g. fig. 4 in][]{Tumlinson_2000}.} The different ionizing continuum in both cases is probably the cause of the smoother temperature profile vs. depth seen in AGN (Fig.~\ref{subfig:temp_nh2_agn}), as compared with the sharp profile in SF models (Fig.~\ref{subfig:temp_nh2_sf}). This is likely due to the rapid recombination in the SF nebula, while the harder continuum in AGN produces a thicker layer.

Overall, CRs have a clear impact in both AGN and SF models by increasing the gas ionization, especially in regions where the ionized radiation cannot reach (Figs. \ref{fig:temp}, \ref{fig:struc_bpt1}, \ref{fig:struc_bpt2}). However, the change produced in the temperature profile vs. depth is stronger in SF models, as shown by Fig. \ref{fig:temp}, and therefore a more prominent increase in the line emissivity profiles is seen in Figs. \ref{fig:struc_bpt1} and \ref{fig:struc_bpt2}.

\section{Discussion}\label{discussion}

\subsection{Thermal stability at high CR ionization rates}\label{thermal}

In Section \ref{subsec:structure_plots} we found that high CR ionization rates create a high temperature layer deep in the simulated cloud (teal-shaded area in Fig.~\ref{fig:temp}), enhancing the emission from lower ionization lines (Fig. \ref{fig:struc_bpt1}). To further study the effects of CRs in the thermal stability of the ionized cloud, it is convenient to adopt the definition of the ionization parameter $\Xi$ from \citet{Krolik_1981}, which probes the radiation to gas pressure ratio:
\begin{equation}\label{eq_Xi}
    \Xi = \frac{F_{\rm ion}}{n_{\rm H} k_B T_{\text{e}} c} = \frac{L_{\rm ion}}{n_{\rm H} r^2 4 \pi k_B T_{\text{e}} c} = \frac{\xi}{4 \pi k_B T_{\text{e}} c}
\end{equation}
where $F_{\rm ion}$ and $L_{\rm ion}$ are the ionizing flux between $1$ and $10^3\, \rm{Ryd}$ and its corresponding luminosity, $r$ is the distance from the cloud to the ionizing source, \( T_{\text{e}} \) is the electron temperature, \( k_B \) is the Boltzmann constant, \( c \) is the speed of light, and \( \xi \) is the ionization parameter definition from \citet{Tarter_1969}.

\begin{figure*}[!!!!!!!!!!!!!!!htb]
    \centering
    \subfigure[ \( T_{\text{e}} \) in AGN models for $n_{\rm H}\, =\, 100 \, \rm{cm^{-3}}$.]{\includegraphics[width=0.4\textwidth]{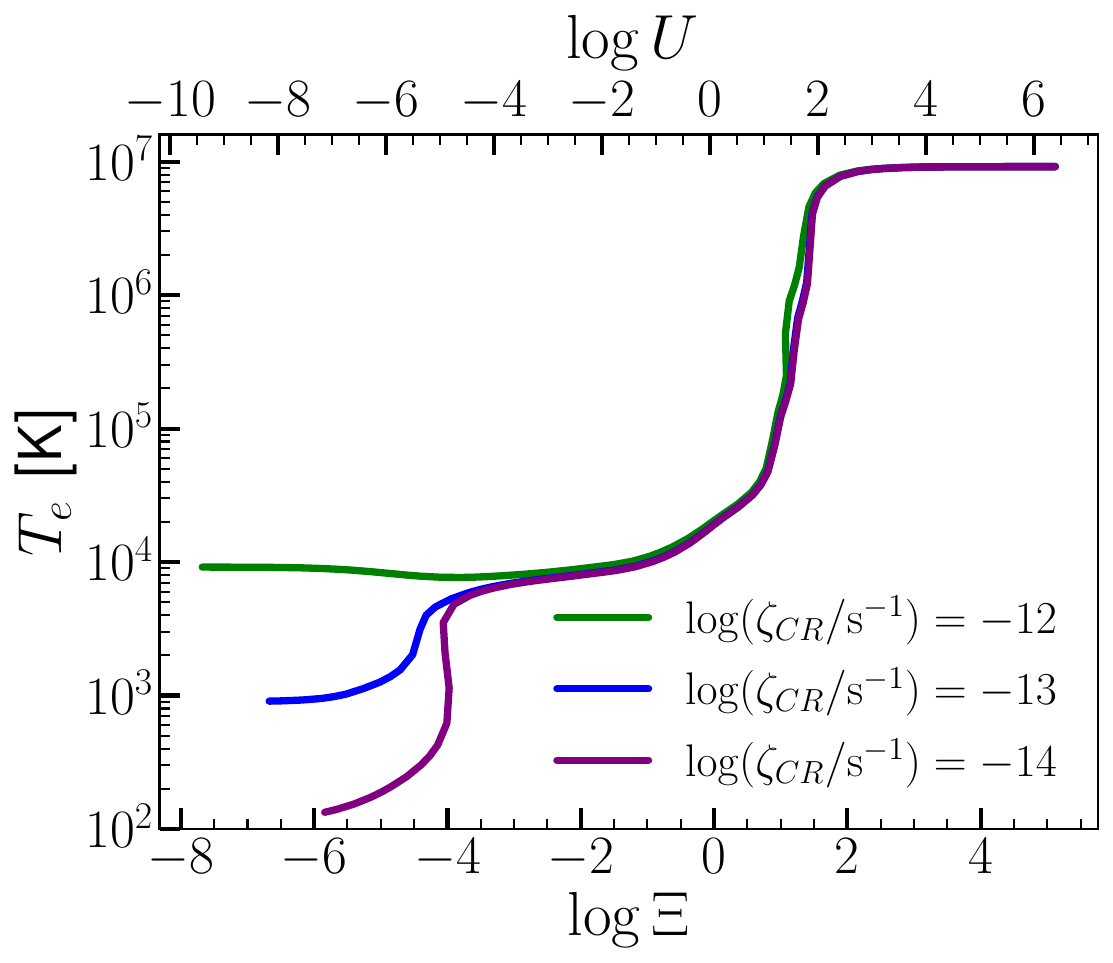}\label{subfig:stab_agn}}~
    \subfigure[ \( T_{\text{e}} \) in SF models for $n_{\rm H}\, =\, 100 \, \rm{cm^{-3}}$.]{\includegraphics[width=0.4\textwidth]{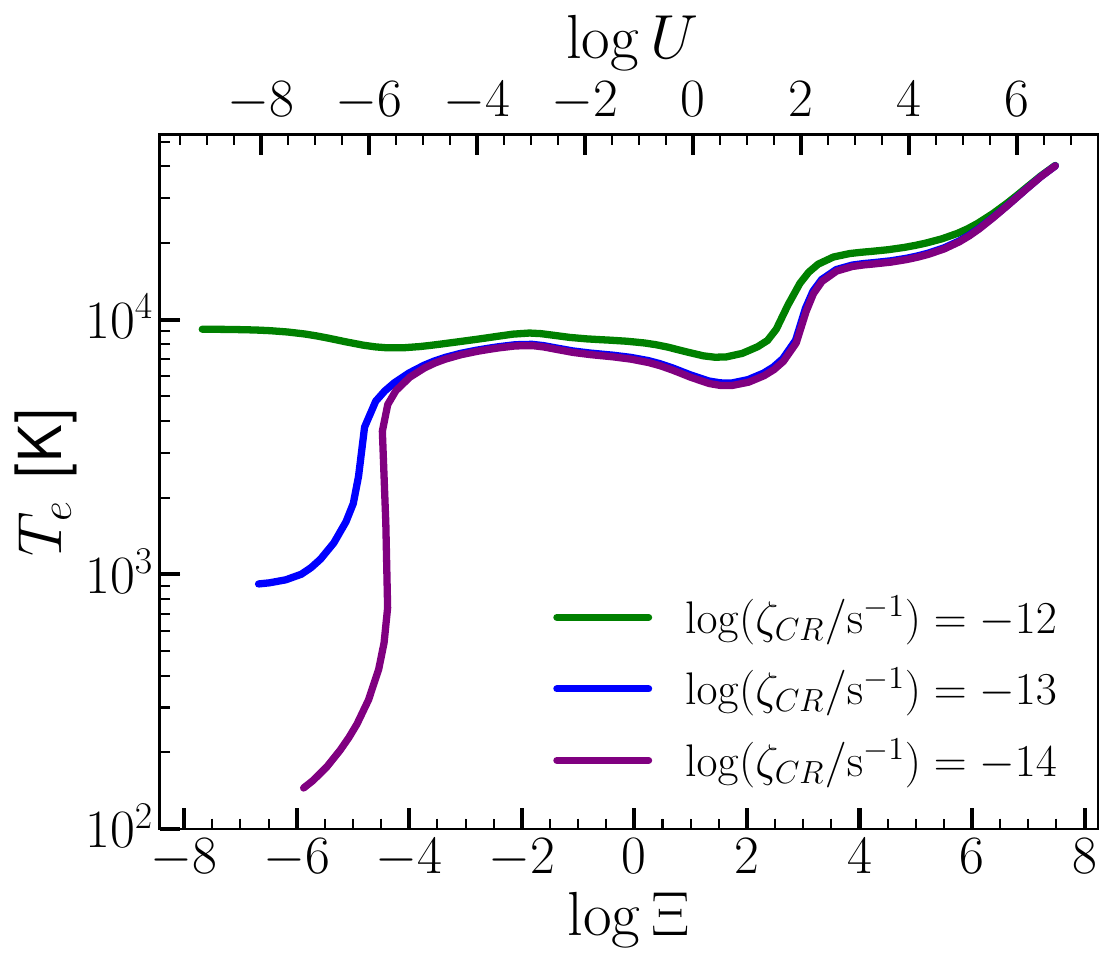}\label{subfig:stab_sf}}
    \caption{Thermal stability plots, depicting electron temperature as a function of the ionization parameter $\Xi$ for different values CR ionization rate, $10^{-14}\, \rm s^{-1},\,10^{-13}\, \rm s^{-1},\,10^{-12}\, \rm s^{-1}$ illustrated with purple, blue and green lines respectively. The upper x-axis shows the corresponding values for the ionization parameter $U$.}\label{fig:stab}
    \vspace{-0.3cm}
\end{figure*}

Thus, following the approach in \citet{Krolik_1981}, we analyze the thermal stability of AGN and SF models including CRs by deriving the dependence of \( T_{\text{e}} \) on \( \Xi \) for different values of the CR ionization rate. For this purpose, we ran additional simulations with the same characteristics as those detailed in Section \ref{Cloudy_models}, but for a wider range of ionization parameter values, for three different CR rates $10^{-14}\, \rm s^{-1},\,10^{-13}\, \rm s^{-1},\,10^{-12}\, \rm s^{-1}$. The results for AGN and SF models are shown in Figs. \ref{subfig:stab_agn} and \ref{subfig:stab_sf}, respectively.

The thermally stable solutions correspond to positive slopes in the $\Xi$--$T_{\text{e}}$ curve, since isobaric perturbations increasing the temperature in these regions correspond also to an increase in the gas cooling efficiency. On the other hand, vertical or negative slopes correspond to thermally unstable solutions, because an increase in temperature will produce a larger contribution of heating processes, thus causing a further increase in the temperature and the amplification of any perturbation. Above $\Xi \gtrsim -4$ ($\log U \gtrsim -5$), the models corresponding to the three different CR rates behave similarly, with negligible differences among the stability curves in Figs. \ref{subfig:stab_agn} and \ref{subfig:stab_sf}. AGN models in Fig. \ref{subfig:stab_agn} show, as expected, two different stable gas phases \citep{Krolik_1981}: ionized gas at $T_{\text{e}} \sim 10^4\, \rm{K}$ for $-4 \lesssim \Xi \lesssim 0.5$, and coronal gas at $\sim 10^7\, \rm{K}$ for $\Xi \gtrsim 1.5$. Similarly, Fig. \ref{subfig:stab_sf} shows a thermally stable solution for ionized gas in star-forming regions in the $-4 \lesssim \Xi \lesssim 2$ range. Higher temperatures ($\gtrsim 2 \times 10^4$) can be reached at extreme ionization parameter values of $\Xi \gtrsim 3$, however the lack of a hard X-ray continuum prevents the coronal stability branch as shown in the case of AGN models.

However, below $\Xi \lesssim -4$, the thermal stability of the gas shows a strong dependence on the CR ionization rate for both AGN and SF models. Models with the highest CR ionization rate value of $10^{-12}\rm s^{-1}$ are able to keep the ionized gas at temperatures of about $10^4\, \rm{K}$ even at very low $\Xi$ values. Thus, their contribution to the gas heating extends the thermal stability of the simulated cloud to regions where the ionization processes are negligible. For an ionization rate of $10^{-13}\rm s^{-1}$, CRs can still keep the gas in a stable phase at $10^3\, \rm{K}$, while values below $\lesssim 10^{-14}\rm s^{-1}$ are no longer able to sustain a stable solution as depicted in Figs. \ref{subfig:stab_agn} and \ref{subfig:stab_sf}. These results are in agreement with the theoretical analysis in \citet{Walker_2016}, who demonstrates that high fluxes of low-energy CRs ($\lesssim 1\, \rm{GHz}$) are able to efficiently heat the warm ionized medium and maintain a high gas temperature against radiative cooling.


The stability diagrams demonstrate that the effect of CRs is significant at high CR ionization rates and low ionization parameter values. This is relevant for ionization-bounded clouds when the flux of UV photons does not reach the deepest layers in the gas clouds, which are shielded from the radiation and thus are heated mainly via CRs. This is consistent with the results shown in the structure plots in Figs. \ref{fig:struc_bpt1} and \ref{fig:struc_bpt2}, in Section \ref{subsec:structure_plots}, where we observe two separate domains in the simulated cloud, one dominated by photoionization, next to the illuminated face of the cloud, and a deeper one dominated by CRs where the electron temperature remains high. The strong dependence of the temperature on the CR ionization rate observed at large cloud depths in Figs. \ref{fig:struc_bpt1} and \ref{fig:struc_bpt2}, and at low $\Xi$ in Figs. \ref{subfig:stab_agn} and \ref{subfig:stab_sf}, is likely associated with the dissociation of H$_2$ gas by the CRs. 

Models developed by \citet{Ferland_2009} show that, as the fraction of free electrons rises due to H$_2$ dissociation and ionization, the gas heating becomes increasingly more efficient, due to the larger probability of CR secondary electrons to collide with the thermal electrons. 
{This indicates that the greatest portion of the CR energy, beyond $10^{-13} \rm s^{-1}$, is directed towards heating rather than H$_2$ dissociation leading to thermal instabilities \citep{Ferland_2009}. This effect,  prominent in the ionized gas,} produces a steep change in the gas temperature predicted above $\zeta_\mathrm{CR} \gtrsim 3\times 10^{-13}\, \rm{s^{-1}}$ (see fig. 8 in \citealt{Ferland_2009}).



\subsection{CR heating and low ionization nebular lines}\label{subsec:NOS_lines}

The temperature boost caused by CR heating deep in the nebula at high $\zeta_\mathrm{CR}$ values (Fig. \ref{fig:temp}) produces an extra contribution to the low-ionization nebular lines. As shown in Section \ref{subsec:structure_plots}, the electron heating by CRs increases significantly the emissivity of the low-ionization collisional lines, that is, [\ion{N}{ii}]$\lambda$6584\AA, [\ion{S}{ii}]$\lambda \lambda$6716,6731\AA, \ and [\ion{O}{i}]$\lambda$6300\AA \ (Figs. \ref{subfig:n28_nh2_agn}--\ref{subfig:n28_nh2_sf} and \ref{subfig:s2_nh2_agn}--\ref{subfig:s22_nh2_sf}). In contrast, the increase in emissivity of high-ionization and recombination lines such as [\ion{O}{iii}]$\lambda$5007\AA, H$\alpha$, and H$\beta$ is much lower (Figs. \ref{subfig:ha_nh2_agn}--\ref{subfig:hb_nh2_sf}). Thus, the different response of the BPT lines to CR heating drives both AGN and SF models to the right side along the horizontal axes in Figs. \ref{fig:cent_BPTS_U}, \ref{fig:1068_BPTS_U}, and \ref{fig:253_BPTS_U}.

Nevertheless, the three line ratios including low-ionization lines discussed here seem to be best reproduced by different $\zeta_\mathrm{CR}$ values. The measured [\ion{N}{ii}]/H$\alpha$ ratios in Centaurus A, NGC 1068 and NGC 253 (Figs. \ref{subfig:U_cent_12}, \ref{subfig:U_1068_12} and \ref{subfig:U_253_12}, respectively) are reproduced by models with $\zeta_\mathrm{CR} = 10^{-12}\, \rm s^{-1}$, while the measured [\ion{S}{ii}]/H$\alpha$, and [\ion{O}{i}]/H$\alpha$ ratios for the same regions tend to be overestimated. These are best reproduced by $\zeta_\mathrm{CR} = 10^{-13}\, \rm s^{-1}$ in Figs. \ref{subfig:U_cent_13}, \ref{subfig:U_1068_13}, and \ref{subfig:U_253_13}, and possibly by $\zeta_\mathrm{CR} \lesssim 10^{-14}\, \rm s^{-1}$ for [\ion{O}{i}]/H$\alpha$ in NGC 1068 (Fig. \ref{subfig:U_1068_14}). These transitions have different ionization potentials, with [\ion{N}{ii}]$\lambda$6584\AA \ showing the largest value ($14.5\, \rm{eV}$), followed by [\ion{S}{ii}]$\lambda \lambda$6716,6731\AA \ ($10.4\, \rm{eV}$) and [\ion{O}{i}]$\lambda$6300\AA \ (neutral gas). This implies that [\ion{S}{ii}]$\lambda \lambda$6716,6731\AA \ and [\ion{O}{i}]$\lambda$6300\AA \ emission can be produced also in neutral gas regions located farther from the ionizing source than the N$^+$ gas, as has been reported in resolved \textsc{H\,ii} regions \citep[e.g.][]{Scowen_1998,Barman_2022}. Additionally, the [\ion{O}{i}]$\lambda$6300\AA \ line is very sensitive to shocks \citep{Dopita_1976} and X-ray radiation \citep{Polles_2021}. Thus, a possible explanation for the lower $\zeta_\mathrm{CR}$ values required by [\ion{S}{ii}]$\lambda\lambda$6716,6731\AA \ and [\ion{O}{i}]$\lambda$6300\AA \ could be related to inhomogeneities in the nebular structure and the different physical conditions in the regions where these lines form, which may receive on average a lower CR flux.
Photoionization modeling of several emission-line ratios with a simple geometric configuration is a challenging task, as shown by the different coverage of the observed distribution in [\ion{N}{ii}]/H$\alpha$, [\ion{S}{ii}]/H$\alpha$  and [\ion{O}{i}]/H$\alpha$, for the same grid of models \citep[e.g.][]{2004bGroves,Feltre_2016,Zhu_2023}. For instance, the models that best match the [\ion{O}{i}]/H$\alpha$ ratio in the BPT diagram typically have lower ionization parameters and lower metallicities when compared to the models that best match the [\ion{N}{ii}]/H$\alpha$ distribution. This suggests that an accurate description of lines formed in the neutral gas region requires a more sophisticated treatment and possibly also a more complex geometric model. 

\subsection{Reconciling models with the observed metallicities}
\label{obs_metal}

\begin{figure*}[!ht]
    \centering
    \subfigure[]{\includegraphics[width=0.31\textwidth]{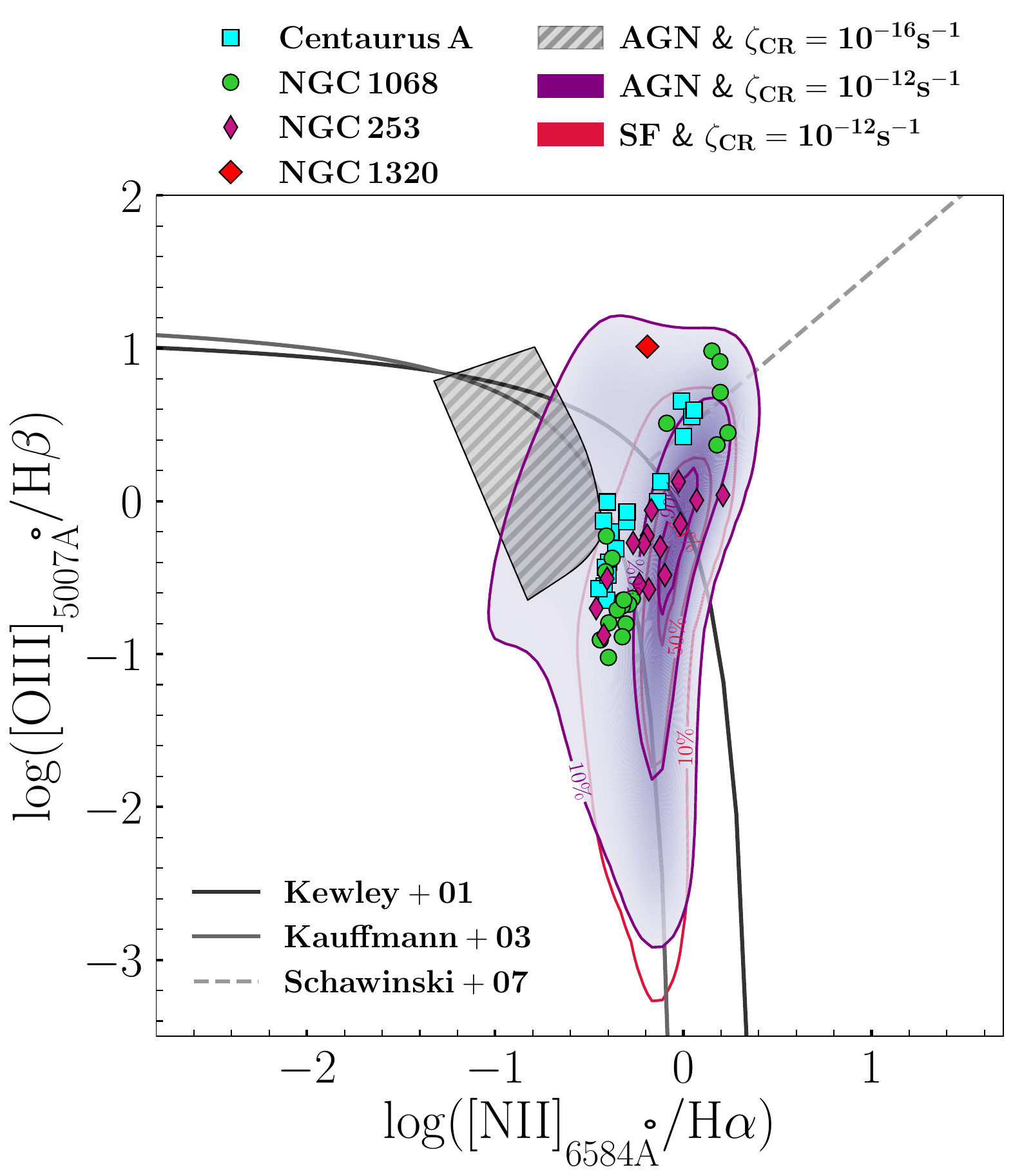}\label{subfig:BPT_ALL_N2}}~
    \subfigure[]{\includegraphics[width=0.31\textwidth]{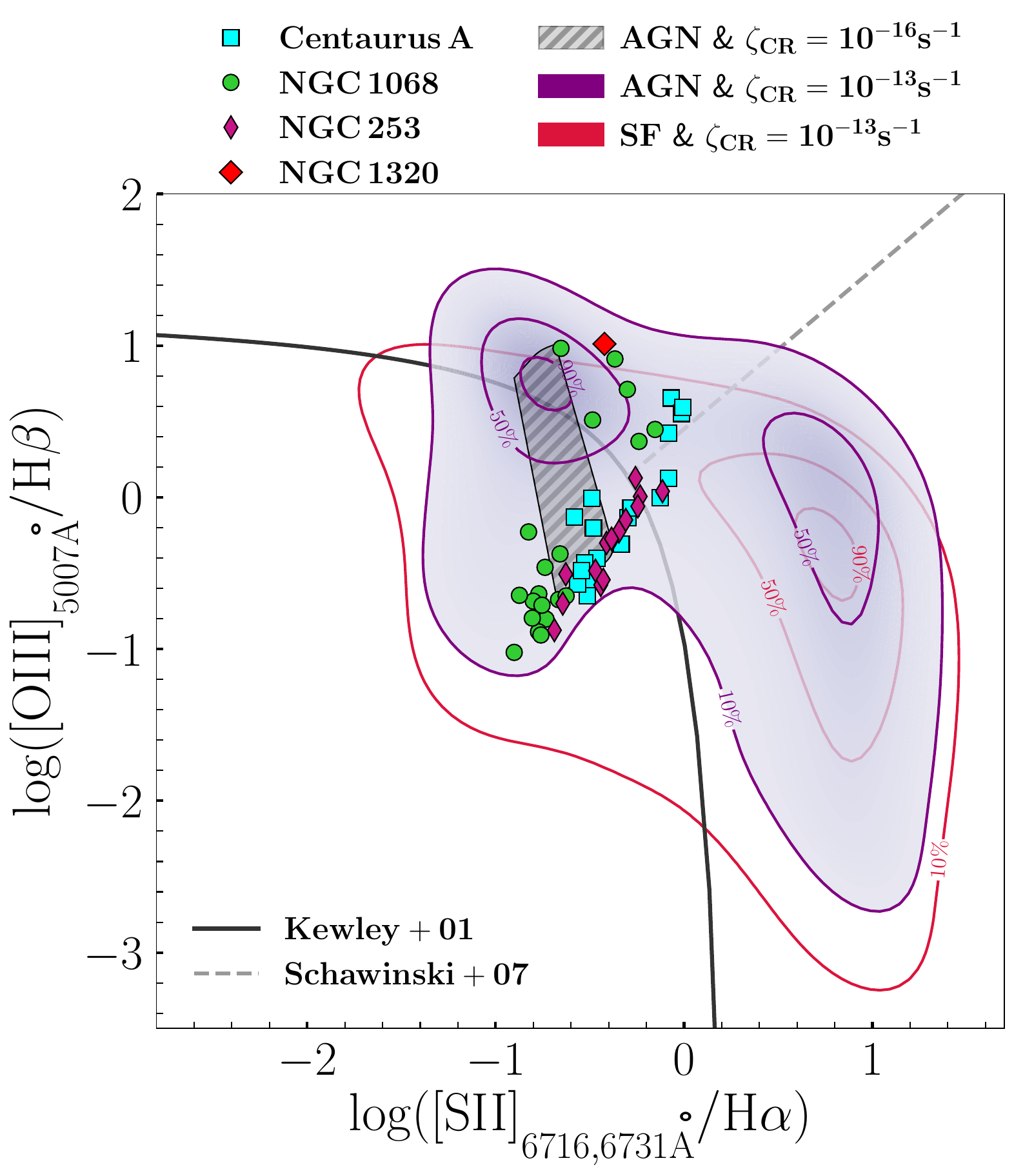}\label{subfig:BPT_ALL_S2}}~
    \subfigure[]{\includegraphics[width=0.31\textwidth]{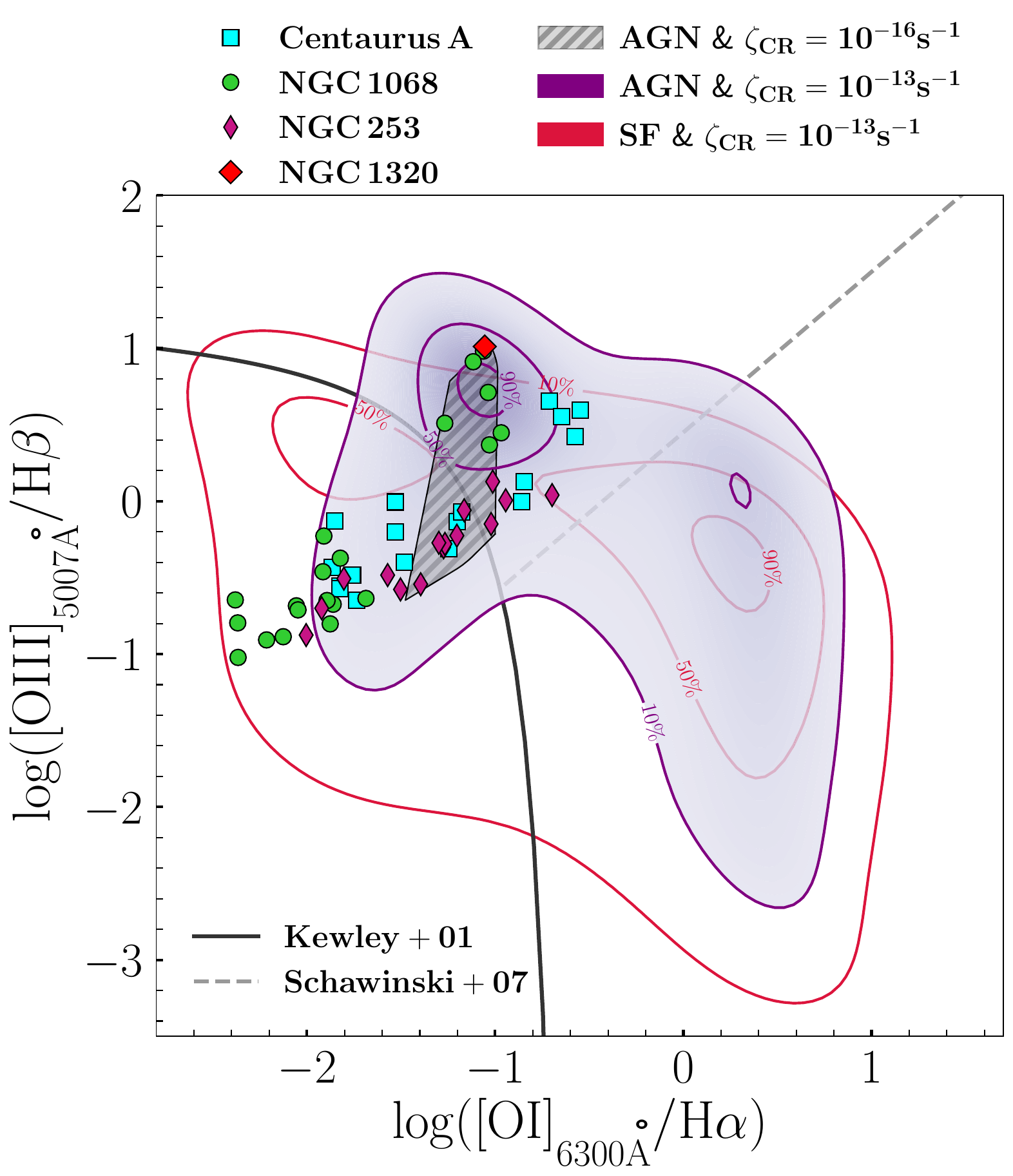}\label{subfig:BPT_ALL_O1}}
    \caption{\eliz{BPT diagrams depicting the area covered by AGN and SF models with solar abundances for  $-3.5 \leq \log U \leq -1.5$, and $1 \leq \log n_\text{H} \leq 3.5$, are shown with purple and red contours, respectively.
    The solid contour lines map regions containing 10\%, 50\%, and 90\% of the models with $\zeta_\mathrm{CR} = 10^{-12}\, \rm s^{-1}$ in the [\ion{N}{ii}] BPT diagram, and $10^{-13}\, \rm s^{-1}$ in both the [\ion{S}{ii}] and the [\ion{O}{i}] panels. The gray hatched area stands for AGN models with $\zeta_\mathrm{CR} = 10^{-16}\, \rm{s^{-1}}$, of the order of the average Galactic CR background \citep{Indriolo_2007}. Cyan squares represent line ratios measured for the regions selected in Centaurus A, 
    green circles for regions in NGC 1068, magenta 
    thin diamonds for regions in NGC 253, and the red diamond correspond to the photoionization-dominated Seyfert 2 nucleus in NGC 1320. The Kewley, Kauffmann, and Schawinski lines correspond to the solid dark gray, the solid medium gray, and dashed light gray lines, respectively.}}
    \label{fig:BPT_ALL}
        \vspace{-0.2cm}
\end{figure*}
One of the main points of tension between models that aim to reproduce the Seyfert/LINER locus in the BPT diagram is the supersolar metallicities ($\sim$$2$--$3\, \rm{Z_\odot}$) that most models require to match the observed distributions in the emission-line ratios \citep[e.g.][]{2004aGroves,2004bGroves,Thomas_2016,Feltre_2016,Zhu_2023}. Such high metallicities are, however, in contrast with recent measurements of the chemical abundances in the NLR of hundreds of Seyfert and LINER galaxies in the local Universe, suggesting that the nebular gas in nearby AGN typically has slightly subsolar to solar-like abundances \citep{Dors_2019,Perez_2019,Dors_2020,Perez_Diaz_2021,Perez_Diaz_2022,Dors_2022,Perez_2023,Perez_Diaz_2024}. These results are supported by different methods applied to emission lines in the UV, optical, and IR ranges.

Photoionization models with solar abundances tend to underestimate the [\ion{N}{ii}]/H$\alpha$ ratio, and to a lesser extent, the [\ion{S}{ii}]/H$\alpha$ ratio \citep[e.g. fig. 3 in][]{Zhu_2023}. \eliz{Including CR ionization allows us to reproduce the LINER/Seyfert domain in BPT diagrams using solar abundances, while previous studies proposed higher than solar metallicities, supersolar N/O ratios, or harder ionizing SEDs with steeper power-law spectral indices \citep{Feltre_2016} or hotter accretion disks \citep{Zhu_2023}.} In this context, the main outcome of the present study is the significant role that CR heating may play in diagnostics based on line ratios, such as the BPT diagram. As discussed in Sections \ref{subsec:structure_plots} and \ref{subsec:NOS_lines}, the heating of electrons caused by CRs deep in the nebula produces an additional contribution to the emissivity of the low-ionization collisional lines, that is, [\ion{N}{ii}]$\lambda$6584\AA, [\ion{S}{ii}]$\lambda \lambda$6716,6731\AA \ and [\ion{O}{i}]$\lambda$6300\AA \ (Figs. \ref{subfig:n28_nh2_agn}--\ref{subfig:n28_nh2_sf} and \ref{subfig:s2_nh2_agn}--\ref{subfig:s22_nh2_sf}). In contrast, the emissivity increase of high-ionization and recombination lines such as [\ion{O}{iii}]$\lambda$5007\AA, H$\alpha$, and H$\beta$ is much lower (Figs. \ref{subfig:ha_nh2_agn}--\ref{subfig:hb_nh2_sf}). Consequently, the different response of the BPT diagnostic lines to CR heating drives AGN models towards the right along the horizontal axes in Figs.~\ref{fig:cent_BPTS_U}, \ref{fig:1068_BPTS_U}, and \ref{fig:253_BPTS_U}, shifting solar-metallicity models into the Seyfert/LINER domain. This result is summarized in Fig.~\ref{fig:BPT_ALL}, which shows the three BPT diagrams including the measured line ratios for all the selected regions in the three sample galaxies presented in Fig.~\ref{fig:chosen_apertures_o3}. The different contours represent the 10th, 50th, and 90th percentiles of the 2D distribution of AGN and SF models. The probability distribution was obtained from the discrete model grids with $1 \leq \log n_\text{H} \leq 3.5$ and $-3.5 \leq \log U \leq -1.5$, using a Gaussian kernel density estimation \citep{Silverman_1986}. The gray-hatched area represents the region occupied by photoionization models including an average Galactic CR ionization rate of $10^{-16}\, \rm{s^{-1}}$ \citep{Indriolo_2007}.

Models with solar abundances and a CR ionization rate of $\zeta_\mathrm{CR} = 10^{-12}\, \rm{s^{-1}}$ in Fig.~\ref{subfig:BPT_ALL_N2} (purple and red contours for AGN and SF models, respectively) are in excellent agreement with the observed distribution in the [\ion{O}{iii}]/H$\beta$ and [\ion{N}{ii}]/H$\alpha$ ratios. The latter are stretched along a relatively narrow stripe from the star formation domain to the Seyfert/LINER division line, which is consistent with an increasing $\log U$ in the models (see Figs.~\ref{subfig:U_cent_12}, \ref{subfig:U_1068_12}, and \ref{subfig:U_253_12}). In contrast, solar metallicity models with the average Galactic $\zeta_\mathrm{CR} = 10^{-16}\, \rm{s^{-1}}$ (gray-hatched area) underestimate the [\ion{N}{ii}]/H$\alpha$ ratio by an order of magnitude. Alternatively, high N/O relative abundances could also explain the enhanced [\ion{N}{ii}]/H$\alpha$ ratios observed \citep{Perez_2019,Ji_2020a}. \eliz{However, the N/O ratios derived with \textsc{HII-CHI-Mistry} for the regions in our sample of galaxies suggest instead values close to the solar abundance ratio of $\log \text{(N/O)}_\odot \sim -0.86\, \rm{dex}$ (see Table~\ref{tab:parameters}; \citealt{Asplund})}. Although CR heating can explain an increase in the emissivity of low-ionization transitions for both AGN and SF models, as shown in Fig.~\ref{fig:BPT_ALL}, other mechanisms such as X-ray heating or shocks may also contribute and cannot be discarded. For instance, X-ray excitation is able to produce bright [\ion{N}{ii}]$\lambda6584$\AA, [\ion{S}{ii}]$\lambda\lambda6716,6731$\AA, and [\ion{O}{i}]$\lambda$6300\AA \ emission \citep[e.g.][]{Polles_2021,Wolfire_2022,Lagos_2022}. Photoionization models with shocks can also explain the high [\ion{N}{ii}]/H$\alpha$, [\ion{S}{ii}]/H$\alpha$, and [\ion{O}{i}]/H$\alpha$ ratios in AGN and SF regions \citep{Contini_2001a,Contini_2001b,Allen_2008,Dors_2021}.

The observed [\ion{S}{ii}]/H$\alpha$ and [\ion{O}{i}]/H$\alpha$ ratios in Figs.~\ref{subfig:BPT_ALL_S2} and \ref{subfig:BPT_ALL_O1}, respectively, are closer to the models with the average Galactic $\zeta_\mathrm{CR}$ value, although an additional contribution to the [\ion{S}{ii}]$\lambda\lambda$6716,6731\AA \ line is still required, as well as for [\ion{O}{i}]$\lambda$6300\AA \ in some of the regions in Centaurus A. A significantly lower $\zeta_\mathrm{CR} = 10^{-13}\, \rm{s^{-1}}$ value than in the [\ion{N}{ii}] case causes a substantial stretch in the model grids along the horizontal axes, while the line ratio distributions tend to favor models with moderate densities ($n_\mathrm{H} \sim 100\, \rm{cm^{-3}}$; see Figs.~B.\ref{subfig:nh_cent_13}, B.\ref{subfig:nh_1068_13}, and B.\ref{subfig:nh_253_13}). Nevertheless, SF models with $\zeta_\mathrm{CR} = 10^{-13}\, \rm{s^{-1}}$ can explain the [\ion{S}{ii}]/H$\alpha$ and [\ion{O}{i}]/H$\alpha$ ratios measured for the regions in NGC 253 that are located on the right side of the Kewley division line, beyond the limit of pure SF photoionization models.

At very low densities ($0 \leq n_\text{H} \leq 1$) with $\zeta_\mathrm{CR} \gtrsim 10^{-13}\, \rm{s^{-1}}$, the models predict that [\ion{S}{ii}]/H$\alpha$ and [\ion{O}{i}]/H$\alpha$ ratios fall within the LINER domain (Figs.~B.\ref{subfig:nh_cent_13}, B.\ref{subfig:nh_1068_13}, and B.\ref{subfig:nh_253_13}) for most of the $\log U$ values (Figs.~\ref{subfig:U_cent_13}, \ref{subfig:U_1068_13}, and \ref{subfig:U_253_13}). This suggests that heating of low-density gas by CRs could also represent an important contribution to the Diffuse Ionized Gas emission in galaxies \citep{Vandenbroucke_2018}, which is characterized by enhanced [\ion{N}{ii}]/H$\alpha$, [\ion{S}{ii}]/H$\alpha$, and [\ion{O}{i}]/H$\alpha$ ratios \citep[e.g.][]{Zhang_2017}.

\subsection{CR effect on metallicity and ionization parameter}\label{CR_impact_Z_U}


\eliz{Line ratios involving [\ion{N}{ii}]$\lambda 6584$\AA \ \citep{Marino_2013,Carvalho_2020, Oliveira_2024}, and [\ion{S}{ii}]$\lambda \lambda 6716,6731$\AA \ \citep{Christensen_1997,Perez_2006} have been used as metallicity tracers for gaseous nebulae. In particular, the index N2 = [\ion{N}{ii}]/H$\alpha$ is widely adopted as metallicity indicator \citep{Marino_2013,Carvalho_2020}, and also used to compute the ionization parameter \citep{netzer}. Given the CR influence on the emission lines from low-ionization species predicted by our models, a significant impact on the metallicity and ionization parameter is expected. }

\eliz{To quantify the effect of CRs on metallicity and ionization parameter estimations, we applied \textsc{HII-CHI-Mistry} using as input the median values of different combinations of emission-line ratios as the CR ionization rate increases from $10^{-16}\rm s^{-1} $ to $10^{-12}\rm s^{-1}$ for both AGN and SF models. When the auroral line [\ion{O}{iii}]$\lambda$4363\AA \, is included, in the \textsc{HII-CHI-Mistry} routines, the temperature is constrained \citep{Perez_2014} and thus the metallicity estimation does not show a clear trend with increasing $\zeta_\text{CR}$, although the derived abundances are subsolar and the associated errors are quite large ($\sim$0.7 dex). The N/O abundance is relatively robust, as CRs affect similarly both the [\ion{O}{ii}]$\lambda 3725$\AA \ and [\ion{N}{ii}]$\lambda 6584$\AA \ lines, whose ratio is the main tracer used by \textsc{HII-CHI-Mistry} to determine N/O. Nevertheless, \eliz{the ionization parameter} tends to show lower values with increasing $\zeta_\text{CR}$ for AGN models, while SF models seem to be more stable.}

\eliz{
Considering only the BPT lines, the enhancement caused by CRs on the [\ion{N}{ii}]/H$\alpha$ line ratio (see Sec. \ref{subsec:NOS_lines}) is not balanced now by the missing [\ion{O}{ii}]$\lambda 3725$\AA, thus leading to underestimated N/O values and uncertain O/H and $\log U$ estimations. The increase in the derived metallicities between $\zeta_\text{CR} = 10^{-16}\, \rm{s^{-1}}$ and $10^{-12}\, \rm{s^{-1}}$ is $\sim 25\%$, and $\sim 65\%$ for AGN and SF models, respectively. This is likely caused by the lower [\ion{O}{iii}]/[\ion{N}{ii}] ratios as CR ionization rate grows larger.}

\eliz{
On the other hand, using [\ion{N}{ii}]/H$\alpha$ as the only input results in larger metallicities by approximately 40\%, and 45\% for AGN and SF models, respectively. This is due to the enhancement of the [\ion{N}{ii}]/H$\alpha$ ratio with increasing $\zeta_\text{CR}$, which is interpreted by \textsc{HII-CHI-Mistry} as a decreasing $\log U$. 
For instance, in our AGN and SF models, the $\log U$ varies by 1 dex and 0.2 dex, respectively. Thus, the resulting metallicities are overestimated in SF models due to the known empirical anti-correlation between $\log U$ and O/H, which \textsc{HII-CHI-Mistry} exploits. Additionally, high [\ion{N}{ii}]/H$\alpha$ ratios for AGN in \textsc{HII-CHI-Mistry} are only explained by models with low $\log U$ and high O/H \citep[see fig. 4 in][]{Perez_2019}.} 

\eliz{
Finally, we applied the metallicity calibrations based on the N2 ratio from \cite{Carvalho_2020}, for AGN, and \cite{Marino_2013}, for star-forming galaxies, to the median N2 values in our AGN and SF models with CR ionization rates in the $10^{-16}\rm s^{-1} $ to $10^{-12}\rm s^{-1}$ range. We find that the predicted metallicities are approximately 5 times larger for $10^{-12}\rm s^{-1}$ models when compared to $10^{-16}\rm s^{-1}$ models in the AGN case, while in the star-forming case the difference is found to be approximately 1.5 times larger.
} \eliz{Extreme models with densities outside the $1 \leq n_\text{H} \leq 10^3\ \rm{cm}^{-3}$ range were excluded, in all aforementioned calculations, to avoid outliers.}

As we incorporated different emission line ratios into \textsc{HII-CHI-Mistry} we gain some insight of their diagnostic power in the metallicity when CRs are present. This drives us to the conclusion that [\ion{N}{ii}]/H$\alpha$ alone, being strongly affected by CRs, can produce metallicity estimations biased in environments where CRs are prevalent. Therefore, having the temperature estimate through [\ion{O}{iii}]$\lambda4363$\AA \ makes the metallicity calibrations more reliable.


\subsection{Comparison with CR rate estimates}\label{subsec:high_CRs}

The average CR ionization rate measured in the Milky Way is of the order of $\zeta_\mathrm{CR} \sim 10^{-16} \rm s^{-1}$ \citep{Indriolo_2007,Indriolo_2009,Neufeld_2017}, while the CR ionization rate at which photoionization models show a noticeable effect on the BPT diagnostics is considerably higher ($\gtrsim 10^{-13}\rm s^{-1}$). Although higher $\zeta_\mathrm{CR}$ values are anticipated for active nuclei and strong SF environments, produced by nonthermal processes in AGN jets and SNRs \citep{Meijerink_2011,Guo_2011,Phan_2024}, a difference of approximately three orders of magnitude raises the question of how realistic this scenario is for the regions under study in this work. Thus, in this Section we compare the values adopted in our models with observational measurements and independent estimates of CR ionization rates in AGN and SF galaxies found in the literature.

CR ionization rates cannot be directly measured, and therefore our comparison is based on available indirect estimates for nearby galaxies. The CR ionization and heating of H and H$_2$ initiates a series of chemical reactions in molecular gas clouds that lead to the formation of molecular ions such as OH$^+$, H$_2$O$^+$, and H$_3$O$^+$ \citep{Herbst_1973,Gonz_2013}. Therefore, the observed transitions from these molecules can be compared with chemical models that predict their relative abundances as a function of $\zeta_\mathrm{CR}$. Following this approach, in Mrk 231, closest hosting quasar galaxy, CR ionization rates $\sim 5 \times 10^{-13}$ in torus, and $\sim 10^{-12}\, \rm{s^{-1}}$ in outflows were found \citep{Gonz_Alf_2018}. Likewise, \citet{Holdship_2022} derives $\zeta_\mathrm{CR} \sim 10^{-13}\rm  s^{-1}$ within the innermost few hundred parsecs in NGC 253. Slightly higher values ($10^{-13}$--$10^{-12}\, \rm{s^{-1}}$) were obtained by \citet{Behrens_2022} by modeling the HCN/HNC ratio for the same galaxy. These values are in agreement with the results found by \cite{Gonz_2013} in the starburst/AGN composite nuclei of NGC~4418 and Arp~220. Finally, values significantly higher than the Galactic average ($10^{-14}$--$10^{-13}\, \rm{s^{-1}}$) have also been inferred in diffuse molecular clouds near the Galactic center \citep{LePetit_2016}, where a high CR flux is expected.


An estimate of the CR ionization rate can also be derived from the nonthermal radio continuum emission \citep[e.g.][]{Yusef-Zadeh_2013,Gabici_2022}. In AGN, the synchrotron continuum measured along the jet can be used as a proxy for the number density of accelerated electrons. The energy distribution of relativistic electrons can be inferred from the slope of the synchrotron emission continuum \citep{Rybicki_Lightman},
\begin{equation}
  N_{\text{e}} dE \propto E^{-p}dE,
\end{equation}
where $N_{\text{e}}$ is the number density of electron with energies in the range $E$ up to $E + dE$, with $E$ being the energy of electron \citep{Rybicki_Lightman}. The magnetic field strength is required to translate the radiated luminosity into a number density of particles \citep{Padovani_2018}, which is done by assuming equipartition between the energy density of relativistic electrons and the magnetic field energy density of the synchrotron emitting area. Therefore, with this method, we infer the electron component of the CRs, by using the corresponding ionization rate from primary and secondary electrons --\,as displayed below\,-- obtained for eq.~(40) in \cite{Gabici_2022}, and we derive an estimate of the CR ionization rate of atomic hydrogen via the assumption that $1.5\,\zeta_{\text{CR}}^{\text{H}_2} \simeq 2.3\,\zeta_{\text{CR}}^{\text{H}}$ \citep{1974Apj}.
\begin{equation}
\zeta_{\text{CR}}^{\text{H}} \simeq 1.3\pi \int_{E_{\min}}^{E_{\max}} dE \, j_{\text{e}}(E) \, \sigma^{\text{ion}} _{\text{H}_2-\text{e}}(E) \left[1+\phi_{\text{e}, \,\text{H}_2}(E) \right],
\end{equation}
where $j_{\text{e}}(E) = c N_{\text{e}}/(4\pi)$, with $c$ being the speed of light, is the differential energy spectrum of CR electrons, representing the number of electrons with energy \(E\) per unit energy, per unit area, per unit time, and per unit solid angle, $\phi_{\text{e}, \,\text{H}_2}(E)$ represents the secondary ionization by electrons, and we adopt the cross-section of H$_2-{\text{e}}^{-}$ interactions by \cite{Padovani_2018}, $\sigma^{\text{ion}} _{\text{H}_2-{\text{e}}}(E) $. Furthermore, we assume $E_{\min} = 0.1\, \rm MeV$ as in \cite{Yusef-Zadeh_2013} and $E_{\max} = 10\, \rm GeV$ as the upper energy limit which only weakly affects the outcome.

Specifically, in Centaurus A we calculate the equipartition magnetic field of the emitting radio source A1A \citep{Goodger_2010} and for an index $p \sim 2.7$ coming from an indicative index $\alpha \sim 0.85$ based on the $\alpha$ indices of table~7 in \citep{Goodger_2010}, we find it to be $300\, \rm{\mu G}$. The SED and the emitting source radius used in this calculation are provided in table 4 of \cite{Goodger_2010}, while the position of each source is depicted in fig.~2 of the same work. We then acquire a CR ionization rate $\sim 2 \times 10^{-10}\, \rm{s^{-1}}$ in this radio knot. We then repeat the calculation, assuming equipartition, and the same index $p$, in the northern lobe of Centaurus A in the area that covers A1A up to A2A radio knots ($1\, \rm{kpc} \times 100\, \rm{pc}$) of \cite{Goodger_2010}, which corresponds to an effective radius of $\sim 300\, \rm{pc}$. The flux used is the integrated flux of radio knots A1A up to A2A (see table~4 in \citealt{Goodger_2010}). Thus, we derive an equipartition magnetic field of $\sim 8\, \rm{\mu G}$ and a CR ionization rate $\sim 4.3 \times 10^{-13}\, \rm{s^{-1}}$.

In an identical fashion, for the emitting source marked as C in in NGC 1068 \citep{Mutie_2024}, we calculate an equipartition magnetic field of $\sim 700\, \rm \mu G$. The SED, and emitting source radius used in this calculation are extracted from table~2 in \cite{Mutie_2024}, while the position of the source can be seen in fig.~2 of that work. Assuming an electron energy index $p \sim 2.8$, derived from a spectral index $\alpha \sim 0.9$ (see table~2 in \citealt{Mutie_2024}), we determine a CR ionization rate of about $\sim 2 \times 10^{-9}\, \rm{s^{-1}}$. For an effective radius of $160\, \rm{pc}$, representative of the surface of the jet between the components NE up to S3 ($500\,\rm{pc} \times 50\, \rm{pc}$), and using the integrated radio fluxes provided by table~2 in \cite{Mutie_2024}, we estimate an equipartition magnetic field of $\sim 36\, \rm{\mu G}$, and a CR ionization rate of $\sim 1.4 \times 10^{-11}\, \rm{s^{-1}}$.

The CR ionization rate values inferred for Centaurus A and NGC 1068 are indicative of the typical values that can be found along the trail of AGN jets. We expect these values to decrease as we move away from the jet trail, but the numbers obtained from the synchrotron continuum suggest that the order of magnitude of the CR rate in these galaxies is in agreement with the range of values explored in this work. It is also crucial to bear in mind that heavier particles such as protons are not traced by the synchrotron continuum, but can also contribute to gas heating and ionization, suggesting a higher value for the total CR rate \citep{Padovani_2018}. Finally, the CR rates estimated from this method could be subject to future refinement and more detailed analysis.

Overall, both indirect measurements in SF galaxies based on the abundances of molecular transitions, and theoretical estimates based on the synchrotron continuum emission in AGN jets suggest that the CR ionization rates in starburst and active nuclei is expected to be about three to four orders of magnitude higher than the Galactic average, in agreement with the $\zeta_\mathrm{CR} = 10^{-14}$--$10^{-12}\, \rm{s^{-1}}$ range probed by our models.

\begin{figure*}[!ht]
    \centering
    \subfigure[]{\includegraphics[width=0.33\textwidth]{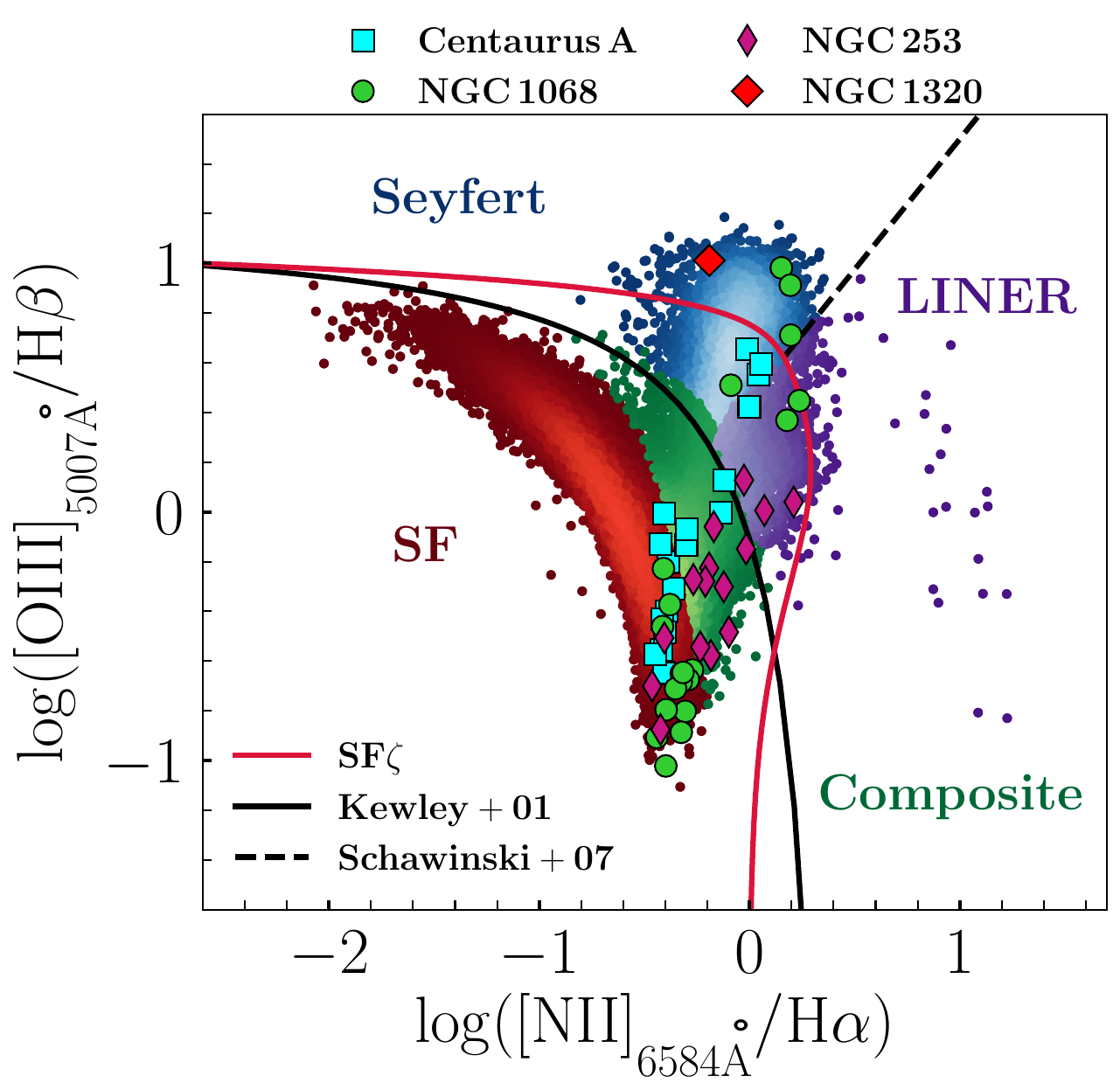}\label{subfig:sfzeta_line_N2}}~
    \subfigure[]{\includegraphics[width=0.33\textwidth]{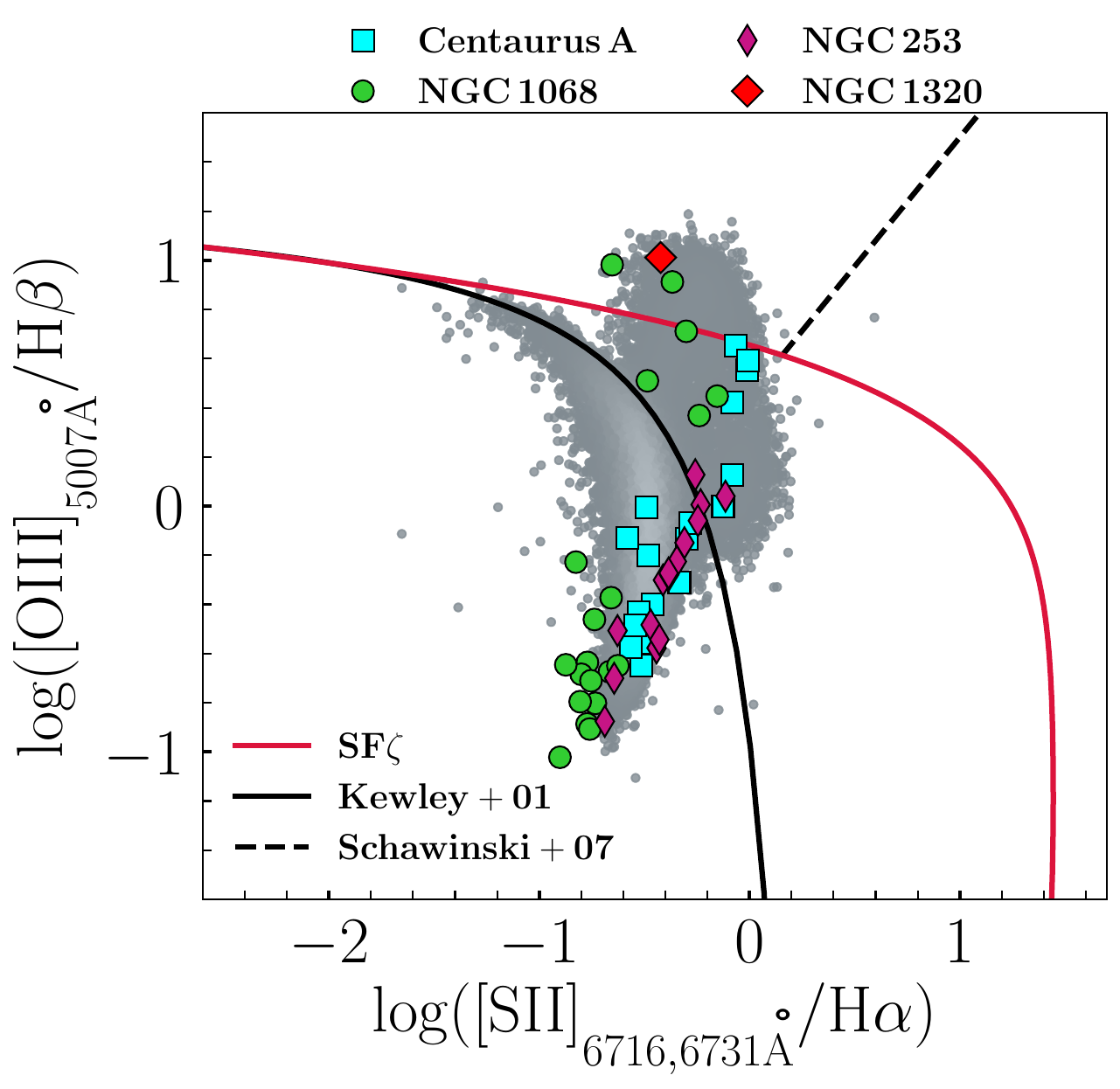}\label{subfig:sfzeta_line_S2}}~
    \subfigure[]{\includegraphics[width=0.33\textwidth]{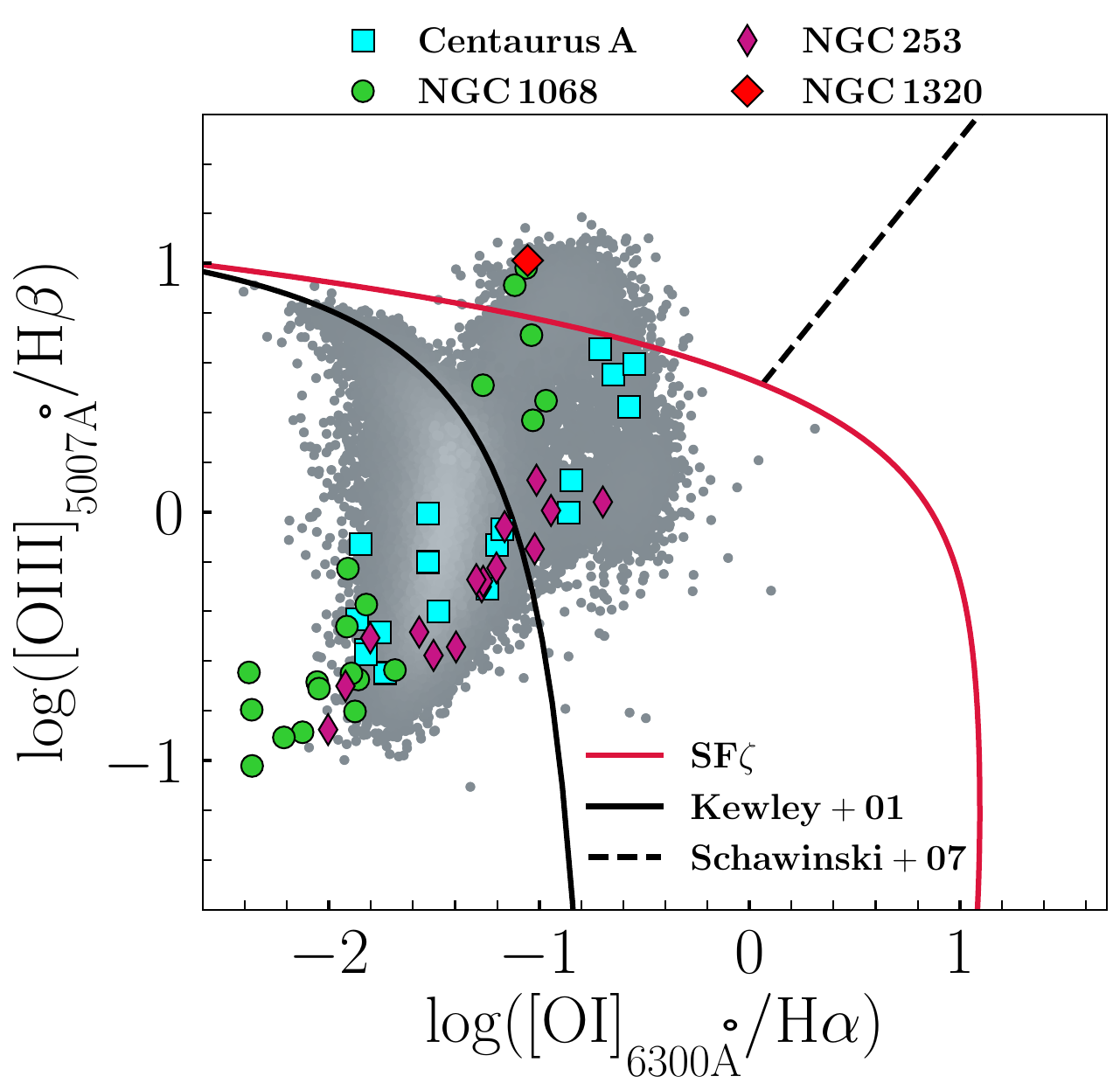}\label{subfig:sfzeta_lineL_O1}}
    \caption{\eliz{BPT diagrams depicting [\ion{N}{ii}]/H$\alpha$, [\ion{S}{ii}]/H$\alpha$, and [\ion{O}{i}]/H$\alpha$ ratios. Our proposed SF\texorpdfstring{$\zeta$}{zeta} maximum starburst line is illustrated in 
    red color, compared with the full observational dataset. Observations from Centaurus A, NGC 1068, NGC 253, and NGC 1320 are marked by cyan squares, 
    green circles, magenta 
    thin diamonds, and a red diamond, respectively. The Kewley and Schawinski lines are indicated with solid and dashed black, respectively, while the SF\texorpdfstring{$\zeta$}{zeta} line is depicted by the red-solid line. In the background we show the line ratios measured for nearby galaxies from the Sloan Digital Sky Survey (SDSS) Data Release 7 \citep{Abazajian_2009}.}}
    \label{fig:sfzeta_line}
    \vspace{-0.2cm}
\end{figure*}

\subsection{A new SF\texorpdfstring{$\zeta$}{zeta} line for starburst galaxies with CRs}\label{sfzeta}

Our analysis shows that certain areas in the LINER or Seyfert domain of classical BPT diagrams, which are beyond the capabilities of pure photoionization star-forming models \citep{2001Kewley}, can instead be explained when typical CR ionization rate values, as those observed in starburst or expected in radio galaxies, are taken into account. For instance, several regions in NGC\,253, Centaurus A, and NGC\,1068 exhibit line ratios consistent with SF plus $\zeta_\text{CR}$ models (red contour in Fig.~\ref{fig:BPT_ALL}). Thus, we propose new maximum starburst lines for BPT diagrams to account for the potential contribution of CRs, which may alter the low-ionization line to H$\alpha$ ratios beyond the boundaries defined by \citet{2001Kewley}.

\eliz{To define the new maximum starburst lines, we used our SF models with densities in the $1 \leq \log n_\text{H} \leq 3$ range, which are typical values observed in star-forming regions \citep[e.g.][]{Perez_2003}. Subsequently, we consider only the area including 90\% of the 2D probability density kernel associated with the model distribution 
(red contour in Fig.~\ref{fig:BPT_ALL}), to exclude extreme values from the definition of the new boundaries. 
The new maximum line for starburst plus CR contribution, hereafter referred as SF\texorpdfstring{$\zeta$}, is obtained by fitting the 90\% contour boundary in each case, taking also into account the \citet{2001Kewley} model values at the lowest [\ion{N}{ii}]/H$\alpha$, [\ion{S}{ii}]/H$\alpha$, and [\ion{O}{i}]/H$\alpha$ ratios. For this purpose, we define the following $X$ and $Y$ variables:
\begin{equation}
    X = \left\{ \rm\frac{[\ion{N}{ii}]}{H\alpha}, \frac{[\ion{S}{ii}]}{H\alpha},\frac{[\ion{O}{i}]}{H\alpha} \right\}, \quad Y = \rm \frac{[\ion{O}{iii}]}{H\beta}
\end{equation}
while the relationship between them is characterized by the equation:
\begin{equation}
    X = \left(a Y^3 + b Y^2 + c Y + d\right) \, \mathrm{e}^{f Y}.
\end{equation}
The best-fit parameters obtained for the different BPT diagrams are listed in Table~\ref{tab:line_parameters}. The proposed SF\texorpdfstring{$\zeta$}{zeta} boundaries, shown by the solid red lines in Fig.~\ref{fig:sfzeta_line}, can be used to distinguish regions that are unambiguously dominated by AGN photoionization from those that could be explained by starburst plus CR ionization, up to values of about $\zeta_\text{CR} \lesssim 10^{-12}\, \rm{s^{-1}}$. The latter scenario is perhaps less likely when the line fluxes are measured within large apertures in relatively quiescent galaxies, as suggested by the lack of SDSS galaxies near the upper side of the SF\texorpdfstring{$\zeta$}{zeta} line in Fig.~\ref{fig:sfzeta_line}. However, the new separation line is more appropriate for line fluxes derived from resolved regions within galaxies, since the emission lines from low-excitation gas in these cases can be significantly affected when the regions are exposed to nearby sources with high CR rates. This is the case for some regions in the center of NGC 253, which are located above the Kewley line in Fig.~\ref{fig:sfzeta_line}, suggesting an additional source of excitation apart from star formation. According to the new SF\texorpdfstring{$\zeta$}{zeta} line, these regions are consistent with starburst plus CR ionization.}

\begin{table}[!!!h] 
\caption{\eliz{Parameters defining the SF$\zeta$ boundaries in BPT diagrams, i.e., the maximum contribution predicted for models including ionization from star-forming regions and CRs.}}\label{tab:line_parameters}
\centering
\setlength{\tabcolsep}{5.pt}
\begin{tabular}{c c c c c c} 
\hline\\[-0.2cm] 
\textbf{\emph{X}} & \textbf{\emph{a}} & \textbf{\emph{b}} & \textbf{\emph{c}} & \textbf{\emph{d}} & \textbf{\emph{f}} \\
\hline\\[-0.3cm]
$\rm [\ion{N}{ii}]/H\alpha$ & -0.600 & 1.272 & -0.980 & 0.273 & 4.313 \\[0.03cm]
$\rm [\ion{S}{ii}]/H\alpha$ &-0.408 &  0.090 & -1.783 &  1.245 &  0.893 \\[0.03cm]
$\rm [\ion{O}{i}]/H\alpha $ & -0.399 &  0.119& -1.568&  0.865&  0.993 \\[0.05cm]
\hline                  
\end{tabular}
\vspace{-0.3cm}
\end{table}




\subsection{Further remarks and considerations}\label{subsec:extra}

Our analysis shows that the addition of CRs to photoionization models can significantly increase the nebular emission from low-ionization species, due to the temperature increase in deep gas layers located beyond the Str\"omgren radius (see Section \ref{subsec:structure_plots}). The predicted temperatures drop rapidly as $\zeta_\text{CR}$ decreases, reaching $\lesssim 10^3\, \rm{K}$ for $\lesssim 10^{-13}\, \rm{s^{-1}}$ (Fig.~\ref{fig:stab}). At such low temperatures, the emissivities of optical nebular lines decrease by orders of magnitude, because the collisions do not have enough energy to populate the energy levels involved in these transitions. However, the infrared nebular lines are produced by transitions from levels located much closer to the ground state, which become populated at low temperatures ($\lesssim 1000\, \rm{K}$), and therefore their emissivities have a negligible dependency with the temperature \citep[e.g. fig.~1 in][]{Fernandez2021}. Thus, IR lines from low-ionization species should be more sensitive to the effect of CR heating. We will further study this aspect in a future work.



Beyond their role on nebular emission lines, CRs also play a significant role in the dynamical stability of gas clouds and the regulation of star formation \citep{Naab_2017,Girichidis_2024}. Considerable progress has been achieved in simulating the influence of CRs as a driving mechanism for both warm and cool outflows \citep{Breitschwerdt_1991,Armillotta_2024,Montero_2024}. Simulations of CR driven winds are nowadays 3D magnetohydrodynamical (MHD) and quite elaborate, taking into account magnetic ﬁelds, radiative cooling, self-gravity, star formation, the dynamical role of CRs injected by SNe, and CR transport processes—anisotropic diffusion and CR streaming along the magnetic ﬁelds \citep{Uhlig_2012,Rosdahl_2015,GRICHIDIS,Peschken_2023,Armillotta_2024}. Simulations like this support that the presence of CRs quenches star formation due to them being able to move cooler gas away from the galactic disk \citep{Jub,Ruszkowski_2017, Peschken_2023}. Our work supports the idea that CRs might be able to sustain ionization in outflowing material farther away, potentially extending the observable impact and physical reach of outflows. We found that CRs deposit energy and sustain higher temperatures in deeper parts of clouds, and by doing so, they possibly assist in their dispersion and therefore in the quenching of star formation processes. 


\section{Summary}\label{summary}

In this work, we use photoionization simulations with \textsc{Cloudy} to investigate the impact of CRs on the ionized gas phase of AGN and starburst galaxies. The models sample a wide range in density ($1$ to $10^4\,\rm{cm^{-3}}$), ionization parameter ($-3.5 \leq \log  U \leq -1.5$), and CR ionization rate ($10^{-15}$ to $10^{-12}\, \rm{s^{-1}}$), and are compared with VLT/MUSE spectroscopic observations of two jet-dominated AGN (Centaurus A and NGC 1068) and a prototypical starburst galaxy (NGC 253).

We find that CR rates of the order of $10^{-13} \,\rm{s^{-1}}$ are able to significantly affect the thermal structure of the ionized gas by creating a secondary low-ionization layer of warm ionized gas deep within the cloud, and far beyond the photoionization-dominated region surrounding the central source. This secondary ionized gas layer produces bright emission from low-ionization transitions such as [\ion{N}{ii}]$\lambda$6584\AA, [\ion{S}{ii}]$\lambda \lambda$6716,6731\AA, and [\ion{O}{i}]$\lambda$6300\AA, whereas [\ion{O}{iii}]$\lambda$5007\AA, H$\alpha$, and H$\beta$ are marginally enhanced. These findings are supported by analyzing the line emissivities of these transitions as a function of cloud depth. Thus, high CR ionization rates can significantly affect classical line-ratio-based diagnostics such as BPT diagrams. CR ionization rates of the order of $10^{-13}\, \rm{s^{-1}}$ have been measured in both jet-dominated AGN and strong starburst nuclei, which is in agreement with the values adopted in our simulations.

In contrast with pure photoionization models for NLR, which require between two and three times the solar metallicity to reproduce the Seyfert loci in BPT diagrams, simulations for AGN photoionization with high CR ionization rates are able to reproduce the Seyfert distribution in BPT diagrams with solar-like metallicities due to the CR boost of low-ionization transitions. Metallicity estimates derived for the regions extracted from the MUSE datacubes in our sample of galaxies suggest solar to subsolar values ($0.4$--$1.2\, \rm{Z_\odot}$; see Section \ref{subsec:param_space} and Table~\ref{tab:parameters}). Additionally, our photoionization simulations of star-forming regions with high CR ionization rates can also explain the presence of non-AGN sources in the LINER domain. \eliz{Consequently, we propose a new maximum starburst SF$\zeta$ boundary (Section~\ref{sfzeta}) that extends the previous limit defined by \citet{2001Kewley} to distinguish regions dominated by AGN photoionization from those that can be explained by star formation plus CR ionization.}

The results of this study suggest that the gas ionization and heating of the ISM can be largely affected in environments with high CR rates. In particular, CRs can play a major role in the excitation of low-ionization nebular gas in environments close to jet-dominated AGN and starburst galaxies, where nonthermal processes are important. SNRs originated by strong star-forming activity and particle acceleration and shocks along AGN jets are the main sources of CRs in galaxies. \eliz{CRs can also significantly impact the metallicity and physical parameters derived from line ratios involving low-ionization and neutral species. Therefore, future studies should consider the potential contribution of CR heating when high rates of these particles are expected.}

\begin{acknowledgements}
The authors are grateful to the anonymous referee for their insightful and constructive comments, for suggesting the inclusion of the SF$\zeta$ line, and for overall improving this work. EK extends heartfelt gratitude to Dr. Steven Tingay and Dr. Emil Lenc for generously providing VLA data of Centaurus A, and Dr. Isaac Mutie for providing the combined e-Merlin plus VLA data of NGC 1068. EK is thankful to Prof. Despina Hatzidimitriou and Prof. Apostolos Mastichiadis for insightful discussions and their guidance and to Dr. Christophe Morisset for all his help regarding \textsc{Cloudy} and py\textsc{Cloudy}. EK acknowledges full financial support by the State Scholarships Foundation (IKY) from the proceeds of the "N. D. Chrysovergis" bequest. JAFO acknowledges financial support by the Spanish Ministry of Science and Innovation (MCIN/AEI/10.13039/501100011033), by ``ERDF A way of making Europe'' and by ``European Union NextGenerationEU/PRTR'' through the grants PID2021-124918NB-C44 and CNS2023-145339; MCIN and the European Union -- NextGenerationEU through the Recovery and Resilience Facility project ICTS-MRR-2021-03-CEFCA. Based on data obtained from the ESO Science Archive Facility with DOI: \url{https://doi.org/10.18727/archive/41}. 
Funding for the Sloan Digital Sky Survey V has been provided by the Alfred P. Sloan Foundation, the Heising-Simons Foundation, the National Science Foundation, and the Participating Institutions. DSS acknowledges support and resources from the Center for High-Performance Computing at the University of Utah. SDSS telescopes are located at Apache Point Observatory, funded by the Astrophysical Research Consortium and operated by New Mexico State University, and at Las Campanas Observatory, operated by the Carnegie Institution for Science. The SDSS website is \url{www.sdss.org}. 
\end{acknowledgements}


%
%

\bibliographystyle{aa.bst} 
\bibliography{name} 


\begin{appendix} 
\section{Galaxies}\label{appendix_gal}
\subsection{Centaurus A}

NGC 5128, commonly known as Centaurus A is a nearby Seyfert galaxy and among the closest AGN, and it is situated at a distance of $\sim$3.84 Mpc \citep{Rejkuba_2004, Harris_2010}. In this AGN ionized, atomic, neutral, and molecular gas is oriented along the dust-lane and is observed at many wavelengths \citep{Morganti_2010}. 

Two subrelativistic jets that span a few kiloparsecs have been detected in radio and X-ray observations of Centaurus A \citep{Hardcastle_2003, Kraft_2008}. The jet structures have been resolved down to subparsec scale by utilizing Very Long Baseline Interferometry (VLBI) and Event Horizon Telescope (EHT) observations \citep{Tingay_1998, Janssen_2021}. The jet's northern side is more prominent than its southern one. From the inner lobes $\sim$5 kpc up to the outer lobes at 250 kpc from the nucleus, the jet expands into radio lobes \citep{Israel_1998, Eilek_2014}.

The jet and the interstellar medium (ISM) interact in a complicated way at this source \citep{Santoro_2015, Salome_2016}. Aligned with the jets' direction, there is a $\sim$500 pc area of ionized gas with  [\ion{O}{iii}]$\lambda$5007\AA\, emission \citep{Kraft_2008, Sharp_2010}. From far-infrared (FIR) emission lines in ionized gas, as well as in neutral atomic and molecular gas phases, an outflow is observed, spanning over $\sim$200 pc  \citep{Israel_2017}.


\subsection{NGC 1068}

NGC 1068 is another one of the closest and brightest 
active galaxies, classified as a Seyfert 2, located at a distance of $\sim$14.4 Mpc. NGC 1068 played a key role in the establishment of the unified model for AGN \citep{Bland_Hawthorn_1997}, which reconciles the two populations of Seyfert galaxies (types 1 and 2) to a simple orientation effect of an obscuring torus, as it shows broad optical permitted lines in its polarized spectrum.  It has a relatively high bolometric luminosity of $\rm L_{bol} \sim 2.6\times 10^{44} erg\, s^{-1}$ \citep{spinoglio2024}. 

From X-ray to radio wavelengths, NGC 1068 is known to contain radio jets \citep{Bland_Hawthorn_1997, Mutie_2024}. Its activity appears to be limited to ionization cones centered on the AGN, which cross the disk plane at an inclination of $\sim45^\circ$ and are roughly aligned with the inner radio jet \citep{Gallimore_1996, Bland_Hawthorn_1997}. This means that the nuclear radiation field immediately affects the interstellar medium of the disk situated in the north-east and south-west direction \citep{Cecil_1990, Bland_Hawthorn_1997, Gallimore_2004}. 

There are evident outflows in the ionized \citep{Cecil_1990, Barbosa_2014}, atomic \citep{Saito_2022} and molecular gas \citep{Garcia_2016, Gallimore_2016, Saito_2022}, possibly driven by the radio jet intercepting the disk \citep{Cecil_1990, Mutie_2024}. Strong starburst activity is seen in the galaxy, centered on a notable starburst ring that has a radius of around 1-1.5 kpc \citep{spinoglio2012, Garcia_2016}.

\subsection{NGC 253}

The nearby starburst galaxy NGC 253 has a distance of $\sim$ 3.94Mpc \citep{Rekola_2005}. NGC 253 is notable for massive dust areas surrounding its core and by an apparent bipolar outflow cone in both X-ray and ionized gas emissions \citep{Strickland_2002}. The outflows in this starburst galaxy are of two kinds, a multiphase outflow due to stellar feedback, also known as a superwind, and a disk wind \citep{Strickland_2002, Heesen_2009a}.

Studies conducted on NGC 253 have been over a broad range of frequencies, including polarization, HI observations and broadband radio continuum \citep{Ulvestad_1997, Carilli_1992, Lenc_2006, Heesen_2011, Lucero_2015}. In the infrared (IR) spectrum hotspots have been identified in the nuclear starburst area \citep{Galliano_2005, beck}. Previous radio observations detected over 60 distinct radio sources \citep{Ulvestad_1997}, most of which were oriented north-east–south-west and most likely connected to a 300 pc nuclear ring \citep{Fernandez2009}.

Thermal emission from electrons colliding with ions in the ionized interstellar medium (ISM) surrounding hot stars and nonthermal synchrotron emission from relativistic electrons traveling around the interstellar magnetic field lines are the primary causes of radio emission in starburst galaxies \citep{Kapiska_2017}. SNRs accelerate CRs and are the main origin of nonthermal emission, and they subsequently create a notable synchrotron radio halo in NGC 253 \citep{Carilli_1992, Kapiska_2017}. A CR rate of $\zeta_\mathrm{CR} \sim 10^{-12}\,\text{s}^{-1}$ (greater than $10^4$ times the average Galactic rate) was recently measured in the central giant molecular clouds (GMCs) of NGC 253, based on measurements of HCN and HNC emission of rotational transitions observed with the Atacama Large Millimeter/submillimeter Array via the ALCHEMI Large Program \citep{Behrens_2022}.

\subsection{NGC 1320}
NGC 1320 is a barred spiral galaxy situated at around $\sim 37.7$ Mpc \citep{Kraemer_2011}. Lying in the constellation of Eridanus, it shows star formation, that is primarily restricted to the narrow, inner Northwest spiral arm \citep{Robertis_1986}. It is a high-inclination galaxy that has been observed with HST in [\ion{O}{iii}]$\lambda$5007\AA \ and [\ion{N}{ii}]$6584$\AA \ \citep{Ferruit_2000}. It also contains a low luminosity, high-ionization, Seyfert type 2 active nucleus ($\rm L_{bol}=6.3\times 10^{43} erg\, s^{-1}$) \citep{Gallimore_2010, spinoglio2024}. It exhibits very narrow emission lines and a faint, featureless continuum \citep{Robertis_1986,Balokovi_2014}. 

\section{BPT diagrams with density notation}\label{appendix_den}

\begin{figure*}[ht]
    \centering
    \subfigure[$\zeta_\mathrm{CR}=10^{-14}\,\rm s^{-1}$.]{\includegraphics[width=\textwidth]{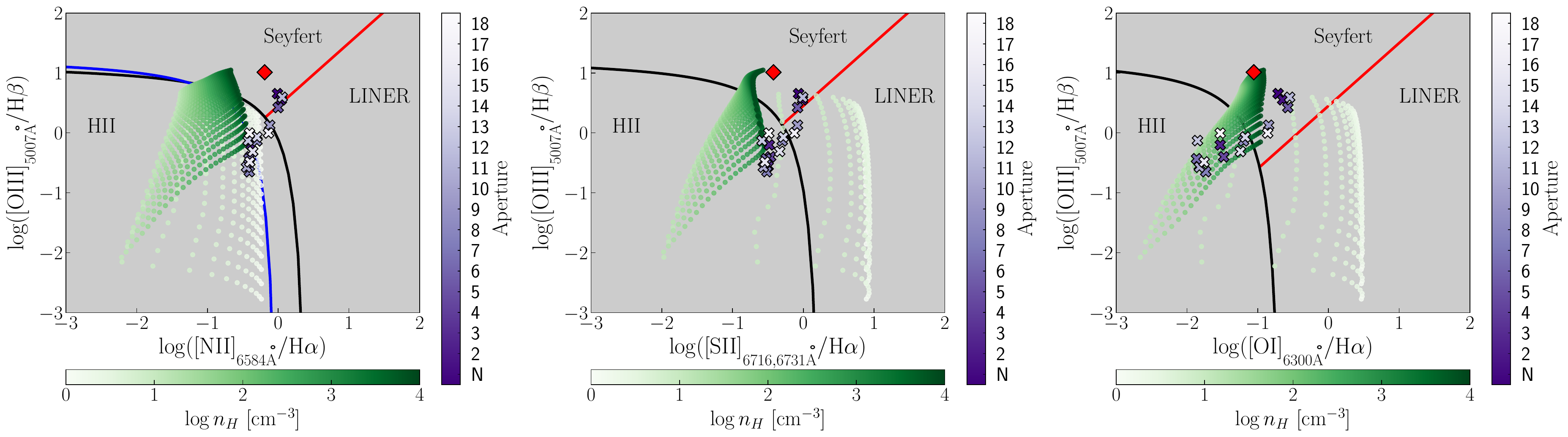}\label{subfig:nh_cent_14}}
    \subfigure[$\zeta_\mathrm{CR}=10^{-13}\,\rm s^{-1}$.]{\includegraphics[width=\textwidth]{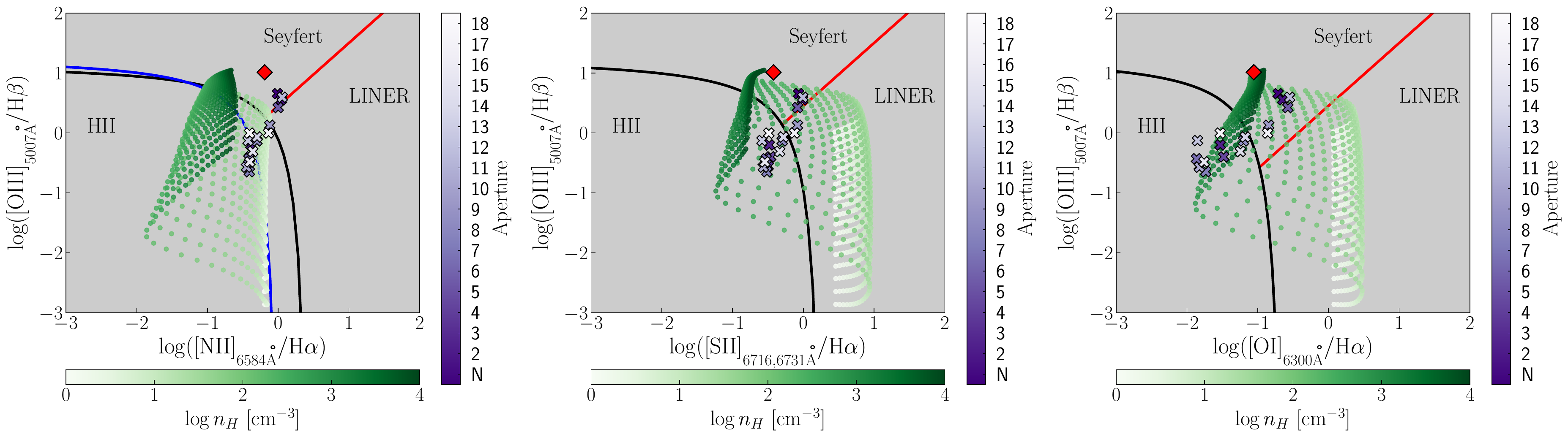}\label{subfig:nh_cent_13}}
    \subfigure[$\zeta_\mathrm{CR}=10^{-12}\,\rm s^{-1}$.]{\includegraphics[width=\textwidth]{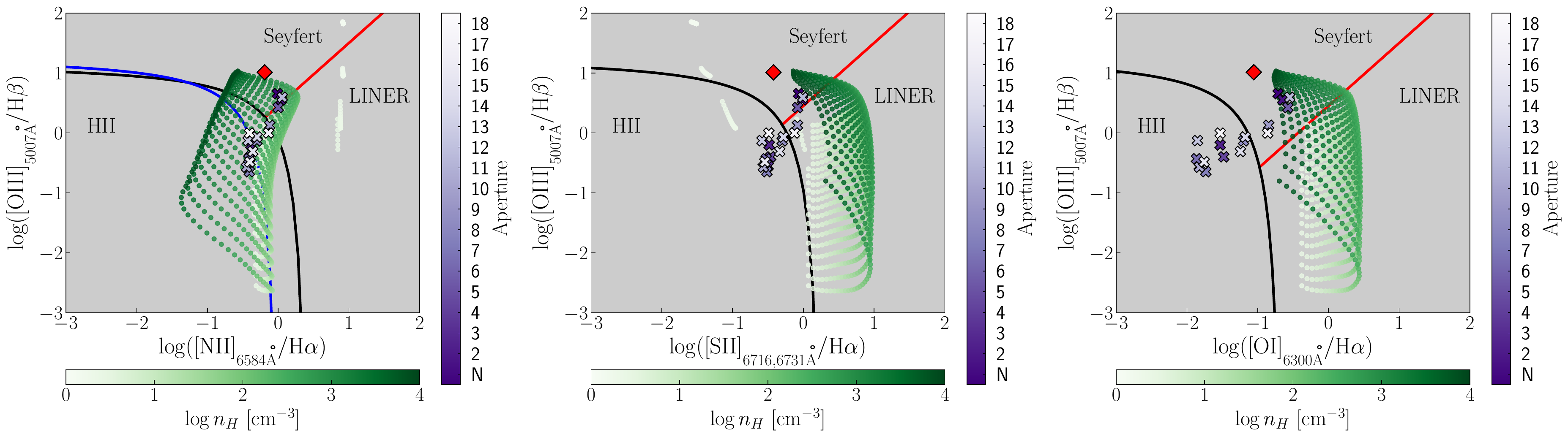}\label{subfig:nh_cent_12}}
    \caption{BPT diagrams with the AGN photoionization models compared with the observations from the selected apertures in Centaurus A (Fig.~\ref{subfig:ha_aper_cent}). The BPT diagrams for [\ion{N}{ii}], [\ion{S}{ii}], and [\ion{O}{i}] are shown on the left, middle, and right, respectively. The different shades of purple going from deep purple to pale lilac/white represent the ascending distance from the nucleus, as also noted with numbers, with "N" being the closest aperture. Also from white to deep green, the different shades of green represent the range of AGN models' densities $1 \leq n_{\rm H}\leq 10^4\,\rm{cm^{-3}}$. All the models shown have solar abundances. The red diamonds represent the measured line ratios for the photoionization-dominated Seyfert 2 nucleus in NGC 1320. The Kewley, Kauffmann, and Schawinski lines correspond to the black, blue, and red solid lines, respectively.} 
    \label{fig:cent_BPTs_nh}
\end{figure*}

\begin{figure*}[ht]
    \centering
    \subfigure[$\zeta_\mathrm{CR}=10^{-14}\,\rm s^{-1}$.]{\includegraphics[width=\textwidth]{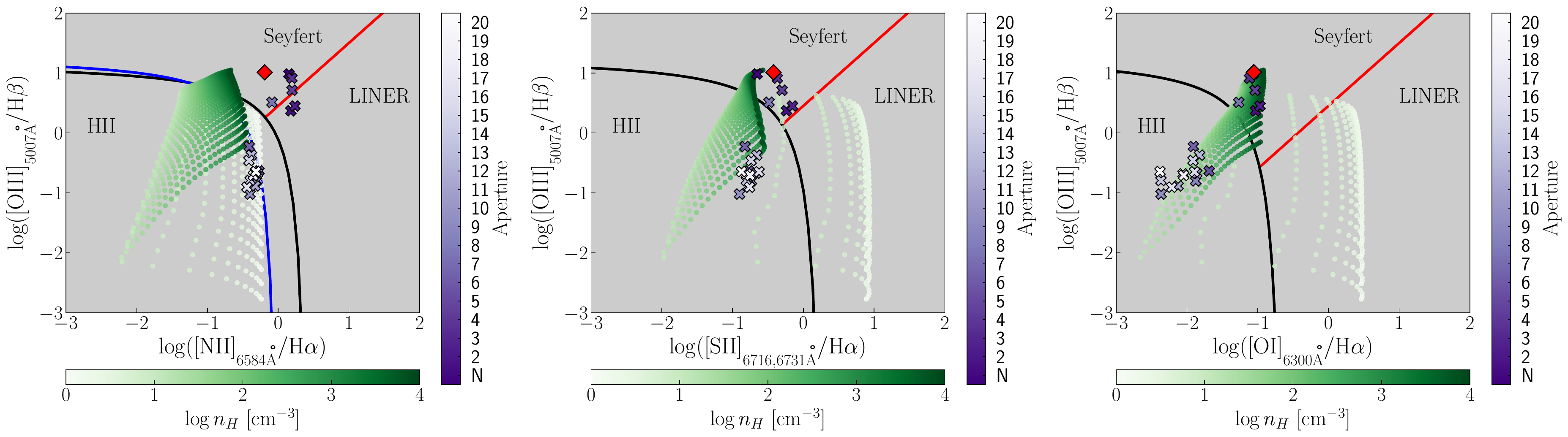}\label{subfig:nh_1068_14}}
    \subfigure[$\zeta_\mathrm{CR}=10^{-13}\,\rm s^{-1}$.]{\includegraphics[width=\textwidth]{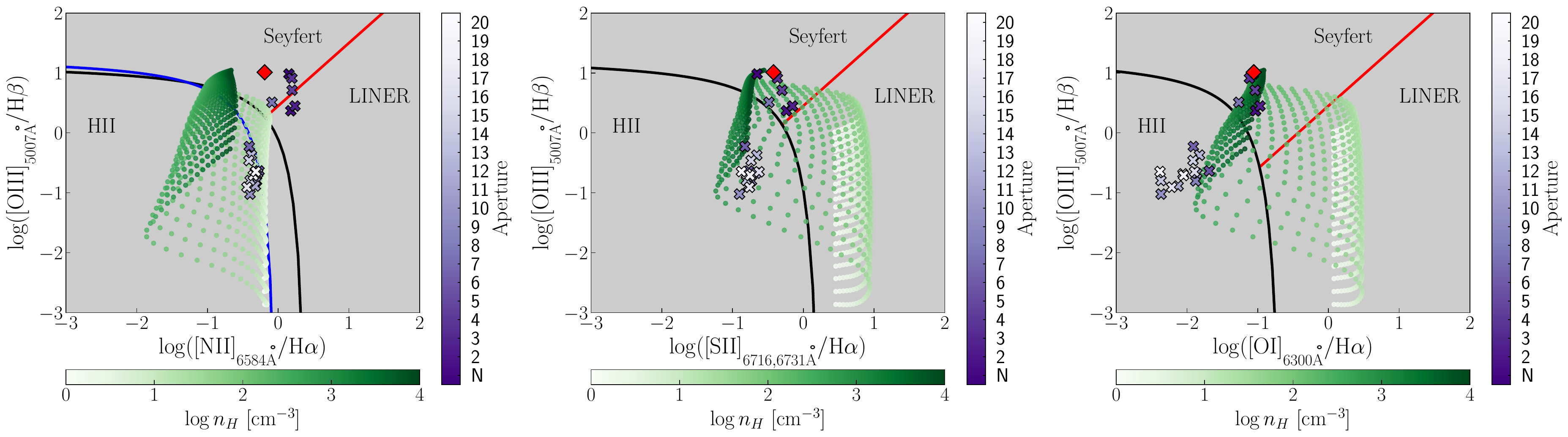}\label{subfig:nh_1068_13}}
    \subfigure[$\zeta_\mathrm{CR}=10^{-12}\,\rm s^{-1}$.]{\includegraphics[width=\textwidth]{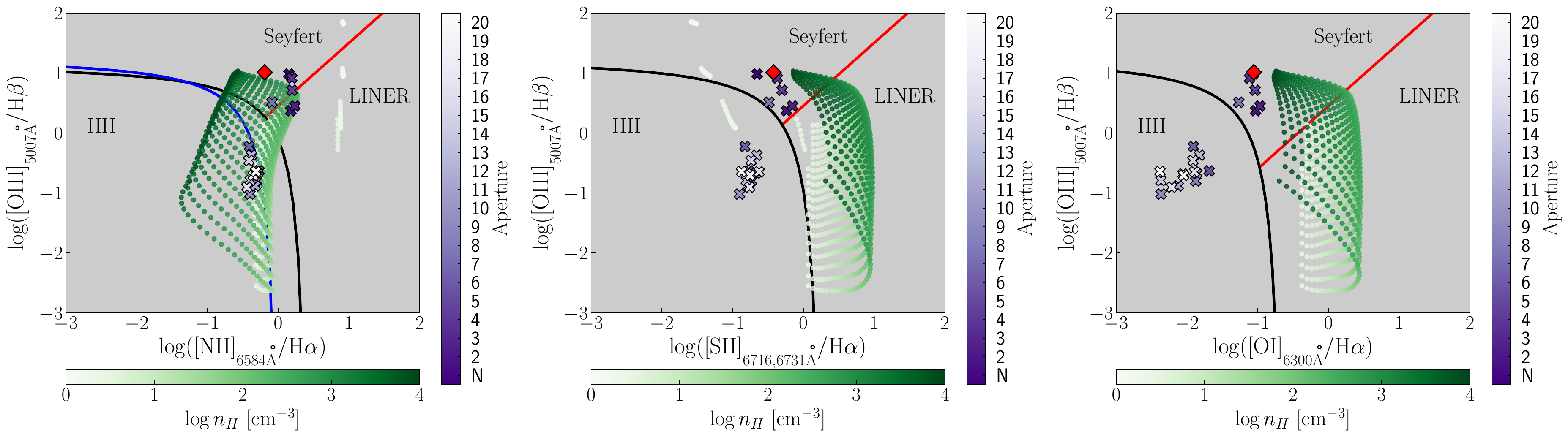}\label{subfig:nh_1068_12}}
    \caption{BPT diagrams with the AGN photoionization models compared with the observations from the selected apertures in NGC 1068 (Fig.~\ref{subfig:ha_aper_68}). The BPT diagrams for [\ion{N}{ii}], [\ion{S}{ii}], and [\ion{O}{i}] are shown on the left, middle, and right, respectively. The different shades of purple going from deep purple to pale lilac/white represent the ascending distance from the nucleus, as also noted with numbers, with "N" being the closest aperture. Also, from white to deep green, the different shades of green represent the range of AGN models' densities $1 \leq n_{\rm H}\leq 10^4\rm{cm^{-3}}$. All the models shown have solar abundances. The red diamonds represent the measured line ratios for the photoionization-dominated Seyfert 2 nucleus in NGC 1320. The Kewley, Kauffmann, and Schawinski lines correspond to the black, blue, and red solid lines, respectively.}\label{fig:1068_BPTs_nh}
\end{figure*}

\begin{figure*}[ht]
    \centering
    \subfigure[$\zeta_\mathrm{CR}=10^{-14}\,\rm s^{-1}$.]{\includegraphics[width=\textwidth]{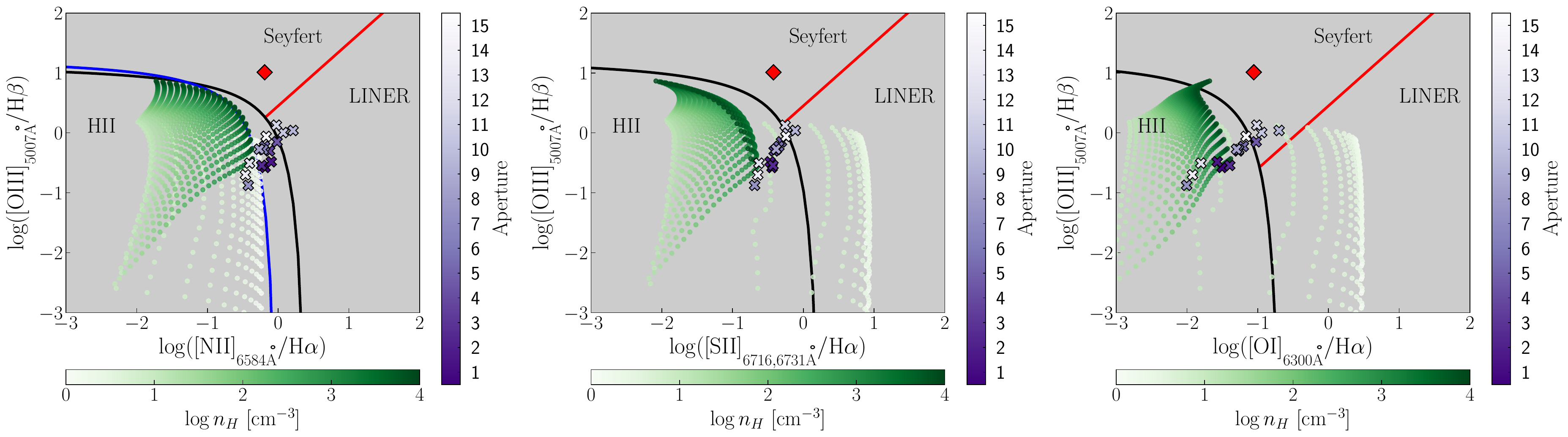}\label{subfig:nh_253_14}}
    \subfigure[$\zeta_\mathrm{CR}=10^{-13}\,\rm s^{-1}$.]{\includegraphics[width=\textwidth]{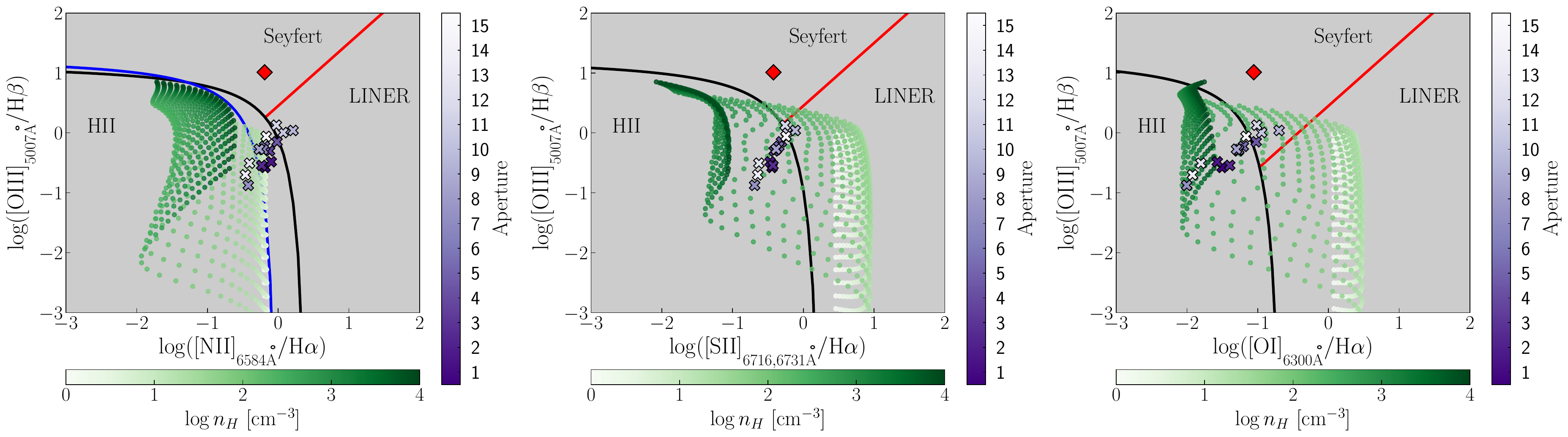}\label{subfig:nh_253_13}}
    \subfigure[$\zeta_\mathrm{CR}=10^{-12}\,\rm s^{-1}$.]{\includegraphics[width=\textwidth]{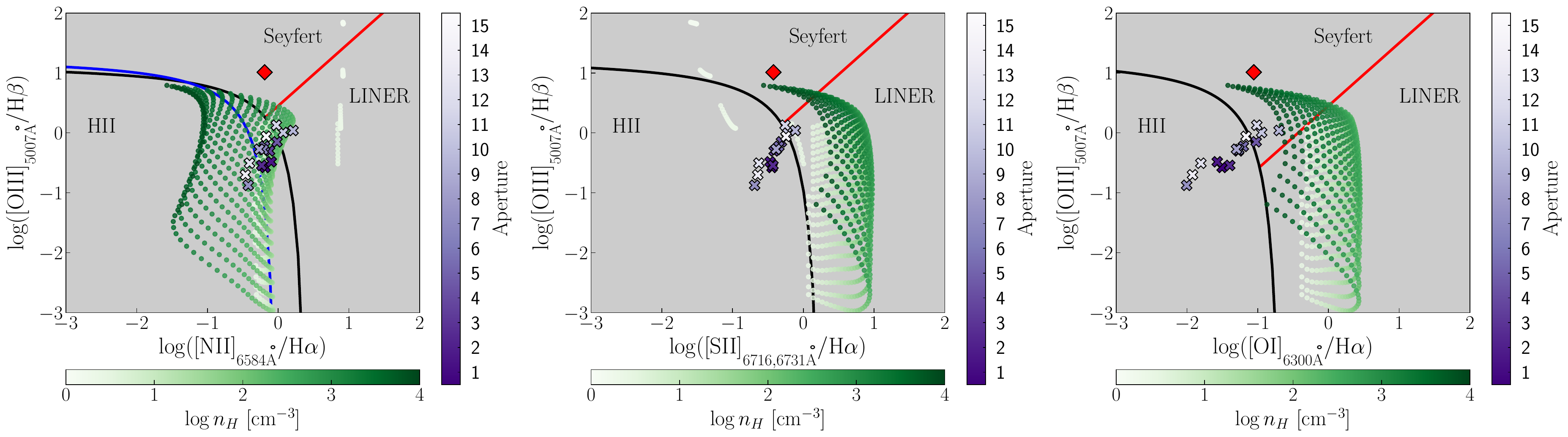}\label{subfig:nh_253_12}}
    \caption{BPT diagrams with the SF photoionization models compared with the observed line ratios from the selected apertures in NGC 253 (Fig.~\ref{subfig:ha_aper_253}). The BPT diagrams for [\ion{N}{ii}], [\ion{S}{ii}], and [\ion{O}{i}] are shown on the left, middle, and right, respectively. The different shades of purple going from deep purple to pale lilac/white represent the ascending distance, as also noted with numbers, with "1" being the most central aperture. Also from white to deep green, the different shades of green represent the range of SF models' densities $1 \leq n_{\rm H}\leq 10^4\rm{cm^{-3}}$. All the models shown have solar abundances. The red diamonds represent the measured line ratios for the photoionization-dominated Seyfert 2 nucleus in NGC 1320. The Kewley, Kauffmann, and Schawinski lines correspond to the black, blue, and red solid lines, respectively.}\label{fig:253_BPTs_nh}
\end{figure*}

Our models span between $0 \leq \log  n_{\rm H}\leq 4$ from white being the smaller initial hydrogen density to forest green being the highest (Figs. \ref{fig:cent_BPTs_nh}, \ref{fig:1068_BPTs_nh}, and \ref{fig:253_BPTs_nh}), and we can also see the range of the same models over the ionization parameter $-3.5 \leq \log  {\rm U} \leq -1.5$ from white being the smaller ionization parameter to deep red being the highest (Figs. \ref{fig:cent_BPTS_U},  \ref{fig:1068_BPTS_U}, and \ref{fig:253_BPTS_U}). The same diagrams are produced for different CR ionization rates $-14 \leq {\rm log~\zeta_\mathrm{CR}\leq -12}$ for all galaxies.

\end{appendix}

\end{document}